\documentclass[12pt,a4paper,twoside,openright]{report}

\usepackage[UKenglish]{babel}
\usepackage[intlimits]{amsmath}
\usepackage{graphicx}
\usepackage{amsfonts}
\usepackage{amssymb}
\usepackage{wasysym}
\usepackage{dsfont}
\usepackage{units}
\usepackage[usenames,dvipsnames]{xcolor}
\usepackage{colortbl}
\usepackage{titlesec}
\usepackage{framed}
\usepackage{psboxit}
\usepackage{cite}
\usepackage[linktocpage=true]{hyperref}
\hypersetup{linkbordercolor=red,citebordercolor=blue,urlbordercolor=cyan}
\usepackage{minitoc}
\usepackage[left=3cm,right=3cm,top=3cm,bottom=2.5cm,nohead,foot=1.2cm]{geometry}
\usepackage[toc,page]{appendix}
\usepackage{ulem}
\usepackage{float}

\newcommand{\be}{\begin{equation}}
\newcommand{\ee}{\end{equation}}
\newcommand{\bea}{\begin{eqnarray}}
\newcommand{\eea}{\end{eqnarray}}
\newcommand{\ba}{\begin{array}}
\newcommand{\ea}{\end{array}}

\newcommand{\dq}{\frac{d^Dq}{(2\pi)^D}}
\newcommand{\sg}{\sqrt{\det\,g_{\mu\nu}}}
\newcommand{\sgb}{\sqrt{\det\,\bar g_{\mu\nu}}}

\newcommand{\csch}{\mathrm{csch} \,}

\def\thebibliography#1{\chapter*{\textcolor{BrickRed}{References}}\list
{[\arabic{enumi}]}{\settowidth\labelwidth{[#1]}\leftmargin\labelwidth
\advance\leftmargin\labelsep
\usecounter{enumi}}
\def\newblock{\hskip .11em plus .33em minus .07em}
\sloppy\clubpenalty4000\widowpenalty4000}

\begin{document}

\baselineskip 16pt plus 2pt minus 1pt

\dominitoc

\begin{titlepage}

~\vspace{0.7cm}

\centering
\begin{tabular}{cc}
\hspace{1cm} \includegraphics[width=3cm]{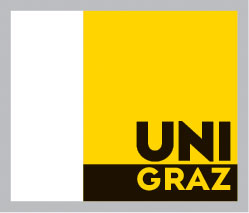} &
\hspace{3cm} \includegraphics[width=5.5cm]{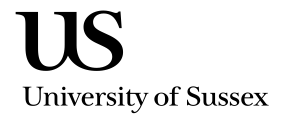} \\
\end{tabular}

\begin{center}
\rule{150mm}{0.8mm}
\end{center}

\begin{center}
\vspace{1cm}
{\bf\LARGE           RENORMALISATION GROUP      }\\ \vspace{0.5cm} 
{\bf\LARGE                	 FOR GRAVITY                 }\\ \vspace{0.5cm} 
{\bf\LARGE                               AND                           }\\ \vspace{0.5cm}
{\bf\LARGE           DIMENSIONAL REDUCTION       }\\ \vspace{3.0cm} 
\end{center}

~\vspace{-4cm}

\begin{center}
\rule{150mm}{0.8mm}
\end{center}

\vspace{1cm}

{\bf\large   MASTER THESIS } \vspace{1cm} 

zur Erlangung des akademischen Grades eines\\
             Master of Science\\
an der Naturwissenschaftlichen Fakult\"at der\\
      Karl-Franzens-Universit\"at Graz\\

\vspace{2cm}

{\Large      Nat\'{a}lia Alkofer, BSc       }\\

\vspace{3.4cm}

\begin{tabular}{rl}
{\large {\it{Advisor:}}}& {\large Prof.\ Dr.\ Daniel F.\ Litim }\\ [1mm]
{\large {\it{Co-advisor:}}}& {\large Priv.-Doz.\ Dr.\ Bernd-Jochen\ Schaefer}\\
\end{tabular}

\vspace{1.2cm}

{\large  2013        }\\
\vspace{0.5cm}

\end{titlepage}

\thispagestyle{empty}


\newpage
~
\newpage

\chapter*{Abstract}

The functional renormalisation group for the Einstein-Hilbert action is investigated
for the case of four infinite (or large) and one compact dimension.

The motivation for this study is given by the suggestion that gravity in more
than four dimensions might be able to solve the so-called hierarchy problem
of the Standard Model of elementary particle physics because the assumed extra 
dimensions allow to unify the true Planck scale with the electroweak scale.
In a first step the Einstein-Hilbert theory for the functional renormalisation group 
treatment of Quantum Gravity is introduced. To set the scene for the following
investigation including a compact dimension first some corresponding
calculations for four and more large dimensions are performed and the results
are found to be  in agreement with already known ones.

The equations determining the renormalisation group flows are then derived for the
case of one compact dimension. These dimensionally reduced renormalisation group
equations are then numerically solved for the case of four large and one compact dimension.
Results for the four- to five-dimensional crossover are then discussed employing 
two forms of the background field flow in
the Einstein-Hilbert theory for the case of one extra compact dimension.
Renormalisation group trajectories allowing for a significant lowering of the true
Planck scale to the electroweak scale are identified. The behaviour of the running 
gravitational coupling at the  four- to five-dimensional crossover  and the true Planck
scale is displayed. Potential phenomenological implications  are briefly discussed.

\thispagestyle{empty}


\newpage

\thispagestyle{empty}

\chapter*{Kurzfassung}

Die funktionale Renormierungsgruppe f\"ur die Einstein-Hilbert-Wirkung wird 
f\"ur den Fall von vier unendlich gro\ss en und einer kompakten Dimension
untersucht.

Die Motivation f\"ur diese Untersuchung liegt in dem Vorschlag begr\"undet, dass
Gravitation in mehr als vier Dimensionen in der Lage ist, das sogenannte 
Hierarchie\-problem des Standardmodells der Elementarteilchenphysik zu l\"osen,
da die hypothetischen extra Dimensionen es erlauben, die tats\"achliche Planck-Skala 
mit der elektroschwachen Skala zu identifizieren. In einem ersten Schritt wird hierzu
die Einstein-Hilbert-Theorie f\"ur die funktionale Renormierungsgruppe  der 
Quantengravitation eingef\"uhrt. Um einen Vergleich zur sp\"ateren
Untersuchung mit einer kompakten 
Dimension zu erm\"oglichen, werden zuerst Berechnungen in vier und h\"oheren
Dimensionen durchgef\"uhrt und die gefundenen Ergebnisse mit bereits be\-kannten 
Resultaten verglichen. 

Die den  Renormierungsgruppenflu\ss \ bestimmenden Gleichungen werden f\"ur den
Fall einer zus\"atzlichen kompakten Dimension  abgeleitet. Diese dimensional reduzierten
Renormierungsgruppengleichungen werden dann f\"ur vier nicht-kompakte und eine
kompakte Dimension numerisch gel\"ost.  Die entsprechenden Ergeb\-nisse f\"ur den 
vier- zu f\"unf-dimensionalen Crossover werden f\"ur zwei Formen des Hintergrundfeldflusses
in der Einstein-Hilbert-Theorie diskutiert.  Es werden
Renormierungsgruppen\-trajektorien pr\"asentient, die eine signifikante Er\-niedrigung der 
tats\"achlichen Planck-Skala bis hin zur elektroschwachen Skala erlauben. 
Das Verhalten der laufenden Gravitationskopplung beim 
vier- zu f\"unf-dimensionalen Crossover und bei der tat\-s\"achlichen Planck-Skala 
wird angegeben. Potentielle ph\"anomenologische Implikationen werden kurz diskutiert.

\thispagestyle{empty}


\newpage

\thispagestyle{empty}
\pagenumbering{roman}
\setcounter{page}{0}

\tableofcontents


\listoffigures


\newpage

\thispagestyle{empty}

\listoftables


\newpage
\thispagestyle{empty}
~
\newpage

\setcounter{page}{1}
\pagenumbering{arabic}

\chapter{{\color{BrickRed}{Introduction}}}
~\vspace{-10mm}

\rightline{\footnotesize{\it{``No question about quantum gravity is more difficult}}}
\rightline{\footnotesize{\it{than the question, ``what is the question?"."}}}
\rightline{\footnotesize{\it{(J.\ Wheeler)}}}
\minitoc

\section{Motivation}

Almost a hundred years ago, in 1915, Einstein proposed a theory of gravity. This theory is 
based on the principle of invariance of general coordinate transformations (diffeomorphism 
invariance), and Einstein called it therefore {\it Allgemeine Relativit\"atstheorie  (General 
Theory of Relativity)}  \cite{Einstein:1916vd}. It is a classical field theory which allows to 
understand gravity only on the basis of the geometry of space-time. In the last decades 
General Relativity (GR) has been experimentally verified from length scales ranging from $\mu
$m to cosmological distances. The corresponding coupling constant,   Newton's gravitational 
constant $G_N$, has been measured in laboratory experiments on scales of about 
50 $\mu$m to 1m, 
see {\it e.g.}, \cite{Hoyle:2004cw} but also the footnote on Table 1.1 in the Particle Physics Booklet July 
2012 \cite{Beringer:1900zz}. Nowadays we know that a further constant, the so-called cosmological 
constant $\Lambda$, has to be considered in the classical theory of gravity. Introduced by 
Einstein in the early twenties to allow for a static universe it was abandoned by him after 
Hubble's discovery of an expanding universe. As recent measurements indicate that this 
expansion is accelerating one reintroduces this constant for a phenomenological description 
of cosmology. The corresponding value of $\Lambda$ is so small that it plays a role only on 
cosmological distances. 

The classical theory of gravity, {\it i.e.}, GR, can be formulated via the Einstein-Hilbert (EH) 
action.  Varying this action with respect to the metric field provides Einstein's 
equation\footnote{Varying the EH action with respect to the metric and the Christoffel symbols 
provides Einstein's equation plus the metric compatibility constraint $D_\rho g_{\mu\nu}=0$ 
(Palatini formalism) and therefore leads to the same classical physics. This makes evident 
that not necessarily the metric coefficients alone are the dynamical degrees of freedom in a 
complete theory of gravity.}.
In this thesis I will choose the EH action, including the term with the  cosmological  constant, 
as a basis. More precisely, as it will be explained below, I will work within the EH theory in 
which the effective action is always projected on two terms \cite{Reuter:2001ag}.

Treating the EH theory as a Quantum Field Theory faces several problems. 
As an exhaustive description of the corresponding challenges is well beyond the scope of this 
thesis let me focus on the problem whose potential solution provides one of the starting points of 
this thesis: The perturbation series derived from the EH action is not renormalisable, {\it i.e.}, 
in every order of perturbation theory new types of divergencies appear which make then the 
perturbation series unpredictive. This can be inferred already from a simple power counting 
argument
(for details on the notation see Appendix~A): Newton's gravitational constant $G_N$ 
possesses in four space-time dimensions a mass dimension -2,  one has $G_N = 1/m_{\mathrm 
Pl} ^2$ with $m_{\mathrm Pl} \approx 10^{19}$ GeV \cite{Beringer:1900zz}.  The enormous size 
of this number tells us that in a standard picture with four space-time dimensions gravity is in 
the classical regime not only at cosmological, astrophysical and every-day length scales but 
also at every current experiment including those at the Large Hadron Collider (LHC). 
Nevertheless, in order to progress on our understanding of gravity and quantum physics a 
quantum theory of gravity is highly desirable. 

Returning to the problem of the potential non-renormalisibility of a gravity action it might 
actually turn out that the above described problem is one of the methods, {\it i.e.}, the 
perturbative expansion, and not one of the theories. 
Already in the 70's Weinberg conjectured that Einstein gravity becomes renormalisable when 
treated non-perturbatively \cite{Weinberg}. Based on different methods, but hereby especially 
on the Functional Renormalisation Group (FRG) \cite{Berges:2000ew,Polonyi:2001se,Litim:
1998nf,Pawlowski:2005xe,Gies:2006wv}, convincing evidence has been gathered for this fact 
which is known nowadays as asymptotic safety  scenario 
\cite{Reuter:1996cp,Lauscher:2001ya,Lauscher:2001rz,Lauscher:2002sq,Niedermaier:2006ns,Niedermaier:2006wt,Percacci:2011fr,Reuter:2012id,Falls:2013bv}. 
Hereby it is important to note that asymptotic safety is quite distinct from the asymptotic 
freedom of four-dimensional quantum gauge field theories which renders them perturbatively 
renormalisable. 
Asymptotic freedom requires the dimensionless running coupling(s)  of the theory to go to 
zero in the ultraviolet (UV). This special property allows then to re-express in the UV ({\it i.e.}, 
in practice well above a characteristic scale as, {\it e.g.}, $\Lambda_{\mathrm QCD}$) the 
perturbative series for a $S$-matrix element such that it contains only finite terms dependent 
on renormalised parameters, {\it i.e.}, all UV divergencies can be removed in a controlled way 
from the perturbative expressions. Therefore gauge field theories as, {\it e.g.}, Quantum 
Chromodynamics (QCD), are perturbatively renormalisable. Asymptotic safety shares with asymptotic 
freedom the requirement of the existence of a non-trivial UV  fixed point. However, in the asymptotic 
safety  scenario the coupling constant(s) can take any finite value. 
Candidate theories for the asymptotic safety scenario 
possess typically (a) coupling constant(s) with negative mass dimensions. Rendering it (them) 
dimensionless by multiplying appropriate powers of the renormalisation scale, and thus 
generating arbitrary numerical values in dependence of the employed units,  makes plain that 
a perturbative treatment close to an UV fixed point cannot work. Therefore we are confronted with 
the quest for an appropriate non-perturbative method.

The non-perturbative method chosen for this thesis is the FRG.\footnote{An approach which 
comes closest to Lattice Quantum Gravity is dynamical triangulation, see, {\it e.g.},  
\cite{Ambjorn:2012jv,Coumbe:2012qr} and references therein.}
Hereby, I will use the background field formalism and two forms of 
the FRG, an exact functional identity  and an approximated background field flow.
For the latter I use proper-time regularisation and for the exact identity, the Wetterich equation
\cite{Wetterich:1992yh}, Litim's optimised regulator 
\cite{Litim:2000ci,Litim:2001up,Litim:2001fd,Litim:2002cf,Litim:2001dt,Litim:2002qn}.  
Both forms of the  
FRG  can be described as a mathematical tool which allows to treat quantum fluctuations 
step-by-step by varying the resolution scale. Hereby one starts with the classical action in the 
UV and solves the differential equations for the $\beta$ functions of the theory, for details see 
Chapters 3 and 4 of this thesis.

Noting that gravity is not at all special in four dimensions (again in contrast to quantum gauge 
field theories which are exactly renormalisable in precisely four dimensions) I will   investigate 
first  the asymptotic safety scenario for a larger number of dimensions, hereby verifying previously obtained 
results \cite{Litim:2003vp,Fischer:2006fz}. These extra dimensions will then be compactified 
such that the corresponding length scales can be continuously decreased from infinity to 
zero. The corresponding observed change of the fixed point values of the coupling constants 
demonstrates the possibility of going continuously from higher-dimensional fixed point values to 
lower-dimensional ones.

However, considering compact extra dimensions has an important other motivation.  As 
mentioned above the enormous size of the Planck mass indicates that gravity will be in its 
classical regime up to truly huge energy scales. Together with today's knowledge on  
elementary particle physics this creates a severe problem in our understanding of the relation 
of gravity to electroweak (EW) physics. As the renormalisation of masses of scalar fields is not protected by 
any known mechanism, the question arises why the mass of the Higgs is only (approximately) 
125 GeV and not close to the Planck mass. This problem is called the hierarchy problem, for 
more details see Subsection~1.2.1 below.  Quite recently there has been an interesting 
speculation \cite{ArkaniHamed:1998rs,ArkaniHamed:1998nn}  which shifts the problem from 
the EW theory to gravity, resp., the geometry of space-time.
At short distances gravity is not tested below $10^{-4}$ m, so there is ample space for "new" 
physics. Especially, the Arkani-Hamed, Dimopoulos and Dvali (ADD) model \cite{ArkaniHamed:
1998rs,ArkaniHamed:1998nn}  suggests that there might exist ``large''  extra dimensions with 
up to a $\mu$m compactification scale. One can easily verify, see Subsection~1.2.3 below, that 
this would allow to unify the true gravity with the EW scale. In case such a scenario would be 
true this would not only solve the hierarchy problem\footnote{More precisely, it is solved on 
the expense that one would like to explain the size of the extra dimensions.} but opens up the 
exciting possibility to probe quantum gravity at the LHC.  For a phenomenological description 
of corresponding cross sections ({\it e.g.}, graviton-mediated Drell-Yan processes) it is vital to 
know how the gravitational coupling runs in the energy domain relevant to the experiment 
\cite{Litim:2007iu,Litim:2007ee,Gerwick:2011jw}. In this thesis it is intended to describe the 
first few steps in a potential calculation of the scale dependence of the gravitational coupling 
constants in a space-time with compactified but ``large'' extra dimension(s).


\section{Extra Dimensions}

Before describing the ADD model \cite{ArkaniHamed:1998rs,ArkaniHamed:1998nn}  and extra 
dimensions one of the original motivations for it will be described in the following subsection.


\subsection{The Hierarchy Problem}

Last year CERN announced the discovery of a particle with a mass of approximately 125 GeV. 
It is generally expected that this particle is the long-searched Higgs boson. If confirmed this 
will be the only fundamental scalar field in the Standard Model (SM) of particle physics. The 
discrepancy between its mass and the Planck or a possible Grand Unified Theory (GUT) scale 
leads to a fine tuning problem which is called the hierarchy problem\footnote{As a matter of 
fact there are several hierarchy problems in physics. However, it is generally agreed upon 
that the question why the EW scale is so much smaller than the cutoff scale given by the 
Planck or GUT scale, is the most prominent hierarchy problem in particle physics.}.

Contrary to the case of Dirac fermions and vector particles where either chiral symmetry or 
gauge invariance limits the divergence of mass renormalisation to a logarithmic one, the mass 
renormalisation of a scalar is quadratic:

\be
m_H^2 \,=\, m_{H,0}^2 + \delta m_H^2 \quad \mathrm{with} \quad \delta m_H^2 \propto 
\Lambda_{UV}^2 \,\,. 
\ee

Therefore to achieve the physical Higgs mass there must be a very precise cancellation 
between $m_{H,0}^2$ and $\delta m_H^2$.

Several explanations have been proposed to solve the hierarchy problem, {\it e.g.}, in 
technicolor models the Higgs is a bound state of TeV scale techniquarks. In this thesis, 
however, I build on another suggestion to circumvent the hierarchy problem, namely large 
extra dimensions.


\subsection{Exploring Extra Dimensions}

The above mentioned extra dimensions are assumed to be large as compared to the EW 
length scale. This is due to the following argument: In $D=4+n$ dimensions (with Euclidean 
signature) the EH action\footnote{For simplicity the cosmological constant is neglected in this 
argument.} is given by   
\be
S_{EH}^{(D)} \,=\, - \frac{M_D^{n+2}}{16\,\pi} \, \int d^{4+n}  x \, \sqrt{\det\, g_{\mu\nu}^{(D)}} \,\, 
R^{(D)} \,=\, - \frac{V_n \, M_D^{n+2}}{16\,\pi} \, \int d^{4}  x \, \sqrt{\det\, g^{(4)}_{\mu\nu}} \,\, 
R^{(4)}
\ee
with $M_D$ being the Planck scale in $D$ dimensions and $V_n$ the volume taken by the $n$
extra dimensions.
Here it was assumed that matter is restricted to our four-dimensional hypersurface, {\it i.e.}, 
the space constituted by the extra dimensions is flat. It is evident that one reproduces 
Einstein gravity if one identifies $M_{Pl}^2 = M_D^{n+2}\,V_n$.

Assuming the simplest compactification, namely a torus with periodic boundary conditions 
such that $x_i = x_i + L$ with $i = 5,6,\ldots, 4+n$, one has $V_n = L^n$. Thus, if the size of the 
extra dimensions is large as compared to the EW scale the fundamental scale of gravity can 
be as low as the EW scale. To make this quantitative we rewrite the above formula as
\be
M_{Pl}^2 = M_D^2 \,(M_D \, L)^n \,\,.
\label{eq:1.3}
\ee

To avoid the hierarchy problem one requires $M_D$ to be of the order of the EW scale, {\it i.e.}, 
1 TeV. As $M_{Pl}\approx 10^{16}$ TeV this implies $(M_D \, L)^n \approx 10^{32}$ and therefore
an enormously large separation of scales, 
\be
1/L \ll M_D\approx 1 {\mathrm {TeV}}  \approx m_{EW} \, .
\ee

From this discussion it is evident  that for $n=1$ there are approximately 32 orders of magnitude
between  $1/L$ and $M_D$.
One recognises then that $n=1$ is excluded because this would imply 
deviations from Newtonian gravity over solar system distances: Using $\hbar  c = 1 =
0.2\cdot 10^{-18} \,{\mathrm {m\,TeV}}$ one estimates then $L\approx 0.2 \cdot 10^{14} {\mathrm m}$.
 Nevertheless, for illustrational 
purposes I will use $n=1$ in this thesis, and leave the cases $n\ge2$ for later studies.


\subsection{The ADD Model}

In the simplest version of the ADD model \cite{ArkaniHamed:1998rs,ArkaniHamed:1998nn} of 
large extra dimensions it is required that the fields of the SM are constraint to a four-
dimensional hypersurface (our visible universe) whereas gravity propagates to additional 
dimensions. This subspace of extra dimensions has to be orthogonal to the four-dimensional 
hypersurface, and it is assumed to be compact. Nevertheless, due to Einstein's equation the 
space spanned by the extra dimensions is flat.

As detailed above the ADD model provides a solution to the hierarchy problem of particle 
physics. However, the prize to be paid is, at least, two further undetermined parameters, 
namely $n$ and $L$. From the theoretical point of view the problem of explaining the fine-
tuning of the Higgs mass has been  reformulated to the question why extra dimensions with a 
typical size smaller than 0.1 mm and larger than 1 fm (depending on the number of compactified 
extra dimensions)  exist. From the experimental point of view the ADD model is highly 
interesting: It leads to measurable consequences for scattering processes at LHC energies, 
see {\it e.g.}, 
\cite{Litim:2007iu,Litim:2007ee,Litim:2011cp,Gerwick:2011jw,Hewett:2007st,Giudice:1998ck,Wang:2012qi}.
 However, details of related observables will depend on the coupling 
strength of gravitons at the EW and the fundamental Planck scale. To this end one is 
interested in the Renormalisation Group (RG) running of Newton's gravitational coupling in the 
ADD model geometry. 

\setcounter{footnote}{3}
\begin{figure}[t]
\centering
\includegraphics[width=\textwidth]{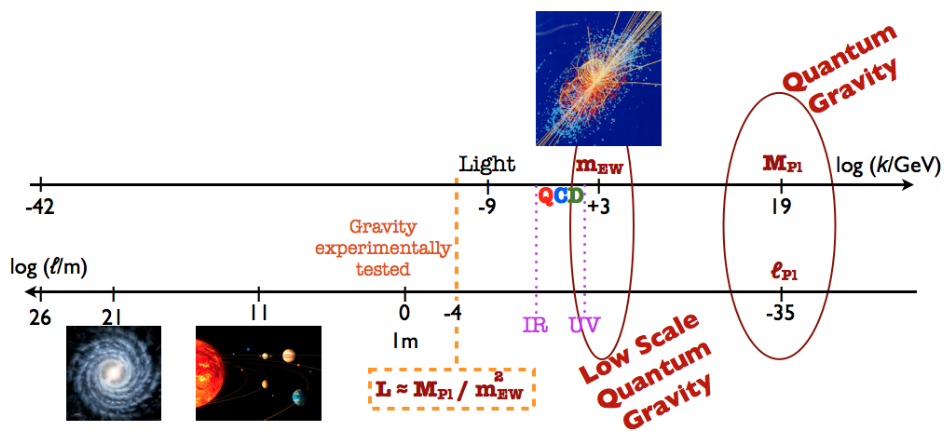}
\caption[An illustration of different scales relevant for gravity.]{An illustration\footnotemark[6] of 
different scales relevant for gravity. On the right side, 
{\it i.e.}, for small distance and large momentum scales one has the (conventional) 
quantum gravity region. On the left side, with the distance scale equal to the Hubble scale 
there is the extreme infrared (IR) region.}
\label{fig:1.1}
\end{figure}
\footnotetext[6]{Images credits: NASA, ESA and CERN.}

The situation is summarised  in Fig.\ \ref{fig:1.1}:  A description of gravity in the conventional 
picture needs to cover at least sixty orders of magnitude from the Planck scale in the UV (in 
the figure on the right side) to the Hubble scale in the IR.  Gravity is experimentally tested 
above scales of approximately 0.1 mm. This allows to introduce additional compact 
dimensions with corresponding sizes denoted by $L$ as displayed in the figure.  (The 
presented formula is valid for $n=2$.) This results in a quantum gravity 
scale of the order of the EW scale $m_{EW}$. This region is marked as ``Low Scale Quantum 
Gravity" in Fig.\ \ref{fig:1.1}.  It will be tested by the LHC in the coming years. An important 
constraint for the purpose of this thesis is the following: To achieve the lowering of the 
quantum gravity scale by 15 - 16 orders of magnitude one needs to have 
\be
1/L \ll m_{EW}.
\ee

\goodbreak


\section{Asymptotic Safety Scenario}

Here I will briefly review a few general aspects of the asymptotic safety scenario and a non-trivial UV fixed point in 
Quantum Einstein Gravity  \cite{Weinberg}.


\subsection{Weinberg's Conjecture and Criteria}

The first and main criterium of the asymptotic safety scenario is the existence of a non-trivial fixed point at short 
distances. Such an UV fixed point can be considered as the generalisation of a non-interacting UV fixed point 
in asymptotically free theories as, {\it e.g.}, QCD. Whereas asymptotically free theories are 
perturbatively renormalisable, the non-vanishing value of the coupling constant(s) at the UV 
fixed point exclude a perturbative treatment. Secondly, it is indispensable that the UV fixed point is 
continuously connected with the IR behaviour of the theory by a well-defined RG trajectory.   
Otherwise, the UV fixed point would be disconnected from the observed semi-classical physics at 
large distances. Thirdly, it is mandatory that the UV fixed point possesses at most a finite number of 
IR unstable directions. Or else, the predictive power would be lost: An infinite number of 
unstable directions would necessitate the input of infinitely many parameters to obtain a 
trajectory reaching the IR limit.

These three criteria being fulfilled the UV fixed point together with a specific RG trajectory provides a 
fundamental definition of the theory. Such a theory is then called asymptotically safe. The 
existence of such theories have been conjectured by Weinberg in the seventies of the last 
century \cite{Weinberg}.


\subsection{Renormalisation Group Flows and Fixed Points}

To move on, the RG flow for Newton's gravitational constant $G_N$, will be discussed. Hereby, 
the number of dimensions $D$ will be kept general. In a first step one introduces the 
renormalised dimensionless coupling as

\be
g \,=\, \mu^{D-2} \, Z_G^{-1} (\mu) \, G_N
\ee 

\noindent
where the momentum scale $\mu$ denotes the renormalisation scale. $Z_G (\mu)$ is the 
graviton wave function renormalisation factor normalised at some $\mu = \mu_0$ such that 
$Z_G (\mu_0) =1$.

The renormalisation scale $\mu$ can be chosen as a typical energy scale $E$ or as a typical 
momentum transfer $p$. However in this thesis I will follow the Wilsonian approach, and $\mu$ 
will be identified with the corresponding momentum cutoff $k$, see below.

\goodbreak
It will turn out that an important quantity is given by the graviton anomalous dimension

\be
\eta_N \,=\, - \mu \, \frac{\partial}{\partial \mu} \, \ln Z_G (\mu) \,\,.
\ee

\noindent
$\eta_N$ is a non-trivial function of $g$ and the other couplings appearing in the theory. The 
RG flow reads then

\be
\mu \, \frac{\partial}{\partial \mu} \, g \,=\, (D - 2 + \eta_N) \, g \,\,.
\label{eq:1.6}
\ee

\noindent
Eq.\ (\ref{eq:1.6}) especially implies the existence of the non-interacting (Gau\ss ian) fixed point $g_
\ast=0$. For this fixed point one has then $\eta_N=0$, {\it i.e.}, classical scaling.  Looking instead for a 
non-trivial fixed point $g_\ast\not=0$ one has to solve

\be
\eta_\ast=2-D \,\,.
\ee

\noindent
Therefore  a non-trivial fixed point of quantum gravity in $D > 2$  leads to a negative integer value for the 
UV limit of the graviton anomalous dimension $\eta_N$. Note that its numerical value is such 
that it precisely counter-balances the canonical dimension of Newton's coupling constant 
$G_N$.

The non-trivial UV fixed point of quantum gravity, if it exists, has two important consequences.
First, Newton's coupling constant $G_N$ scales as 

\be
 G_N (\mu) = Z_G^{-1} (\mu) \, G_N (\mu_0) \,  \xrightarrow{\mu \to \infty} \frac {g_\ast}{\mu^{D-2}}
\ee

\noindent
when the UV fixed point is approached. This implies that Newton's coupling constant $G_N(\mu)$ 
becomes infinitesimally small. Second, the spectral dimension\footnote[7]{The spectral 
dimension is defined via the average return probability in a random walk, a concise 
explanation can be, {\it e.g.}, found in Appendix~A.1 of  \cite{Reuter:2012id}.} of space-time at 
very short distances becomes equal to two, see, {\it e.g.}, \cite{Reuter:2012id} and references 
therein.

Equipped with this general background I will present in the next two chapters the  knowledge 
on EH quantum gravity and on the corresponding use of the FRG relevant 
for this thesis. In Chapter~4 dimensional reduction is introduced and then applied to EH 
quantum gravity. The derivations and results presented in this chapter constitute the main 
part of this thesis. In Chapter~5 some potential phenomenological implications are described as
well as some concluding remarks and an outlook are presented. A number of technical 
details is deferred to six appendices. 


\chapter{{\color{BrickRed}Renormalisation Group for Einstein-Hilbert Quantum 
Gravity}}
~\vspace{-10mm}

\rightline{\footnotesize{\it{``As far as extra dimensions are concerned, very tiny extra dimensions}}}
\rightline{\footnotesize{\it{wouldn't be perceived in everyday life, just as atoms aren't: We see 
many}}}
\rightline{\footnotesize{\it{atoms together but we don't see atoms individually."}}}
\rightline{\footnotesize{\it{(E.\ Witten)}}}
\minitoc

\bigskip

In this chapter the use of non-perturbative RG methods for EH gravity will be shortly reviewed. 
The RG is well suited for such an investigation for several reasons: Firstly, the investigation of 
gravity should cover approximately sixty orders of magnitude, see Fig.\ \ref{fig:1.1}, which 
makes a continuum method preferential. Secondly, as explained in Chapter~1 gravity is 
perturbatively non-renormalisable and might be renormalisable non-perturbatively. Therefore 
a non-perturbative method is required. Thirdly, the asymptotic safety scenario 
requires an UV fixed point, a notion directly referring to the RG. 

Most of the material presented in this chapter follow closely the seminal paper by Reuter 
\cite{Reuter:1996cp}. For further use in the following chapters the presentation has been 
reordered and new definitions have been introduced.


\section{Functional Renormalisation Group}

In this as in the following two chapters I will employ two forms of the FRG: 
An exact functional identity and an approximated background field flow.
For the FRG there exist several versions of exact functional identities. 
In this thesis I will use the  Wetterich
equation \cite{Wetterich:1992yh} with the optimised  regulator of \cite{Litim:2000ci}, called
Litim regulator from here on. Both versions are functional differential equations for the 
effective action.  Before providing
expressions specific to EH gravity I will summarise the basic idea of the FRG, see 
Appendix~B for a  brief summary of corresponding derivations.

Denoting by  ``$\mathrm{Tr}$'' all occurring integrals and sums 
leads to the compact notation for the Wetterich equation \cite{Wetterich:1992yh}
\be
\partial _t \Gamma_k = \frac 1 2 {\mathrm{Tr}} 
{\Bigg(\left( \Gamma_k^{(2)} 
+ R_k  \right)^{-1} \partial_t R_k \Bigg)  } 
\label{eq:2.2}
\ee
with 
\be
\Gamma_k^{(2)} 
:= \frac{\delta^2 \Gamma_k [\bar \Phi]}{\delta \bar \Phi (q)\,\delta \bar \Phi (-q) }
\,\,.
\ee
The exact functional identity (\ref{eq:2.2}) is the basic equation to be used in this thesis.
As already mentioned, I will use the Litim regulator
\be
R_ {k} (q^2) = ({k}^2 - q^2)\, \Theta \, ({k}^2 - q^2) \,\,.
\label{eq:2.4}
\ee 
Its scale derivative is given by 
\be
\partial_t R_ {k} (q^2) = 2 \,{k} ^2 \,\Theta ({k} ^2 - q^2)  \,\,.
\ee
As we will see later in this thesis the step function $\Theta (k^2 - q^2)$ will allow for some 
significant simplifications in the appearing integrals.

In this thesis a background field formalism will be used. To this end one introduces a 
non-propagating background field $\Phi_0$ into the effective action, 
$ \Gamma_k [\bar \Phi] \to \Gamma_k [\bar \Phi, \Phi_0]$, and couples the fluctuating field
$\bar \Phi - \Phi_0$ to the external current and to the regulator. When deriving the flow the 
background field acts as kind of ``spectator'', and one obtains Eq.\ (\ref{eq:2.2}) with the 
replacement
\be
\displaystyle{\Gamma_k^{(2)} [\bar \Phi] \to 
 \frac{\delta^2 \Gamma_k [\bar \Phi , \Phi_0]}{\delta \bar \Phi (q)\,\delta \bar \Phi (-q) }},
 \ee
 see {\it e.g.}, \cite{Litim:2010tt} and references therein. Eventually, the background field 
 $\Phi_0$ is identified with the physical mean field, $\Phi_0 = \bar \Phi = \langle \Phi \rangle _J$
 thereby leading to an effective background independence. In this way one has arrived at
 the background field flow for an effective action $\Gamma_k [\bar \Phi] \equiv
 \Gamma_k[\bar \Phi , \bar \Phi]$.  The main difference between standard and background 
 flows is given by the presence of the background field at an intermediate step of the calculation
 which amounts to a re-organisation of the flow. 
 
 In addition, it will prove useful to substitute  $q^2 \to  \Gamma_k^{(2)} [\Phi_0 , \Phi_0 ]  (q^2)$ in 
 the regulator function $R_k(q^2)$. Note that $R_k$ is chosen to depend on the background field 
 and to be independent of  the fluctuating field. As we will see in the following sections this will 
 introduce the
 covariant background Laplacian and the wave function renormalisation function into the 
 regulator.
 
 The background field formalism has also been used for a derivation of generalised proper-time 
 flows  \cite{Litim:2001hk,Litim:2002xm}, a concise summary (followed here)
 is given in \cite{Litim:2007jb}.  Using the notation $x:=\Gamma_k^{(2)} [\bar \Phi , \bar \Phi ] $ and
$x_0:= \Gamma_k^{(2)} [\Phi_0 , \Phi_0 ] $ background-dependent regulators of the form 
$R_k(q^2)\to x_0 \, r(x_0)$ are introduced. With a special choice of an one-parameter family of 
regulator functions $r_m(x)$ one can derive the background field flow
\be
\partial_t {\Gamma_k} = {\mathrm {Tr}}  \left( \frac {k^2}{k^2+x/m} \right)^m 
+\frac 1 2  {\mathrm {Tr}}  \left(   \frac{r_m}{x(1+r_m)}-  \frac 1 x \left( \frac {k^2}{k^2+x/m}
\right)^m\right) \partial_t x
\ee
with $m\in [1,\infty]$.\footnote{Convergence of the integrals require a dimension-dependent 
and in general more restrictive lower bound on the parameter $m$, see below.}
If the term $\sim \partial_t x$ is neglected ({\it i.e.}, the additional terms due to the scale 
dependence in the regulator function are dropped), and if one uses for the numerator of the first 
term a proper-time integral representation the exact functional flow Eq.\ (\ref{eq:2.2})
reduces to the proper-time flow of \cite{Liao:1994fp}:
\be
\partial_t {\Gamma_k}  [\Phi]= - \frac{1}{2} \, {\mathrm {Tr}} \int_0^{\infty} \, \frac{ds}{s} \,\, 
\partial_t f_k^m (s) \,\, e^ {-s \,  {\Gamma}_k^{(2)}}  + {\cal O} \left( 
\partial_t {\Gamma_k^{(2)}}  \right) \,\,.
\label{eq:2.6new}
\ee
Hereby the regulator functions $f_k^m(s)$ take the form as defined in Appendix~B.
This way of deriving the approximate background field flow equation (\ref{eq:2.6new})
make it plausible that already this simpler equation, as compared to the exact functional 
identity (\ref{eq:2.2}), provides qualitatively correct and quantitatively almost correct results.

In Appendix~B also an alternative way \cite{Bonanno:2004sy}
how to arrive at the background field flow 
equation (\ref{eq:2.6new}) is given.  Taking into account (non-dynamical) ghost contributions 
the background field flow reads then
\be
\partial_t {\Gamma_k}  [\Phi]= - \frac{1}{2} \, {\mathrm {Tr}} \int_0^{\infty} \, \frac{ds}{s} \,\, 
\partial_t f_k^m (s) \,\, \Big(e^ {-s \,  {\Gamma}_k^{(2)}} -2 \,\, e^{-s \, S_{gh}^{(2)}}\Big) \,\,.
\label{eq:2.1}
\ee
Here $S_{gh}^{(2)}$ is the second variation of the tree-level ghost action. 
The form (\ref{eq:2.1})
of the background field flow equation (with ghost quantum corrections neglected, see also 
Subsection~2.2.2) will be used in the following section for EH gravity.

\goodbreak


\section{Einstein-Hilbert Quantum Gravity}

\subsection{Background Field Formalism for Gravity}

Throughout this thesis it is assumed that the metric $g_{\mu \nu}$ represents the 
dynamical degrees of freedom in gravity. To be more precise, from the $\frac 1 2 D(D+1)$ 
coefficients of the symmetric rank-2 tensor $g_{\mu \nu}$ only $\frac 1 2 D(D-3)$  
represent propagating degrees of freedom. The remaining $2D$ coefficients have to be dealt 
with by adding a gauge fixing and a Faddeev-Popov ghost term to the action. 

The generating functional is represented by an integral over all metrics 
$\gamma_{\mu \nu}$, {\it e.g.}, for vanishing sources one has
\be
Z= \int {\cal D} \gamma_{\mu \nu} \exp \left( -S [\gamma_{\mu \nu} ]\right) \,\,,
\ee
where $S$ is an arbitrary diffeomorphism-invariant classical action. In the background field
formalism one splits now the metric into a background $\bar g_{\mu \nu}$ and a fluctuation
$h_{\mu \nu}$ which is not necessarily small \cite{Reuter:1996cp}:
\be
\gamma_{\mu \nu}  = \bar g_{\mu \nu} + h_{\mu \nu} \,\,.
\ee
Correspondingly, the gauge fixing term $S_{gf}[h;g]$ needs to be of the background type,
 {\it i.e.}, it is invariant under a {\it combined} transformation of $\bar g_{\mu \nu}$  and
$h_{\mu \nu}$. This invariance requirement  ensures that the effective action will be
invariant under diffeomorphisms. 

The Faddeev-Popov ghosts are vector fields, and I will denote them by $c_\mu$ and the 
antighosts by $\bar c_\mu$. With $S_{gh}$ being the ghost action the functional integral 
reads then
\be
Z= \int {\cal D} h_{\mu \nu} {\cal D} c_\mu {\cal D} \bar c_\mu
\exp \left( -S [\bar g_{\mu \nu}  + h _{\mu \nu}] - S_{gf} - 
S_{gh} [h _{\mu \nu},c_\mu,\bar c_\mu ; \bar g_{\mu \nu}  ]\right) \,\,.
\ee

In a next step one introduces a source coupled to the fluctuating part of the metric and 
performs  the Legendre transform of the connected generating functional
$\ln Z$ to obtain the effective action
$\Gamma [\bar h_{\mu \nu} ; \bar g_{\mu \nu} ]$ where $\bar h_{\mu \nu} := \langle 
h_{\mu \nu}  \rangle$ denotes the expectation value of the fluctuating part of the metric.
The expectation value of the total metric $\gamma_{\mu \nu} $ is called from here on 
$g_{\mu \nu}$, and one has
\be
g_{\mu \nu} := \langle  \gamma_{\mu \nu}  \rangle =\bar g_{\mu \nu} +
 \langle  h_{\mu \nu}  \rangle = \bar g_{\mu \nu} +\bar h_{\mu \nu} \,\,.
 \label{eq:2.9}
\ee

Usually the effective action $\Gamma$ is considered to be a functional of $g_{\mu \nu}$ 
and $\bar g_{\mu \nu}$ instead of $\bar h_{\mu \nu}$ and $\bar g_{\mu \nu}$. 
It can be shown \cite{Reuter:1993kw,Litim:1998nf,Reuter:1996cp,Freire:2000bq,Litim:2002hj} 
that the effective action in the usual 
formalism, {\it i.e.}, the generating functional of the one-particle irreducible Green 
functions,
$\Gamma [g_{\mu \nu} ]$, is obtained by setting $g_{\mu \nu} = \bar g_{\mu \nu}$ or, 
equivalently, $\bar h _{\mu \nu} =0$.

A further advantage of the background formalism is the fact that the background metric can 
be used in the following way in the regulator function: One constructs first the background 
covariant Laplacian $-\bar D^2$ and use its eigenmodes\footnote{These eigenmodes are only 
used implicitly 
because the operator traces will be evaluated within a heat-kernel expansion in first order.} to 
define the cutoff at a momentum scale $k^2=-\bar D^2$. This implies that the regulator 
function $R_k$ depends on the background metric. The background is then eliminated by 
identifying it with the physical average metric  in the final equations. This dynamical 
adjustment of the background metric implements in an approximate way the background 
independence required for a theory of quantum gravity. 


\subsection{Einstein-Hilbert Theory}

The EH action including a cosmological constant term is given by
\be
S_{EH} = \int d^Dx\, \sg \Bigg(\frac{-R+2{\Lambda}}{16\pi{G_N}} \Bigg) \,\,,
\label{eq:2.29}
\ee
where $R(g_{\mu \nu})$ is the Ricci scalar, $\Lambda$ denotes the cosmological constant 
(with canonical mass dimension $[\Lambda]=2$), and $G_N$ is Newton's coupling constant. 
Its canonical mass dimension is $[G_N]=2-D$. For later use I define 
\be
\kappa = (32\,\pi\, G_N)^{-1/2} \,\,.
\ee

The Wilsonian effective action of Quantum Einstein Gravity can be written as
\be
\Gamma_k = \Gamma_{k,EH} + \Gamma_{k,gf} + \Gamma_{k,gh} + \Gamma_{k,matter} \,\,.
\ee
In the truncation employed in this thesis the matter part is completely neglected and quantum corrections to the 
ghosts are not considered, {\it i.e.}, the anomalous dimension of the ghosts is set to zero, $
\eta_{gh}=0$.\footnote{Dynamical ghosts have been considered in 
\cite{Groh:2010ta,Eichhorn:2010tb,Eichhorn:2013ug}.} 
The gauge fixing term will be discussed in the next subsection. 

Under the RG flow higher order interactions in the metric field will be generated. In the EH 
theory these are neglected and only the couplings in Eq.\ (\ref{eq:2.29}) become running 
couplings and thus functions of the momentum scale $k$. As long as this scale $k$ is much 
smaller than the $D$-dimensional Planck scale $M_D$ gravity is well approximated by the EH 
action with a slowly running $G_k$. This justifies the use of the EH theory for the purpose 
of this thesis.

Introducing the running of the cosmological constant and of Newton's constant one writes
\be
\Gamma_{k,EH}=
2 \, \kappa^2 \, \mathrm Z_{Nk} \, \int d^D x \, \sg\,\, (- R(g) + 2 \, \bar \lambda_k) \,\,.
\label{eq:2.32}
\ee
In this notation the running gravitational coupling constant is reexpressed as
$G_k =  G_N / \mathrm Z_{Nk}$ where $\mathrm Z_{Nk}$ is the running graviton wave function 
renormalisation constant.

In addition, as usual in this context, an additional approximation will be used: To simplify the 
expression for the second variation of the effective average action I will employ for the 
background metric a maximally symmetric space in which the Riemann and the Ricci
tensors can be expressed in function of the background metric and the Ricci scalar.
Note that
a space, respectively, space-time is maximally symmetric if and only if the Riemann tensor 
can be written as
\be
\bar {R}_{\rho \mu \nu \sigma} = \frac {1}{D (D - 1)} \, ( \bar {g}_{\rho \nu} \, \bar {g}_{\mu \sigma} - 
\bar {g}_{\rho \sigma} \, \bar {g}_{\mu \nu} ) \, \bar {R} \,\,,
\label{eq:2.33}
\ee
and the Ricci tensor is then given by
\be
\bar {R}_{\mu \nu} = \frac {1}{D} \, \bar {g}_{\mu \nu} \, \bar {R} \,\,,
\label{eq:2.34}
\ee
with the Ricci scalar being defined as usual as 
$\bar {R}=\bar {g}^{\mu \nu} \bar {R}_{\mu \nu}$, {\it cf.}, Appendix~A.


\subsection{Gauge Fixing and DeDonder Gauge}

Following Ref.\ \cite{Reuter:1996cp} in this thesis I will use 
deDonder gauge. The corresponding gauge fixing functional is given in Appendix~A. 
This results in the ghost action
\be
 \Gamma_{gh} [h,c,\bar {c};\bar {g}] = - \sqrt{2} \, \int d^{D} x \, \sgb
 \,\, \bar  {c}_{\mu} \,\, {\cal M} [\gamma;\bar {g}]^{\mu}_{\,\,\,\nu} \, c^{\nu} 
\ee
where the Faddeev-Popov operator  is given by
\be
{\cal M} [\gamma;\bar {g}]^{\mu}_{\,\,\,\nu} = (\bar {g}^{\mu \beta} \, \bar {g}^{\alpha \gamma} \, \bar 
{D}_{\gamma} \, \gamma_{\beta \nu} \, D_{\alpha} \, + \bar {g}^{\mu \beta} \, \bar {g}^{\alpha 
\gamma} \, \bar {D}_{\gamma} \, \gamma_{\alpha \nu} \, D_{\beta} \, - \bar {g}^{\mu \lambda} \, \bar 
{g}^{\sigma \rho} \, \bar {D}_{\lambda} \, \gamma_{\rho \nu} \, D_{\sigma} ) \,\,.
\ee


\section{$\beta$-Functions and Analytic Flows}
For deriving the $\beta$-Functions I followed for the general part
Ref.~\cite{Reuter:1996cp}, for the background  field flow in proper-time 
regularisation  Ref.~\cite{Bonanno:2004sy} and for the
Wetterich equation (except the usage of the regulator function) again
Ref.~\cite{Reuter:1996cp}.  
All technical details are given in Appendix~D.

Before going to the detailed expressions a few definitions are in order. 
One introduces the dimensionless Newton's constant
\be
g_k \equiv k^{D-2} \,\,\, G_k \equiv  k^{D-2} \,\,\, \mathrm Z_{Nk}^{-1} \,\,\,  {G_N} 
\label{eq:gk}
\ee
and the dimensionless cosmological constant
\be
\lambda_k \equiv  k^{-2} \,\,\, \bar {\lambda}_k 
\label{eq:lambdak}
\ee
as well as the anomalous dimension
\be
\eta_{Nk} \equiv - \, \partial_t \, \ln \, \mathrm Z_{Nk} \,\,.
\label{eq:etak}
\ee

As remarked at the end of Sect.~2.1 the approximate background field flow equation 
(\ref{eq:2.6new}), although being the simpler equation as compared to the exact functional 
identity (\ref{eq:2.2}), is expected to provide qualitatively correct and quantitatively almost 
correct results. In the following always the simpler approximate background field flow equation
will be treated first in order to allow a focus on the important qualitative features.

\bigskip

\noindent
{\large{\uuline{\bf{Background Field Flow}}}}
~
\bigskip

Following the explicit steps as detailed in Appendix~D 
one finds 
\bea
\eta_{Nk} &=& \; g_k \,\, B_0 \, (\lambda_k) \,\,\, \quad \mathrm {with} \label{eq:2.62}\\
B_0 \, (\lambda_k) &=& \; 4 \, (4 \, \pi)^{-D/2 \, + 1} \,\,\, \frac {\Gamma (m+2 - D/2)}{\Gamma (m+1)} 
\,\,\,
\times \nonumber \\
&& \hspace{1mm}  \times
\Bigg ( \frac {1}{( 1 - 2 \, {\lambda}_k )^{(m+2 - D/2)}} \,\,\, \frac {1}{12} \,\,\, (-5\, D + 7) \, D \, - \frac {1}
{3} \,\, (D + 6) \, \Bigg ).
\label{eq:2.63} 
\eea
For the two $\beta$-functions in the background field flow one finally obtains
\bea
\partial_t \, \lambda_k &=& \; \beta_{\lambda} \, ( g_k , \lambda_k ) =
 \eta_{Nk} \, \lambda_k - 2 \, \lambda_k + 
g_k \,  A_0 \, (\lambda_k) \label{eq:2.64} \\ [2mm]
\quad \mathrm {with} \nonumber \\
A_0 \, (\lambda_k) &=& \; (4 \, \pi)^{-D/2 \, + 1} \,\,\, 
\frac {\Gamma (m+1 - D/2)}{\Gamma (m+1)} \,\,\, \Bigg ( \frac {D \, ( D+1 )}
{( 1 - 2 \, {\lambda}_k )^{(m+1 - D/2)}} \, - 4 \, D \, \Bigg ) \qquad \label{eq:2.65} \\ [2mm]
\quad \mathrm {and} \nonumber \\
\partial_t \, g_k &=& \beta_g \, ( g_k , \lambda_k ) = ( D - 2 + \eta_{Nk} ) \, g_k \,\,.
\label{eq:2.66}
\eea

\bigskip
~

\noindent
{\large{\uuline{\bf{Exact Functional Identity}}}}
~
\bigskip

As already mentioned several times in the context of the exact functional identity 
(\ref{eq:2.2}) I am using exclusively 
the Litim regulator which for a scalar field theory is given by Eq.\ (\ref{eq:2.4}).
As  one sees in the discussion on the background field flow 
above for EH gravity one employs the cutoff 
on the modes of the background covariant Laplacian, $-\bar D^2$. The details, including 
the employed heat kernel expansion,  are given again in Appendix~D, see also 
Eqs.\ (\ref{eq:C.13}) and (\ref{eq:C.14}).  

The exact functional identity  for EH theory is given in Eq.\ (\ref{eq:C.17}).
Introducing as before dimensionless constants one arrives from this equation at
\bea
\partial_t \, \lambda_k &=& \beta_{\lambda} \, ( g_k , \lambda_k ) = \eta_{Nk} \, \lambda_k - 2 \, 
\lambda_k + g_k \,\,\, 
\Big( A_0 \, (\lambda_k) - \eta_{Nk} \,\, A_1 \, (\lambda_k) \Big) \,\,\, \label{eq:2.68}  
\qquad \\
\quad \mathrm {with}  \qquad  \quad \nonumber \\
A_0 \, (\lambda_k) &=& 8 \, \pi \,\,\, \frac {(4 \, \pi)^{-D/2}}{\Gamma \, (D/2)} \,\,\, \Bigg (\, 
\frac {(D + 1)}{(1 - 2 \, \lambda_k)} \, - 4 \, \Bigg ) \,\,\,\label{eq:2.69}\\
 \quad \mathrm {and} \qquad  \quad \nonumber \\
A_1 \, (\lambda_k) &=& 8 \, \pi \,\,\, \frac {(4 \, \pi)^{-D/2}}{\Gamma \, (D/2)} \,\,\, \Bigg (\, \frac {(D + 
1)}{(D + 2)} \,\,\,
 \frac {1}{(1 - 2 \, \lambda_k)} \, \Bigg ) \,\,, 
 \label{eq:2.70}
\eea
where
\be
\partial_t \, g_k = \beta_g \, ( g_k , \lambda_k ) = ( D - 2 + \eta_{Nk} ) \, g_k \,\,.
\label{eq:2.71}
\ee
Using a Hartree-Fock type resummation, one obtains for the anomalous dimension
\begin{equation}
\eta_{Nk} = \frac {g_k \,\,B_0 (\lambda_k)}{1 + g_k \,\,
 B_1 (\lambda_k)} \,\,,
\label{eq:2.72}
\ee
with
\bea
B_0 (\lambda_k) &=& 16 \, \pi \,\,\, \frac {(4 \, \pi)^{-D/2}}{\Gamma \, (D/2)} \,\,\, 
\Bigg (  - (D - 1) \,\,\, \frac {1}{(1 - 2 \, \lambda_k)^2}\, + 
\frac {D \, (D + 1)}{12} \,\, \frac {1}{(1 - 2 \, \lambda_k)} 
\nonumber \\ && \qquad  \qquad  \qquad  \qquad  \qquad - \, \bigg ( \frac {4}{D} + 
\frac {D}{3} \bigg ) \Bigg ) 
\label{eq:2.73} \\
 {\mathrm {and}}&& \nonumber \\ 
B_1 (\lambda_k) &=& - \, 16 \, \pi \,\,\, 
\frac {(4 \, \pi)^{-D/2}}{\Gamma \, (D/2)} \,\,\, 
\Bigg ( \frac {(D - 1)}{(D + 2)} \,\,\, \frac {1}{(1 - 2 \, \lambda_k)^2} \, - 
\frac {(D + 1)}{12} \,\, \frac {1}{(1 - 2 \, \lambda_k)} \Bigg ) . \qquad \quad 
\label{eq:2.74}
\eea
 
Before using these expressions in the next chapter to discuss the fixed points and the phase
 diagrams it is worthwhile to compare the two different versions of the $\beta$-functions.
 
 \goodbreak
 
 \bigskip

\noindent
{\large{\uuline{\bf{Comparison of the $\beta$-Functions and discussion of the}}}}\\[2mm]
 {\large{\uuline{\bf{ RG flows}}}}
~
\bigskip

When comparing the two forms of $\beta_\lambda$  (\ref{eq:2.64}) and (\ref{eq:2.68}) 
one recognises  that they are quite similar, the difference is the term
 $-g_k\eta_{Nk} A_1(\lambda_k)$ on the r.h.s.\ of the $\beta$-function following from
 the exact functional identity. In addition, the flow of the cosmological constant 
$\lambda_k$ is of smaller significance than the flow of the coupling constant $g_k$
for the physics along a phenomenologically acceptable RG trajectory. 
Due to this one may conclude that the differences between the two types of
RG flow for $\beta_\lambda$ are of minor importance. 
 
At first sight Eqs.\ (\ref{eq:2.66}) and (\ref{eq:2.71}) for $\beta_g$ look identical, 
however, the expressions 
(\ref{eq:2.62}) and (\ref{eq:2.72}) for the anomalous dimension $\eta_{Nk} $ are 
distinctively different. Whereas the expression within the approximate background field  
method for the anomalous dimension $\eta_{Nk} $
is strictly linear in $g_k$ this is not the case for the $\eta_{Nk} $ derived from the exact
equation. The origin of the denominator appearing in the expression (\ref{eq:2.72}) 
can be directly traced back to the fact that (as anticipated from the derivation of the 
background field flow) the scale derivative of the regulator 
function contains a term proportional to $\eta_{Nk} $, see Eq.\ (\ref{eq:C.14}). Neglecting
the first term on the r.h.s. of Eq.\ (\ref{eq:C.14}) would result in vanishing functions 
$A_1(\lambda_k)$ and  $B_1(\lambda_k)$ as can be verified straightforwardly from the
steps of the calculation which are explicitly given in Appendix~D. As the scale derivative
of the  approximate background field regulator function, see Eq.\ (\ref{eq:B.6}), 
does not contain a term 
proportional to the anomalous dimension $\eta_{Nk} $ the appearance of a simpler
structure in the respective  $\beta$-functions than in the ones from the exact equation 
is evident.

Due to the non-linear dependence of $\eta_{Nk} $ on $g_k$ in the case of the
exact equation one might expect some significant differences in the flows generated 
by the approximate and the exact functional equation. As will be discussed in the next chapter
it turns out that the differences in the two RG flows are restricted to regions in the
$(\lambda_k, g_k)$ plane close to a line where the anomalous dimensions $\eta_{Nk} $ diverges.
(A detailed discussion will be given in Sect.~3.2.) For all physical RG trajectories discussed in this 
thesis neglecting the functions $A_1(\lambda_k)$ and $B_1(\lambda_k)$ is qualitatively and
semi-quantitatively good approximation. 

As already stated above the limit $\lambda_k\to 0$ has no significant impact on the 
$\beta$-function for $g_k$ and thus the qualitative behaviour of the coupling and the
anomalous dimension are not changed. On the other hand, the limit $\lambda_k\to -\infty$
is an interesting one. As the functions $A_1(\lambda_k)$ and $B_1(\lambda_k)$  go in this case to 
zero the discussion applies to the two different forms of the flow alike.  Due to the terms 
originating from the ghost action the functions $A_0(\lambda_k)$ and $B_0(\lambda_k)$
take a finite non-vanishing limit. In addition, it is important to note that $B_0$ is also in this
limit negative, therefore the anomalous dimension will be non-positive and goes to  
zero only for $g_k\to 0$. As also $A_0$ is in this limit negative one arrives at the consistent 
conclusion that $\beta_\lambda$ is always negative and thus the absolute value of 
$\lambda_k$ increases towards  the IR. Furthermore, a finite non-vanishing value of $g_k$    
leads to a contradiction. This leaves the possibilities $g_k\to 0^+$ and $g_k\to +\infty$.
(Note that negative and thus unphysical values of $g_k$ are discarded anyhow.)
In the first case, the anomalous dimension vanishes and both, $g_k$ and $\lambda_k$,
 approach  with classical scaling laws ($k^{D-2}$ and $k^{-2}$, respectively) this 
 IR fixed point when $k$ is  lowered. 
 In the second case, $\eta_{Nk}\to -\infty$ and the dominant term in $\beta_g$ is $B_0g_k^2$.
 The resulting approximate differential equation has the solution
 \be
 g_k=\frac 1 {const.-B_0 \ln k}
 \ee  
and one recognises that the flow will stop at a finite value of $k$, $k=e^{const./B_0}$, because
both, $g_k$ and $\eta_{Nk}$ diverges. As $\beta_\lambda$ diverges then too, also $\lambda_k$ 
diverges but now with an infinite scale derivative. The limits discussed here will be numerically 
verified in Sect.~3.2, the interested reader may peak ahead to Fig.~\ref{fig:3.2}.

At this point a remark is in order: The $\beta$-functions become singular when 
$\lambda_k\to \frac 1 2$. In addition, as already mentioned 
the expression (\ref{eq:2.72}) for the anomalous dimension diverges at a boundary line which 
in the case of the exact functional identity also restricts the values of $\lambda_k$ for 
RG trajectories starting at negative or small positive values of   $\lambda_k$.\footnote{A
RG trajectory cannot cross a line where the $\beta$-functions diverges.}
Therefore  in this thesis all trajectories with $\lambda_k>\frac 1 2 $ (or to the right of the 
above mentioned boundary line) will be not considered.

Last but not least, the limit $g_k\to 0^+$ will be discussed here. Then the anomalous dimension also 
goes to zero, $\eta_{Nk} \to 0^-$, and the $\beta$-functions contain basically only the canonical
dimensions. Considering now a flow from the UV to the IR  one has then three possibilities to
obtain $g_k\to 0$:
\begin{itemize}
\item There is exactly one RG trajectory, the so-called separatrix, for which both $g_k$ and 
$\lambda_k$ will vanish with the classical scaling law for $g_k$.  To flow into this IR fixed point
a precise cancelation  in $\beta_\lambda$ is required which makes plain that one and only one 
RG trajectory ends up in this IR fixed point. As all the couplings vanishes in this fixed point it is
called a Gau\ss ian fixed point.
\item In the IR the RG trajectory may end up in $g_k\to 0^+$ and $\lambda_k\to -\infty$, see the 
discussion above.
\item As the flows end at a singular boundary line there is one RG trajectory with $g_k=0$ not 
yet covered by the previous cases: If one starts a RG trajectory with $g_k=0$ and $0<\lambda_k<1/2$ this trajectory will stop at $(g_k,\lambda_k) = (0,1/2)$ which is then a 
degenerate fixed point. 
\end{itemize} 


\newpage
\thispagestyle{empty}

\chapter{{\color{BrickRed}Quantum Gravity in $D \ge 4$}}
~\vspace{-10mm}

\rightline{\footnotesize{\it{``If I take the theory as we have it now, literally, I would conclude that 
extra}}}
\rightline{\footnotesize{\it{dimensions really exist. They're part of nature. We don't really know how
big}}}
\rightline{\footnotesize{\it{they are yet, but we hope to explain that in various ways."}}}
\rightline{\footnotesize{\it{(E.\ Witten)}}}
\minitoc

\bigskip

In the last chapter the expressions for the $\beta$-functions and the graviton  
anomalous dimension have been derived for two forms of the RG. In this chapter I 
am 
going to discuss the RG trajectories in $D=4$ and higher dimensions\footnote{For a brief
description of the numerical treatment see Appendix~F.} to set the scene 
for the investigation of the ADD model scenario with a ``large'' compact dimension.
Besides confirming results already known in the literature, the presentation in this chapter is 
intended to provide the background for a meaningful comprehension of the novel results to be 
presented and discussed in Chapter~4.

Before 
discussing the flow it is instructive to discuss the fixed point structure.  


\section{Fixed Points in $D \ge 4$}

Fixed points are zeros of the $\beta$-functions. As thus they are sources and sinks of RG trajectories.
Defining the flow to run from large to small $k$ an UV fixed point is a source whereas an IR fixed point is a sink 
of RG trajectories. Having UV and IR fixed points special trajectories connecting an UV to an IR fixed point exist,
those trajectories are called separatrices. From the discussion at the end of Chapter~2 it 
is evident that both versions of the
$\beta$-functions as derived in the last chapter possesses always a Gau\ss ian fixed point at 
$(g_\ast , \lambda _\ast ) = (0,0)$ for every value $D$ of the number of space-time
dimension. For the cases
discussed here ($D\ge 4$) this is always an IR fixed point. Therefore in the following we will 
concentrate on the discussion of the non-trivial fixed point which will turn out to be an UV fixed 
point. Note also that for $D>2$ there can be no non-trivial fixed point in the classical regime 
because $\beta_g$ will vanish for $g_k\not =0$ only if $\eta_{Nk} =2-D$ and thus 
the absolute value of the anomalous  dimension and therefore quantum effects are large.

The $\beta$-function for $g_k$ reads:
\be
\partial_t \, g_k = \beta_g \, ( g_k , \lambda_k ) = ( D - 2 + \eta_{Nk} ) \, g_k \,\,,
\ee
for both forms as derived from the approximate and the exact functional identity.
For a non-trivial fixed point, {\it i.e.}, for $g_\ast\not = 0$ a necessary condition is
\be
\eta_\ast = 2 -D \,\,.
\label{eq:3.2}
\ee
Therefore this relation and 
\be
\left. \beta_\lambda \right|_{\eta_{Nk}=2-D}  =0
\label{eq:3.3}
\ee
will be used to find the non-trivial UV fixed point.

In addition, I will compute  the scaling exponents at the UV fixed point  from the stability matrix
\be
B = 
\begin{pmatrix}
\hspace{2mm} \frac {\partial \, \beta_{\lambda}} {\partial \, \lambda} & 
\frac {\partial \, \beta_{\lambda}} {\partial \, \mathrm {g}} \hspace{2mm} \vspace{5mm} \\ 
\hspace{2mm} \frac {\partial \, \beta_{\mathrm {g}}} {\partial \, \lambda} & 
\frac {\partial \, \beta_{\mathrm {g}}} {\partial \, \mathrm {g}} \hspace{2mm} 
\hspace{2mm}
\end{pmatrix}
\hspace{1cm} \mathrm {evaluated ~ at} \hspace{1cm} ( \, g_{\ast} \, , \, \lambda_{\ast} \, ) \,\,.
\ee
The scaling exponents are the real and imaginary part of the negative eigenvalues of $B$ ({\it 
i.e.}, $B \, V = - \theta \, V$) 
\be
\theta =\theta^\prime \pm i \theta^{\prime \prime} \,\,.
\ee
Solving the linearised flow equations  one obtains 
\be 
\begin{pmatrix} g_k\\  \lambda_k  \end{pmatrix}=
\begin{pmatrix} g_\ast\\  \lambda_\ast  \end{pmatrix} + 
\begin{pmatrix}  \sum_{l=1}^2 c_l V_{1l} (k_0/k)^{\theta_l}\\   
\sum_{l=1}^2 c_l V_{2l} (k_0/k)^{\theta_l}\ \end{pmatrix} 
\ee
with some appropriate constants $c_l$ and $k_0$ being some reference scale. 
One recognises that if the real part of the scaling exponent is
positive then the coupling constants approach with  corresponding power laws
 the UV fixed point values. 
Therefore a positive value of $\theta^\prime $ indicates an attractive fixed point. On the other
hand, a negative value would indicate a repulsive fixed point in contradiction to the
anticipated  result. Thus, it is important to check 
that the real parts of the scaling exponents come out positive. As one sees below one obtains
for the cases considered in this thesis
a complex-conjugate pair of scaling exponents, {\it i.e.}, only one value for the real part.
The imaginary part of the scaling 
exponent leads to oscillations around the fixed point values, the RG trajectory `spirals' into the
fixed point. 

It is also important to note that a change of the regulator function leaves the scaling exponents 
invariant, see {\it e.g.}, Sect.~5 of Ref.\ \cite{Lauscher:2002sq}. Quantities which are unchanged
under a variation of the cutoff scheme, like the scaling exponents, are called universal.

\goodbreak

\bigskip
~

\noindent
{\large{\uuline{\bf{Approximate Background Field Flow}}}}
~
\bigskip

Using Eqs.\ (\ref{eq:2.62})  and  (\ref{eq:2.64})  for the background field flow
Eqs.\ (\ref{eq:3.2})  and  (\ref{eq:3.3}) can be written as 
\bea
2-D &=&  \, g_\ast \, \frac 1 3 (4 \, \pi)^{-D/2 \, + 1} \,\,\,
\frac {\Gamma (m+2 - D/2)}{\Gamma (m+1)}  
\Bigg ( \frac {(-5\, D + 7) \, D }{( 1 - 2 \, {\lambda}_\ast )^{(m+2 - D/2)}} \,\,\,  - 4 \,\, (D + 6) \, \Bigg )  
\nonumber \\ 
&=& g_\ast B_0 (\lambda_\ast) \\
D\, \lambda_\ast &=& g_\ast \, (4 \, \pi)^{-D/2 \, + 1} \,\,\, 
\frac {\Gamma (m+1 - D/2)}{\Gamma (m+1)} 
 \Bigg ( \frac {D \, ( D+1 )}{( 1 - 2 \, {\lambda}_\ast )^{(m+1 - D/2)}} \, - 4 \, D \, \Bigg )
 \nonumber \\ 
&=& g_\ast A_0(\lambda_\ast) \,. 
\eea
The resulting condition 
\be
g_\ast = (2-D)/B_0(\lambda_\ast) = D\, \lambda_\ast /A_0(\lambda_\ast) 
\label{eq:cond1}
\ee
is graphically displayed for $D=4$ and $D=5$ ($m=2$) in Fig.\ \ref{fig:3.1}. 
\begin{figure}[H]
\centering
\includegraphics[width=0.85\textwidth]{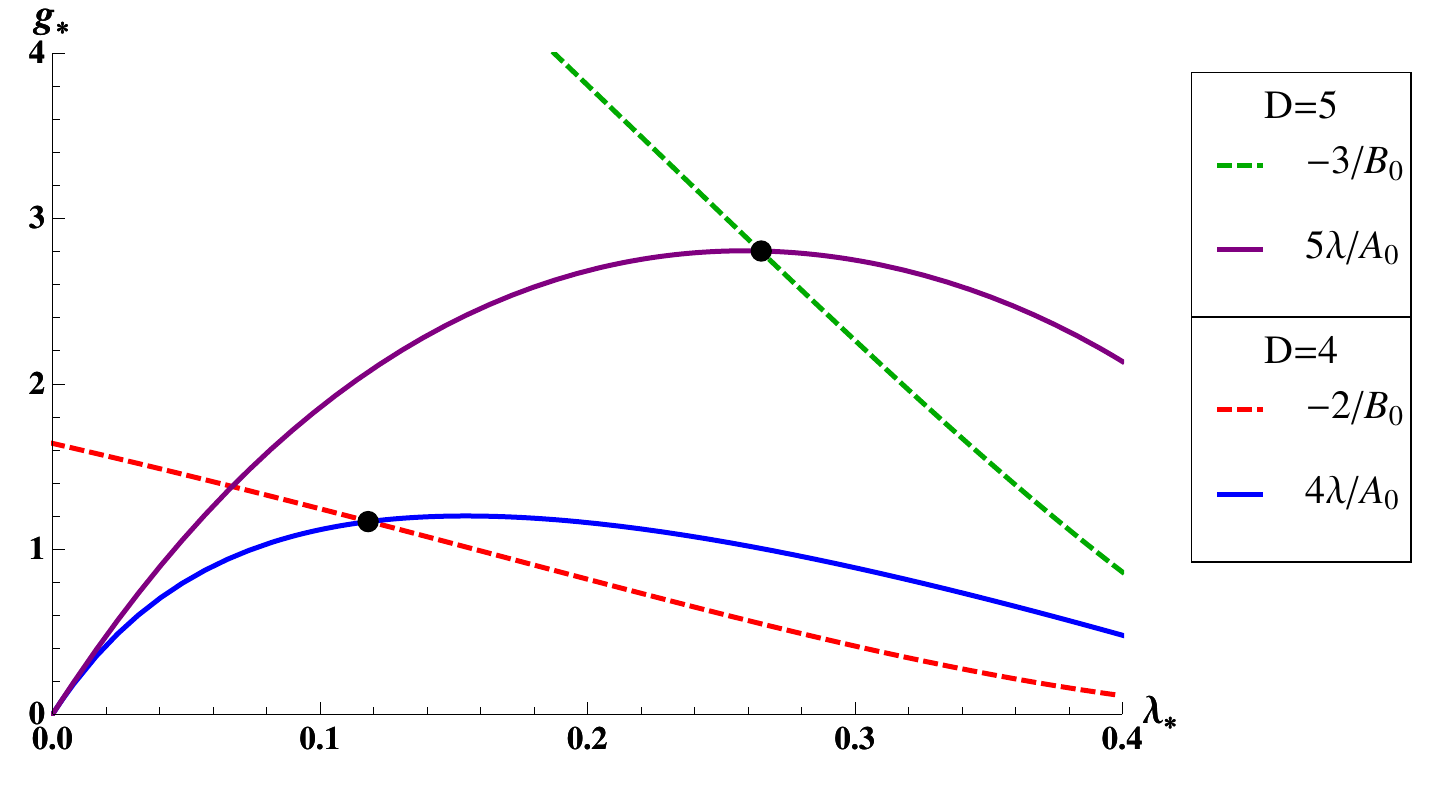}
\caption[Condition for the UV fixed point  values for $D=4$ and $D=5$
for the background field flow.]{\label{fig:3.1} 
Condition (\ref{eq:cond1}) for the UV fixed point  values for $D=4$ and $D=5$ 
for the background field flow ($m=2$).
The crossings of the curves, marked by the black points, provides the fixed point values
$g_\ast$ and ${\lambda}_\ast$.}
\end{figure}

In Table \ref{tab:3.1} some  results for the resulting UV fixed point in $D=4$ using several values
of $m$ are given.
As the fixed point values $g_\ast$ and $\lambda_\ast$ are scheme dependent they cannot possess 
a direct physical meaning.  On the other hand, in $D=4$ 
$g_\ast \, \lambda_\ast$ is an universal, {\it i.e.}, scheme-independent, 
quantity,  and  the product 
 $g_k \, \lambda_k =$ $(k^{D-2} \,\,\, \mathrm Z_{Nk}^{-1} \,\,\, 
\bar {G})$  $(k^{-2} \,\,\, \bar {\lambda}_k) \propto k^{D-4}$ is independent of the RG scale $k$. 
For general $D$ one generalises  this consideration and defines the $k$-independent
product 
\be
\tau_\ast = \lambda_\ast g_\ast^{2/(D-2)}
\ee
which can be shown to be an universal quantity. Therefore in an exact treatment the value of 
$\tau_\ast$ has to be the same for all cutoff schemes. Deviations for different regulators are thus
an indication of truncation errors.

In Tables \ref{tab:3.2} and \ref{tab:3.3}  results for the UV fixed point values in
$D=5$ and $D=8$ are displayed. In the three considered 
cases $D=4$, $D=5$ and $D=8$ one sees a  reasonably weak dependence 
of the universal quantity $\tau_\ast$  indicating the robustness of physical 
quantities under a change of the cutoff function. 

\begin{table}[h]
\centering
\begin{tabular}{|c| c |l| l|}
\hline\hline
$m$ & $g_\ast$ & $\lambda_\ast$ & $\tau_\ast$ \\
\hline
3/2 & 0.764 & 0.193 & 0.147 \\
2 & 1.166 & 0.118 & 0.138 \\
3 & 1.890 & 0.0663 & 0.125 \\
4 & 2.589 & 0.0460 & 0.119 \\
5 & 3.281 & 0.0352 & 0.115 \\
6 & 3.971 & 0.0285 & 0.113 \\
10 & 6.719 & 0.0162 & 0.109 \\
40 & 27.27 & 0.00381 & 0.104 \\
\hline\hline
\end{tabular}
\caption{\label{tab:3.1}
$\mathbf{D=4}$:  UV fixed points within the 
background field flow 
for various values of the cutoff parameter $m$.}
\end{table}
\begin{table}[h]
\begin{minipage}[b]{0.48\linewidth}
\centering
\begin{tabular}{|c| c| l| l|}
\hline\hline
$m$ & $g_\ast$ & $\lambda_\ast$ & $\tau_\ast$ \\
\hline
2 & 2.803 & 0.265 & {0.527} \\
3 & 7.254 & 0.141 & {0.528} \\
4 & 12.40 & 0.0958 & {0.513} \\
5 & 18.27 & 0.0726 & {0.504} \\
6 & 24.80 & 0.0585 & {0.497} \\
10 & 56.55 & 0.0329 & {0.485} \\
40 & 479.1 & 0.00767 &{0.470} \\
\hline\hline
\end{tabular}
\caption{\label{tab:3.2}
$\mathbf{D=5}$: UV fixed points within the 
background field flow 
for various values of the cutoff parameter $m$.}
\end{minipage}
\hfill
\begin{minipage}[b]{0.48\linewidth}
\centering
\begin{tabular}{|c| c| l| l|}
\hline\hline
$m$ & $g_\ast$ & $\lambda_{\ast}$ & $\tau_\ast$ \\
\hline
4 & 862.3 & 0.235 &{2.24} \\
5 & 2474 & 0.156 & {2.11} \\
6 & 5217 & 0.117 & {2.03} \\
10 & 33432 & 0.0584 & {1.88} \\
40 & 2.875 $\times 10^6$ & 0.0123 & {1.75} \\
\hline \hline
\end{tabular}
\caption{\label{tab:3.3}
$\mathbf{D=8}$:  UV fixed points within the 
background field flow 
for various values of the cutoff parameter $m$.}
\end{minipage}
\end{table}

For some selected cases the scaling exponents are given in Table \ref{tab:3.4}. The sizeable
imaginary part, $\theta^{\prime\prime}$, leads to the fact that the RG flows will ``spiral'' 
into the fixed point, see also next section.

\begin{table}[h]
\centering
\begin{tabular}{|l| c |c|}
\hline\hline
 & $\theta^\prime$ & $\theta^{\prime\prime}$ \\
\hline
$D=4$, $m=3/2$ & 2.004 & 1.659 \\
$D=4$, $m=2$ & 1.835 & 1.299 \\
$D=5$, $m=2$ & 3.595 & 2.896 \\
\hline\hline
\end{tabular}
\caption{\label{tab:3.4}
The scaling exponents for some values of the cutoff parameter $m$ and number of 
dimensions $D$.}
\end{table}

\goodbreak

\bigskip
~

\noindent
{\large{\uuline{\bf{Exact Functional Identity}}}}
~
\bigskip

Using Eqs.\ (\ref{eq:2.68})  and  (\ref{eq:2.72}), Eqs.\ (\ref{eq:3.2})  and  (\ref{eq:3.3}) can be written as
\bea
D \, \lambda_\ast &=& 8 \, \pi \, g_\ast \,\,\,
\frac {(4 \, \pi)^{-D/2}}{\Gamma \, (D/2)} \,\,\, \Bigg (\, \frac {(D + 1)}{(D + 2)} \,\,\, 
\frac {2 D }{(1 - 2 \, \lambda_\ast)} \, - 4 \, \Bigg ) \,\,, \\
2-D&=& \frac {g_\ast \,\,B_0 (\lambda_\ast)}{1 + g_\ast \,\,
 B_1 (\lambda_\ast)} \,\,,
\eea
with the functions $B_0(\lambda_k)$ and $B_1(\lambda_k)$ given in Eqs.\ (\ref{eq:2.73})
and (\ref{eq:2.74}), respectively. The condition for the UV fixed point can be also written as
\be
g_\ast = \frac {2-D}{B_0 (\lambda_\ast) + (D-2) B_1 (\lambda_\ast)} = \frac{D\lambda_\ast}
{A_0 (\lambda_\ast) + (D-2) A_1 (\lambda_\ast)} \, .
\label{eq:cond}
\ee

\begin{figure}[H]
\centering
\includegraphics[width=\textwidth]{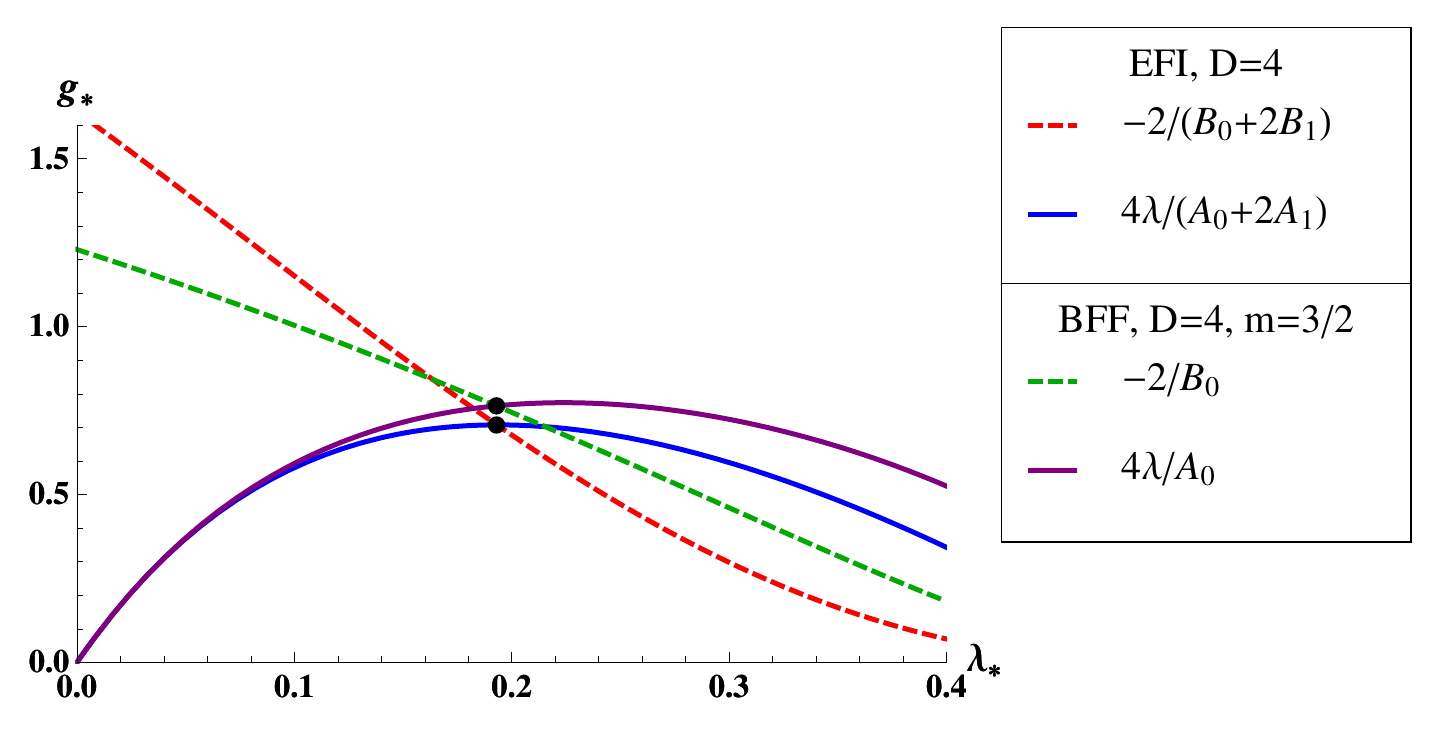}
\caption[Conditions for the UV fixed point  values for $D=4$
for the exact functional identity and the background field flow with $m=3/2$.]{\label{fig:3.1a} 
Conditions (\ref{eq:cond}) for the UV fixed point  values for $D=4$ for the exact functional identity
(dashed red and full blue lines) and   (\ref{eq:cond1}) for 
the background field flow with $m=3/2$ (dashed green and full purple lines).
The crossing of the curves, marked by the black points, provides the fixed point values
$g_\ast$ and ${\lambda}_\ast$.}
\end{figure}

A comparison with the results for the non-trivial UV fixed point shows that the corresponding 
values are almost identical to the one of the approximate background field flow 
with the minimally possible half-integer 
value of $m$, {\it cf.}, Table \ref{tab:3.5}. Note that convergence of the integrals
requires $m > (D/2) - 1$ , and therefore $m = (D-1)/2$ is the smallest half-integer 
number such that the integrals are well-defined.

\begin{table}[h]
\renewcommand{\arraystretch}{1.5}
\begin{center}
\begin{tabular}{|c| c c| c c| c c|}
\hline\hline
D & \multicolumn{2}{|c|}{4} &\multicolumn{2}{|c|}{5} & \multicolumn{2}{|c|}{8} \\
\hline
& $g_*$ & $\lambda_*$ & $g_*$ & $\lambda_*$ & $g_*$ &$\lambda_*$ \\
\hline
Exact functional identity & 0.707 & 0.193 & 2.853 & 0.237 & 461.4 & 0.284 \\
\hline
Background flow, $m=(D-1)/2$& 0.764 & 0.193 & 2.803 & 0.265 & 379.9 & 0.316 \\
\hline \hline
\end{tabular}
\end{center}
\caption{\label{tab:3.5}
UV fixed points for different number of dimensions for the two considered forms of the FRG.}
\end{table}
\begin{table}[h]
\centering
\begin{tabular}{|l| c | c |c|}
\hline\hline
 & $\tau_\ast $&$\theta^\prime$ & $\theta^{\prime\prime}$ \\
\hline
$D=4$ & 0.137 & 1.475 & 3.043 \\
$D=5$ & 0.478 & 2.687 & 5.154 \\
\hline\hline
\end{tabular}
\caption[The $k$-independent product $\tau_\ast$ and the scaling exponents
for $D=4$ and $D=5$.]{\label{tab:3.6}
The $k$-independent product $\tau_\ast$ and the scaling exponents derived from the
exact functional identity  for 
$D=4$ and $D=5$.
These values are identical to the ones given in \cite{Fischer:2006fz,Fischer:2006at}.
}
\end{table}


\goodbreak

\section{Phase Diagrams in $D \ge 4$}

In this section the RG flows in the $(\lambda_k,g_k)$ plane will be displayed for $D=4$ and 
$D=5$ for both considered forms of the FRG. 
Such ``phase portraits''  are called phase diagrams. They will 
be shown here because the comparison to them will make the $5D$-$4D$ crossover evident in 
the latter studied case of one compactified dimension, see the next chapter.
 
\bigskip\bigskip

\noindent
{\large{\uuline{\bf{Background Field Flow}}}}
~
\bigskip

A selection of results obtained from the approximate background field flow 
will be displayed first. 
As $m=2$ will be used for $D=4$ and $D=5$ as well as 
in the case of dimensional reduction I will specialise to this case.

For ${\bf D = 4}$ and ${\bf m = 2}$ one obtains for the functions $A_0$ (\ref{eq:2.65}) and $B_0$ 
(\ref{eq:2.63}):
\be
A_0 \, (\lambda_k) = \frac {-4 \, + \, 5 \, c_k}{2 \,\, \pi} \quad \mathrm {and} \quad B_0 \, (\lambda_k) 
= \frac {- 10 \, - \, 13 \, c_k^2}{6 \,\, \pi} 
\label{eq:A0B0}
\ee
with
\be 
c_k = \frac {1}{(1 - 2 \, \lambda_k)} \,\,. 
\label{eq:3.11}
\ee

For ${\bf D = 5}$ and ${\bf m = 2}$  one obtains for the functions $A_0$ (\ref{eq:2.65}) and $B_0$
(\ref{eq:2.63}):
\be
A_0 \, (\lambda_k) = \frac {-10 \, + \, 15 \, \sqrt{c_k}}{8 \,\, \pi} \quad \mathrm {and} \quad
B_0 \, (\lambda_k) = \frac {(- 11 / 3) \, - \, 7.5 \, \sqrt{c_k}}{8 \,\, \pi}  
\ee
with $c_k$ as defined in Eq.\ (\ref{eq:3.11}).

Note that due to the singularities in the $\beta$-functions the line $\lambda_k = \frac{1}{2}$  
provides a boundary line which cannot be 
crossed by the flows. To make the  flows visible in the full physically relevant range $
\lambda_k \in (-\infty, 1/2)$ and $g_k\in (0,\infty)$ in Fig.\ \ref{fig:3.2} the phase diagram is 
shown with compactified axes. This allows to identify several interesting points in the phase 
diagram.\footnote{As this type of plot looks very similar for the cases $D=4$ and $D=5$ 
the phase diagram is displayed only for $D=4$. Furthermore, the notation for these points 
follows the one in Ref.\ \cite{Litim:2012vz}.}

In Fig.\ \ref{fig:3.2} the point 
{\bf{A}} marks the previously discussed UV stable fixed point with coordinates 
$(g_{\ast}\,,\, \lambda_{\ast}) = (1.166\,,\, 0.118)$ in 4D (see Fig.\ \ref{fig:3.2}) and 
$(g_{\ast}\,,\, \lambda_{\ast}) = (2.803\,,\, 0.265)$ in 5D. 
{\bf{B}} is the already mentioned IR fixed point with coordinates 
$(g_{\ast}\,,\, \lambda_{\ast}) = (0\,,\, 0)$ present for all $D>2$. Furthermore,   
{\bf{C}} is a degenerate IR fixed point at $(g_{\ast}\,,\, \lambda_{\ast}) = (0\,,\, \frac{1}{2})$. 
And {\bf{D}} is an IR attractive fixed point located at negative cosmological constant with 
coordinates $(g_{\ast}\,,\, \lambda_{\ast}) = (0\,,\, -\infty)$ .
Finally, {\bf{E}} is a degenerate IR fixed point located at $(g_{\ast}\,,\, \lambda_{\ast}) = (+\infty\,,
\, -\infty)$ .

\begin{figure}[H]
\includegraphics[width=\textwidth]{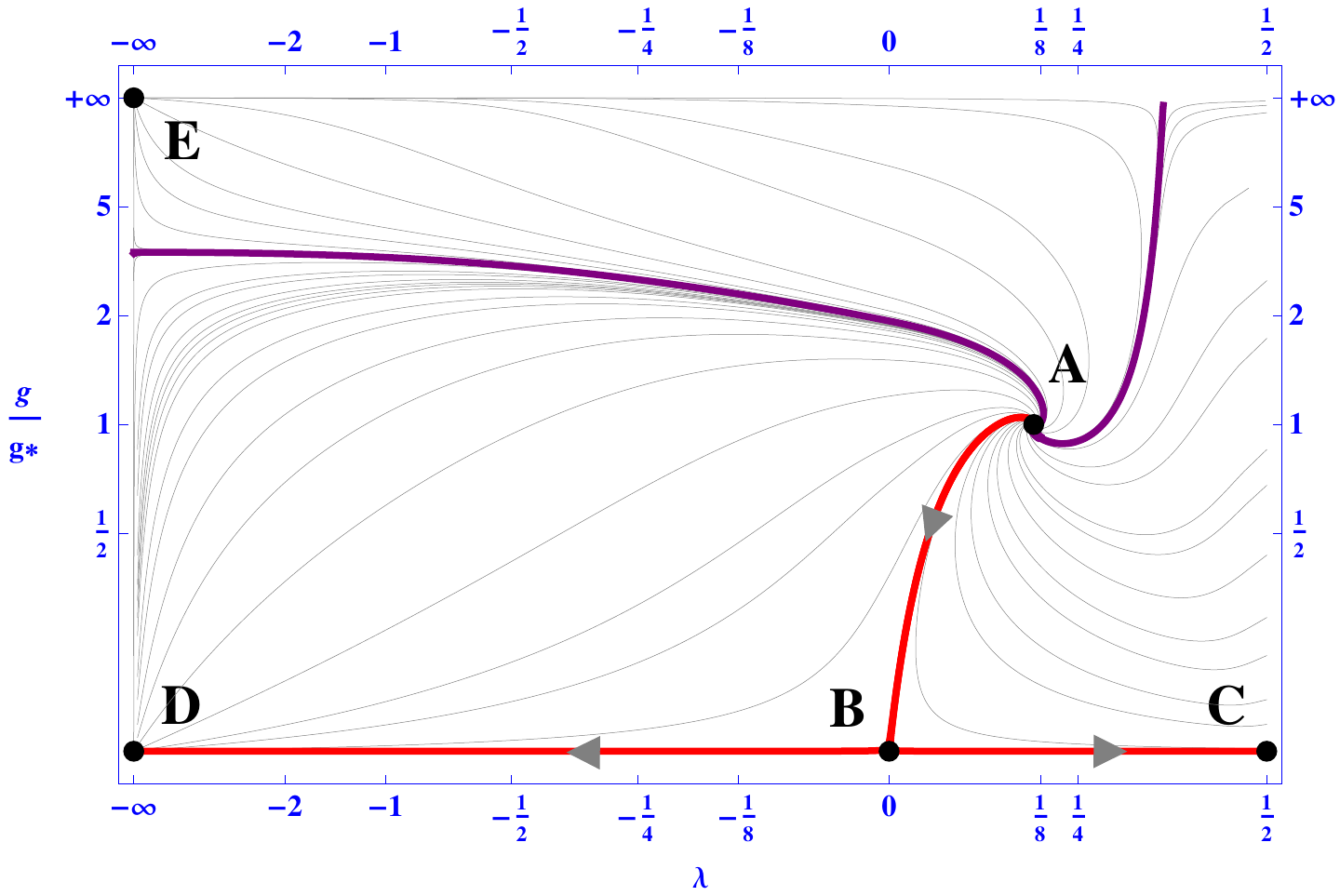}
\caption[The extended phase diagram for the background field flows in $D=4$.]
{\label{fig:3.2} The 
extended phase diagram for the background field
flows in $D=4$. For the meaning of the different points 
see the text above. To obtain this plot the differential equations (\ref{eq:2.64}) and (\ref{eq:2.66})
have been solved with the functions (\ref{eq:A0B0}) as input. The different trajectories are 
generated  by  different starting conditions. }
\end{figure}

The existence of these fixed point leads to several dividing lines or separatrices. 
Note that a physical RG trajectory is expected to pass close to the separatrix from 
{\bf{A}} to {\bf B}. Further separatrices run from {\bf B} to {\bf C} and from {\bf B} to {\bf D}.
The purple lines originating from {\bf A}  are the limits of a region where the RG trajectories
are completely disconnected from a weak-coupling and thus semi-classical regime.\footnote{A 
phase diagram where the strong-gravity region is shielded by separatrices from regions which 
admit an extended semi-classical regime has been obtained recently from
a truncation of the exact functional identity 
based on the graviton propagator \cite{Christiansen:2012rx}.}



\noindent
{\large{\uuline{\bf{Exact Functional Identity}}}}
~
\bigskip

Here the corresponding phase diagrams following from the exact functional identity
 are discussed.

For ${\bf D = 4}$  Eqs.\ (\ref{eq:2.69}-\ref{eq:2.70}) and  (\ref{eq:2.73}-\ref{eq:2.74}) simplify to
\begin{align}
A_0 \, (\lambda_k) & = \, \frac {5 \, c_k - 4}{2 \,\, \pi} \,\,,  &  A_1 \, (\lambda_k) =& \, \frac {5 \, c_k}
{12 \,\, \pi} \,\,, \quad  
\nonumber \\ \label{eq:3.13}~\\ 
B_0 \, (\lambda_k) & = \, \frac {- 9 \, c_k^2 + 5 \, c_k - 7}{3 \,\, \pi} \,\,, &
B_1 \, (\lambda_k) =& \, \frac {- 6 \, c_k^2 + 5 \, c_k}{12 \,\, \pi} \,\,, \nonumber
\end{align}
with $c_k$ as defined in Eq.\ (\ref{eq:3.11}).

The resulting flows are shown in Fig.\ \ref{fig:3.5}. Here an additional remark is in order: 
Whereas for the approximate background field flow
the $\beta$-functions became singular for $\lambda_k=1/2$ 
the corresponding boundary is, based on the exact functional identity, 
another line. It is determined from the 
divergence of the anomalous dimension:
\be
\frac {1}{\eta_{Nk}} = 0 \,\, \Longrightarrow \,\, g_{\mathrm{bound}} \, (\lambda_k) = 21 \, \pi^2 \,\, 
\frac {(1 - 2 \, \lambda_k)^2}{1 + 14 \, \lambda_k}
\label{eq:3.14}
\ee
and displayed in Fig.\ \ref{fig:3.5} as the dotted line.
\begin{figure}[H]
\centering
\includegraphics[width=0.98\textwidth]{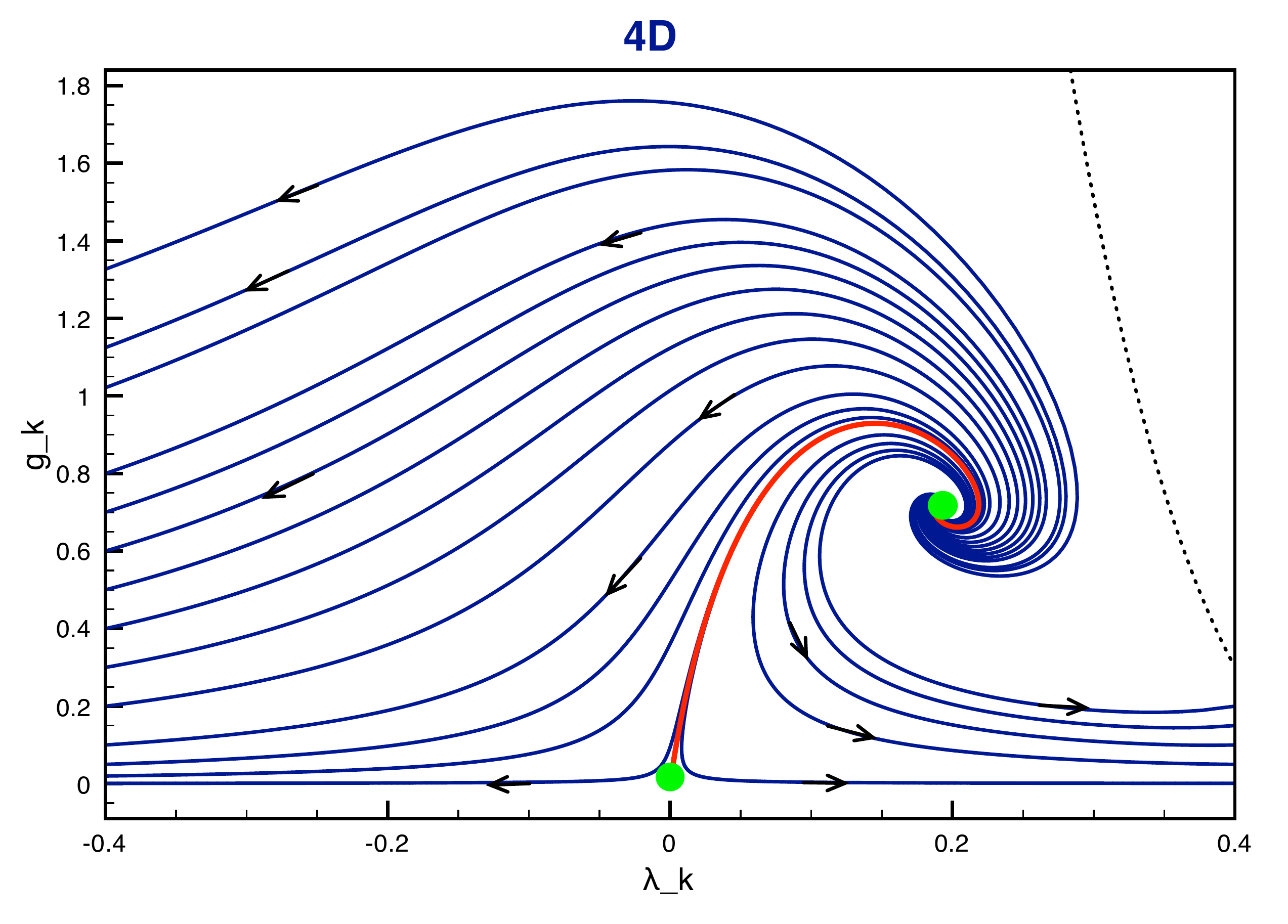} 
\vspace{-7mm}
\caption[The phase diagram for D=4 from the exact functional identity.]
{\label{fig:3.5}  Shown is the phase diagram for  the running gravitational 
coupling~$g_k$ and the cosmological constant $\lambda_k$ in four dimensions 
from the exact functional identity. The RG trajectories are solutions 
for different starting conditions of the differential
equations (\ref{eq:2.68}) and (\ref{eq:2.71}) with the functions (\ref{eq:3.13}) as input.
The separatrix  (full red line) connects the UV and the IR fixed points marked by green dots. 
Arrows indicate the direction of the RG flow with decreasing $k \to 0$. The dotted black line indicates the boundary $g_{\mathrm{bound}} 
\, (\lambda_k)$ where $1/\eta_{Nk} = 0$, see Eq.\ (\ref{eq:3.14}). }
\end{figure}

For ${\bf D = 5}$ the corresponding formulae read
\begin{align}
A_0 \, (\lambda_k) &= \, \frac {6 \, c_k - 4}{3 \,\, \pi^2} \,\,,  & A_1 \, (\lambda_k) &= \, \frac {2 \, c_k}
{7 \,\, \pi^2} \,\,, \quad 
\nonumber \\ 
~ \label{eq:3.15}\\
B_0 \, (\lambda_k) &= \, \frac {- 8 \, c_k^2 + 5 \, c_k - 74/15}{3 \,\, \pi^2} \,\,, & B_1 \, (\lambda_k) &= 
\, \frac {- 8 \, c_k^2 + 7 \, c_k}{21 \,\, \pi^2} \,\,, \nonumber
\end{align}
with $c_k$ as defined in Eq.\ (\ref{eq:3.11}).
Furthermore,
\be
\frac {1}{\eta_{Nk}} = 0 \,\, \Longrightarrow \,\, g_{\mathrm{bound}} \, (\lambda_k) = 21 \, \pi^2 \,\, 
\frac {(1 - 2 \, \lambda_k)^2}{1 + 14 \, \lambda_k}
\ee
which leads then to a phase diagram very similar to the four-dimensional case. 

In both cases one can make an interesting observation for the 
RG trajectories which pass close to the boundary line.  
\begin{figure}[H]
\includegraphics[width=\textwidth]{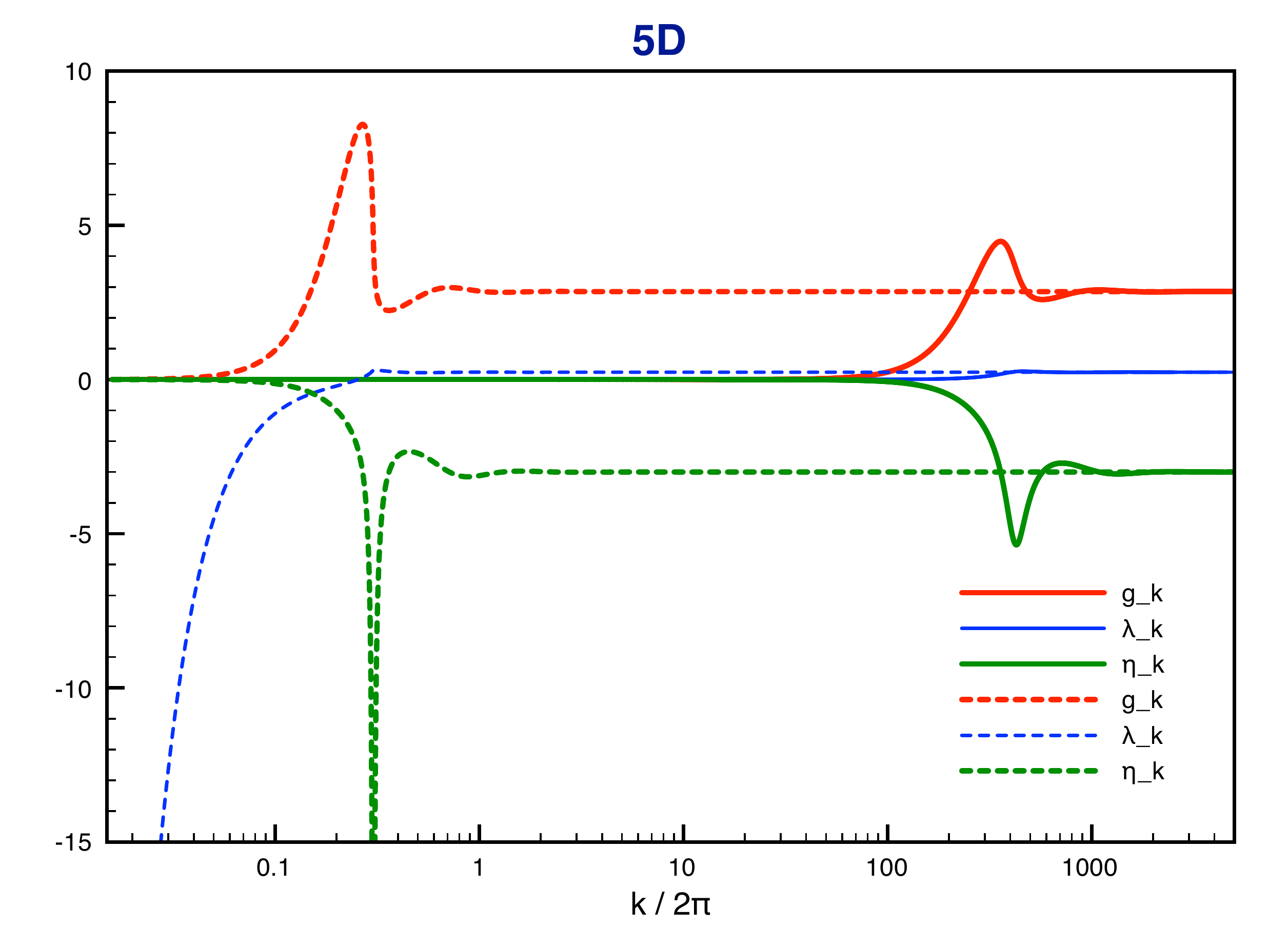} 
\vspace{-7mm}
\caption[$g_k$, $\lambda_k$ and $\eta _k$ for two different trajectories 
in D=5 for the exact functional identity.]
{\label{fig:3.6} Shown are $g_k$, $\lambda_k$ and $\eta _k$ in $D=5$ for two different trajectories,
{\it i.e.}, for two  different starting conditions of the differential
equations (\ref{eq:2.68}) and (\ref{eq:2.71}) with the functions (\ref{eq:3.15}) as input. The 
anomalous dimension $\eta_{Nk}$ is calculated using Eq.~(\ref{eq:2.72}).
For further details see the text below.}
\end{figure}
The runnings of $g_k$, $\lambda_k$ and $\eta _k$ are shown 
for two different trajectories in Fig.\ \ref{fig:3.6} in the case of five dimensions. Needless to say
that the same kind of comparison is possible (and actually has been performed) for the case 
of four dimensions. For the starting conditions $g_{k_{min}}=10^{-12}$ and 
$\lambda_{k_{min}} =10^{-6}$ at $k_{min}=0.1$ the trajectory (which is a trajectory practically 
identical to the separatrix) never comes close to the boundary 
line. Correspondingly, $g_k$ and $\lambda_k$ stay very small until the fixed point region is reached at 
approximately $k/2\pi\approx 500$, see the full lines in Fig.\ \ref{fig:3.6}. The anomalous dimension 
stays very small until the fixed point region is reached where it then quickly oscillates in to the fixed point value 
$\eta_{\ast}=2-D=-3$. The situation is very different from the RG trajectory corresponding to the 
dashed lines in Fig.\ \ref{fig:3.6}.  Here the starting conditions are $g_{k_{min}}=4\cdot 10^{-3}$ and 
$\lambda_{k_{min}} =-45$ at $k_{min}=0.1$. At a value $k/2\pi \approx 0.3$ the RG trajectory passes
close to the boundary line. At this point $g_k$ rapidly decreases and the anomalous dimension 
becomes strongly negative. Very soon after this the fixed point region is reached and all quantities 
oscillate with a very small and further decreasing amplitude around the fixed point values. 

Therefore one concludes that strongly negative values of the anomalous dimension are assumed
on RG trajectories passing close to the boundary line. In order to test whether this might be 
a truncation and regulator artefact the case has been investigated in which the function 
$B_1(\lambda_k)$ has been neglected. The boundary line becomes then 
the vertical line $\lambda_k=1/2$ (as is also the case for the approximate background field flow). 
Using now the same 
starting conditions in the IR the RG trajectory comes the close to the vertical boundary line
with a very similar effect on the anomalous dimension. On the other hand, except for regions
very close to the boundary line the value and the running of the cosmological constant is of 
only minor importance. 

\bigskip

This concludes  the summary  of the RG flows for the two investigated forms of the flow
equation. 


\newpage
\thispagestyle{empty}

\chapter{{\color{BrickRed}Dimensional Reduction for Einstein-Hilbert Gravity}}
~\vspace{-10mm}

\rightline{\footnotesize{\it{``The theory has to be interpreted that extra dimensions beyond the}}}
\rightline{\footnotesize{\it{ordinary four dimensions, the three spatial dimensions plus time,}}}
\rightline{\footnotesize{\it{are sufficiently small that they haven't been observed yet."}}}
\rightline{\footnotesize{\it{(E.\ Witten)}}}
\minitoc

\bigskip

In this chapter, which constitutes the main part of this thesis,  the  RG for EH gravity for the
case of four extended and one compact dimensions will be treated.
The main objective is to provide an explicit example for the realisation of the ADD model.
Although the case of only one extra compact dimension is excluded by observation  
(because then the extension of the extra dimension would be of the size of the solar system)
such a scenario is a theoretically sound model.
Note, however,  that in this simplest version of the ADD model one needs to have 
an enormously large separation of scales, 
\be
1/L \ll M_D \quad {\mathrm {with}} \quad M_D \approx  m_{EW}
\ee
being the Planck scale in $D$ dimensions. 
From Eq.\ (\ref{eq:1.3}) it is evident 
that for $n=1$ there are approximately 32 orders of magnitude between 
$1/L$ and $M_D$.

The first step in the treatment will be to perform the functional trace in a setting where
one dimension is chosen compact, {\it i.e.}, periodic boundary conditions apply in one 
direction. 
The  field modes propagating along the compact dimension are 
called Kaluza-Klein modes, a suitable  formalism to include them in the trace 
is the Matsubara mode sum known from thermal quantum field theory.  As the
corresponding formulae are the basis of the dimensionally reduced RG equations to be derived below
I will here give a more detailed representation of the relevant expressions for the Matsubara 
sums although their treatment is quite technical.

\section[Matsubara Formalism and Kaluza-Klein modes]{Matsubara Formalism and Kaluza-Klein\\
 modes}

The compact dimension implies periodic boundary conditions on the metric 
field
\be
g_{\mu\nu} (x_1, \ldots x_{D-1}, x_D) = g_{\mu\nu} (x_1, \ldots x_{D-1}, x_D+L)
\label{eq:4.1}
\ee
where without loss of generality it was assumed that $D$th dimension is the 
compact one. The maximal extension along this dimension has been called $L$
(in agreement with the discussion in Subsection~1.2.2).

To proceed one should recall the calculations described in Chapter~2.
The traces appearing in the FRG equations have been performed by applying 
a heat kernel expansion to the traces of a function of the covariant background 
Laplacian, keeping the first two terms and integrating then them
over the covariant background field momenta.  It is obvious that when considering a
compact dimension one has to correspondingly adapt this procedure.

Considering $D$ dimensions with periodicity as in Eq.\ (\ref{eq:4.1}) implies that 
one has then $D-1$ continuous 
components of the momentum vector and one discrete one. The values for the discrete
momentum component resembles formally the Matsubara frequencies for bosons:
\be
q_{D,n} = \frac{2\pi n}{L} \,\, ,
\label{eq:4.2}
\ee
such that 
\be
q_D^2\to q_{D-1}^2 +(2\pi n /L)^2,
\ee
and correspondingly the $D$-dimensional integrals split into a $(D-1)$-dimensional
integral and a sum,\footnote{Such Matsubara sums have been introduced  very recently 
({\it i.e.}, after the calculations of this thesis have been completed)
within Quantum Gravity in Ref.\ 
\cite{Rechenberger:2012dt} where the Einstein-Hilbert theory has been investigated using 
the Wetterich equation. 
The objective of this work is, however, to consider foliated space times.}
\be
\int \frac{d^Dq}{(2\pi)^D} \rightarrow \frac{1}{L} \, \sum_n \, \int \frac{d^{D-1}q}{(2\pi)^{D-1}} \,\,.
\label{eq:4.3}
\ee
Physically, Eq.\ (\ref{eq:4.2}) describes the possible momenta of gravitons propagating along 
the $D$th dimension (Kaluza-Klein modes). On a formal level, Eqs. (\ref{eq:4.2}) and (\ref{eq:4.3}) are 
the basis for describing dimensional reduction. The appearing length scale $L$ will lead to 
a separation: For $k\gg 1/L$, {\it i.e.}, in the UV, the behaviour of the RG trajectories will be 
according to $D$ dimensions, in the opposite limit according to $D-1$ dimensions. Hereby the 
implementation of the substitution (\ref{eq:4.3})  into the formalism will  allow to derive 
$\beta$-functions and to calculate RG flows which describe the crossover from a
$(D-1)$- to a $D$-dimensional  characteristic.

The modifications in the calculations of the functional traces related to the 
substitution (\ref{eq:4.3}) are discussed in Appendix~D.3.  A recapitulation of one aspect of
the previously described formalism is here in order:  The angular integrals can be performed
leading to one-dimensional integrals, the so-called threshold functions $l_m^D(w)$. 
For completeness, in Appendix~E their definition is given.  In the same way one can do
the angular part of the $D-1$-dimensional momentum integrals. This leads to generalised
threshold functions $l_m^D(w,L)$ which are expressed as a Matsubara sum of one-dimensional
integrals, for their definition see Appendix~E, where also their limits $L\to \infty$ and $L\to 0$ are
given. In addition, for the Litim regulator the dependence on $L$ factorises.

However, there is one subtle issue related to the heat kernel expansion.  In the integrand of the term
linear in the Ricci  scalar one additional factor $1/q^2$ appears as compared to the ${\cal O}(\bar R^0)$ 
term. The existence of such a term (which will later on prove to be the origin of a substantial complication)
is evident also from a dimensional analysis because $\bar R$ has mass
dimension two. The full expressions resulting from evaluating the trace with the help of a heat kernel 
expansion are given in Appendix~D.3, here in this section I want to focus the description on the 
problematic term. The substitution (\ref{eq:4.3}) implies for such a term
\bea
\int \frac{d^Dq}{(2\pi)^D}  \frac 1 {q_D^2} \ldots \bar R
& \rightarrow&  \frac{1}{L} \, \sum_n \, \int \frac{d^{D-1}q}{(2\pi)^{D-1}} \frac 1 {q_{D-1}^2 +(2\pi n /L)^2}
\ldots \bar R
\nonumber \\
&\propto&   \frac{1}{L} \, \sum_n \, 
 \int_0^{\infty} \, dz \,\, z^{D/2 - 3/2} \,\,  \frac 1 {z+(2\pi n /L)^2} \ldots \bar R 
\eea
As will become evident below the term $\displaystyle{\frac 1 {z+ (2\pi n /L)^2}}$ in 
the part of this expression linear in the Ricci scalar leads to quite complicated
sums over $n$. Lowering for this part the dimensionality of the trace simplifies
the heat kernel expansion sufficiently such that the sums and integrals can be
done in a closed form. The approximate formula corresponding to the last line 
above reads
\be
\frac{1}{L} \, \sum_n \, 
 \int_0^{\infty} \, dz \,\, z^{D/2 - 3/2} \,\,  \frac 1 {z+(2\pi n /L)^2} \ldots \bar R  \rightarrow
\frac{1}{L} \, \sum_n \,  \int_0^{\infty} \, dz \,\, z^{D/2 - 5/2} \ldots \bar R 
 \ee
with the full expressions for the heat kernel expansions given in Eqs.\ (\ref{eq:4.9})
and (\ref{eq:4.10}), respectively. 
Although for completeness also the formulae following 
from the heat kernel expansion (\ref{eq:4.9}) will be presented
in the next section, the evaluation of the traces will be done using the modified
heat kernel expansion (\ref{eq:4.10}).  


\section{Dimensionally Reduced Renormalisation Group Equations}

\noindent
{\large{\uuline{\bf{Background field flow}}}}
~
\bigskip

\noindent
To obtain the following expression I start from Eq.\ (\ref{eq:C.2}) and use 
Eq.\ (\ref{eq:4.10}) for the evaluation of the traces. Some explicit steps of this 
calculation are given in Appendix~D.  To proceed let me note that with a few 
trivial modifications Eq.\ (\ref{eq:C.22}) can be rewritten as
\be
\left.\hspace{-3cm} 
2 \, \kappa^2 \, \int d^D x \, \sgb\, 
\Big(\partial_t \, \mathrm Z_{Nk} \, (- \bar {R}(\bar {g}) + 2 \, \bar \lambda_k )
 + 2 \, \mathrm Z_{Nk} \, \partial_t \,\bar \lambda_k \Big) \right.=  
\label{eq:4.11} \\ 
 \ee
\be
\left.
\hspace{-9cm} 
 = (4 \, \pi)^{-(D-1)/2} \,\,\, \frac{k^{D-3}}{\Gamma \, (m+1)} \,\,\, \times  \right.
\nonumber
\ee
\bea
& \times&  \!\!\! \Bigg\{
\int d^D \, x \,\, \sgb\,\,\, k^2 \,\,\,
\Bigg[\frac{1}{L} \,\,\, \sum_n \,\,\, \Gamma \, \bigg(-
\frac{(D-3)}{2} + m\bigg) \,\,\, \times \nonumber \\
&&\times\Bigg( \frac{1}{2} \,\, D \, (D+1) \,\, \big( \omega_n^2 + 1 - 2\, \lambda_k \big)^{(D-3)/2 \,-\, 
m} - 2 \, D \,\, \big( \omega_n^2 + 1 \big)^{(D-3)/2 \,-\, m} \Bigg) \Bigg]
\nonumber \\
&+& \int d^D \, x \,\, \sgb \,\, \bar R \,\,\,\Bigg[\frac{1}{L} \,\,\, \sum_n \,\,\, 
\Gamma \, \bigg(-\frac{(D-5)}{2} + m\bigg) \,\,\, \times \nonumber \\
&&\times  \Bigg( \frac{-\,D\,\,(5\,D^2 - 23\,D + 20)}{12 \, (D-3)} \,\, \big( \omega_n^2 + 1 - 2\, 
\lambda_k \big)^{(D-5)/2 \,-\, m} 
\nonumber \\ 
&&  \quad +\Big( \frac{-D^2 - 4D + 18}{3 \, (D-3)} \Big) \,\, \big( \omega_n^2 + 1 \big)^{(D-5)/2 \,-\, m} 
\Bigg) \Bigg]  \Bigg\} 
\eea
\noindent
with
\be
\omega_n^2 := \Big(\frac{2\,\pi\,n}{k\,\,L}\Big)^2 \,\,\,.
\label{eq:4.13}
\ee

\noindent
From this one infers that the $\beta$-functions take the form Eq.\ ({\ref{eq:2.64}) 
and Eq.\ ({\ref{eq:2.66}) with the anomalous dimension as in Eq.\ (\ref{eq:2.62}) the only 
difference being that the functions  $A_0 \, (\lambda_k \,;\, kL)$ and 
$B_0 \, (\lambda_k \,;\, kL)$ are deduced from 
\bea
&&k^D \,\,\, A_0 \, (\lambda_k \,;\, kL) = 8 \,\, \pi \,\, (4 \, \pi)^{-(D-1)/2} \,\,\, 
\frac{\Gamma \, (-(D-3)/2 + m)}{\Gamma \, (m+1)} \,\,\, \frac{k^D}{kL} \,\,\, \times
\nonumber \\
&&\times \sum_n \,\,\, \Bigg[ \frac{1}{2} \,\, D \, (D+1) \,\, 
\big( \omega_n^2 + 1 - 2\, \lambda_k \big)^{(D-3)/2 \,-\, m} - \, 2 \, D \,\, 
\big( \omega_n^2 + 1 \big)^{(D-3)/2 \,-\, m} \Bigg] \,\, ,\nonumber  \\ 
&& {\mathrm{and}} \nonumber \\
&&k^{D-2} \,\,\, B_0 \, (\lambda_k \,;\, kL) = 16 \,\, \pi \,\, (4 \, \pi)^{-(D-1)/2} \,\,\, 
\frac{\Gamma \, (-(D-5)/2 + m)}{\Gamma \, (m+1)} \,\,\, \frac{k^{D-2}}{kL} \times
\nonumber\\
&&\times 
\sum_n \,\,\, \Bigg[ \frac{-\,D\,\,(5\,D^2 - 23\,D + 20)}{12 \, (D-3)} \,\, 
\big( \omega_n^2 + 1 - 2\, \lambda_k \big)^{(D-5)/2 \,-\, m} \nonumber \\
&&\qquad \qquad+ \Bigg( \frac{-D^2 - 4D + 18}{3 \, (D-3)} \Bigg) \,\,
 \big( \omega_n^2 + 1 \big)^{(D-5)/2 \,-\, m} \Bigg] \,\,. 
\label{eq:4.14}
\eea

\goodbreak 

\noindent
Putting $\bf D=5$ and $\bf m=2$ one can perform the sums which results in:
\bea
A_0 \, (\lambda_k \,;\, kL) &=& \frac 1 {8 \,\, \pi} \Bigg [ \frac {15}  {\sqrt{ 1 - 2\, \lambda_k}} \,\, 
\coth \,\, \Bigg(\frac{k\,\,L  \sqrt{1 - 2\, \lambda_k}}2 \Bigg) \,\,
- \,\, 10 \,\, \coth \,\, \Bigg(\frac{k\,\,L}{2}\Bigg) \Bigg ]  \nonumber \\
\label{eq:4.15} \\
B_0 \, (\lambda_k \,;\, kL) &=& - \,\, \frac 1 {64 \,\, \pi}  \Bigg [ 
\frac{25}{1 - 2\, \lambda_k}  \,\, \csch^2 \,\, 
\Bigg(\frac{k\,\,L \sqrt{1 - 2\, \lambda_k}}2 \Bigg) \,\, 
\nonumber \\
&\times& \Bigg( k\,\,L \,\,
 + \,\, \frac 1 {\sqrt{1 - 2\, \lambda_k}} \,\, \sinh \,\, \Bigg(k\,\,L \sqrt{1 - 2\, \lambda_k}
\Bigg) \Bigg) 
 \nonumber \\
&&  \qquad \qquad + 18 \,\, \csch^2 \,\, \Bigg(\frac{k\,\,L}{2}\Bigg) \,\, \Big( k\,\,L \,\, + \,\, \sinh \,\, ( k
\,\,L ) \Big) \Bigg ] \,\,.
\label{eq:4.16}
\eea

\noindent
To obtain closed expressions for the functions  $A_0 \, (\lambda_k \,;\, kL)$ and 
$B_0 \, (\lambda_k \,;\, kL)$ it was necessary to employ heat kernel expansion
in  the form (\ref{eq:4.10}).  Had one used instead the expansion (\ref{eq:4.9}) one 
would obtain an unchanged $A_0 \, (\lambda_k \,;\, kL)$  but a different
\bea
B_0 \, (\lambda_k \,;\, kL) &=& \; \frac {1}{\pi} \,\,\, \frac {1}{(1 - 2\, \lambda_k)^{3/2}} \,\,\, 
\frac {1}{k \, L \,\, \sqrt {1 - 2 \, \lambda_k}} \,\,\,
 \nonumber \\
 &&\sum_n \, \Big \{  - \,\, \frac {5}{(\Omega_n^2 + 1)^2} \,\, +
  \,\, \frac {15}{4} \,\, \Big ( 1 - \frac {1}{2} \,\, \frac {1}{\Omega_n^2 + 1} \,\, - 
  \,\, \Omega_n^2 \,\, \ln \,\, \Big (1 + \frac {1}{\Omega_n^2} \Big ) \Big ) \Big \} 
 \nonumber \\[2mm]
&-& \; \frac {1}{\pi} \,\,\, \frac {1}{k \, L} \,\,\, \sum_n \, 
\Big \{ \,\, \frac {1}{(\omega_n^2 + 1)^2} \,\, 
 \nonumber \\ 
 && \qquad  \qquad  +\,\, \frac {5}{2} \,\, 
\Big ( 1 \,\,-\,\, \frac {1}{2} \,\, \frac {1}{\omega_n^2 + 1} \,\, - 
\,\, \omega_n^2 \,\, \ln \Big ( 1 \,\, + \,\, \frac {1}{\omega_n^2} \Big ) \Big ) \Big \}\,\,\,, 
 \nonumber \\[2mm]
&&\quad \mathrm {with} \quad \omega_n := \frac {2 \, \pi \, n}{k \, L} \quad
 \mathrm {and} \quad \Omega_n^2 = \frac {\omega_n^2}{1 - 2 \, \lambda_k}\,\,\,,
 \label{eq:4.17}
\eea
for which no closed expression could be found.

\vspace{10mm}

\noindent
{\large{\uuline{\bf{Exact Functional Identity}}}}
~
\bigskip

As it is evident from the discussion in Appendix~D the step function contained in the Litim 
regulator leads to an upper limit in the appearing integrals over the square of the 
integrated momentum, $z=q^2$. Without dimensional reduction the relevant integrals
are of the form ({\it cf.}, Eq.\ (\ref{eq:C.16}))
\be
\int_0^{\infty} \, dz \,\, z^{D/2 - 1}  z^m \,\, \Theta \,\, \Bigg ( 1 - \frac {z}{k^2} \Bigg ) =
\int_0^{k^2} \, dz \,\, z^{D/2 - 1+m}    \,\,,
\ee
with $m=-1,0,1$.  Now, with the compactification of one dimension the corresponding 
sums plus integrals read:
\bea
&&\frac{1}{L} \,\,\, \sum_{n=-\infty}^{\infty}  \int_0^{\infty} \, dz \,\, z^{D/2 - 3}  z^m 
\Big( z + \big({2\,\pi\,n/L}\big)^2\Big)^l \,\,  
\Theta \,\, \Bigg ( 1 - \frac {z+ \big({2\,\pi\,n/L}\big)^2}{k^2} \Bigg ) = 
\nonumber \\
&&=\frac{1}{L} \,\,\,   \sum_{n= -[k L/2 \pi]}^{[k L/2 \pi]}  \int_0^{k^2- \big({2\,\pi\,n/L}\big)^2} 
dz \, z^{D/2 - 3}  z^m  \Big( z + \big({2\,\pi\,n/L}\big)^2\Big)^l    \,\,,
\eea
with appropriate values of $m$ and $l$.
Due to the step function the upper limit of the integral depends on $n$ but note that
also $n$ cannot exceed $kL/2\pi$ due to the regulator.  This constraint implies finite 
sums with $n_{\mathrm min} = - [kL/2\pi]$ and $n_{\mathrm max} =  [kL/2\pi]$ where 
$[\ldots ]$ denotes Gau\ss 's bracket, resp., the floor function.

The sums are now finite, but again the heat kernel expansion 
(\ref{eq:4.10}) leads to closed expressions for the functions
$A_0$, $A_1$, $B_0$ and $B_1$,  whereas the heat kernel expansion  (\ref{eq:4.9}) 
does not.  I will use both forms in the following.
Hereby one profits from the Litim regulator: 
For small and intermediately large values of $kL$ the sums can be performed easily 
numerically. For large values of $kL$ the sums are accurately approximated by the
leading order using the Euler-MacLaurin formula, see also Subsection~4.4.

The derivation of the $\beta$-functions is then straightforward although quite lengthy,
an intermediate step is given once more in Appendix~D.  
Not surprisingly the flow equations 
have  again the form (\ref{eq:2.68}), (\ref{eq:2.71}) and (\ref{eq:2.72}), however, with the
functions $A_0$, $A_1$, $B_0$ and $B_1$ given by
\bea
A_0 (\lambda_k;kL) &=& \frac 1{2 \, \pi} \,\,\, \frac {1}{k \,\, L} \,\, 
s_2\left( \frac{k L}{2 \pi}\right) 
\,\,\, \Bigg ( \frac {15}{2} \,\, \, \Big ( \frac {1}{1 - 2 \, \lambda_k} \Big ) - 5 \Bigg ) \,\,, 
\label{eq:4.18} \\
A_1 (\lambda_k;kL) &=&  \frac 1{2 \, \pi} \,\,\, \frac {1}{k \,\, L} \,\, 
s_3\left( \frac{k L}{2 \pi}\right) 
\,\,\, \frac {5}{4} \,\, \, \Big ( \frac {1}{1 - 2 \, \lambda_k} \Big ) \,\,, 
\label{eq:4.19} \\
B_0 (\lambda_k;kL) &=& \frac 1 { \pi} \,\,\, \frac {1}{k \,\, L} \,\,  \Bigg[
s_2\left( \frac{k L}{2 \pi}\right)  \,\,\, 
\Bigg ( - 5 \,\,\, \Big ( \frac {1}{1 - 2 \, \lambda_k} \Big )^2 - 1 \Bigg ) \nonumber \\
&&~~~~~~~~~~~~+
s_1\left( \frac{k L}{2 \pi}\right) \,\, 
\Bigg ( \frac {15}{4} \,\, \Big ( \frac {1}{1 - 2 \, \lambda_k} \Big ) - \frac {5}{2} \Bigg ) \Bigg ] \,\,,  
\label{eq:4.20} \\
B_1 (\lambda_k;kL) &=& \frac 1 \pi  \,\,\, \frac {1}{k \,\, L} \,\,   \Bigg[
s_3\left( \frac{k L}{2 \pi}\right)  \,\,\, 
\Bigg ( - \frac {5}{6} \,\,\, \Big ( \frac {1}{1 - 2 \, \lambda_k} \Big )^2 \Bigg ) 
\nonumber  \\
&&~~~~~~~~~~~~+
\frac {15}{16} \, s_2\left( \frac{k L}{2 \pi}\right) \,\, 
\Big ( \frac {1}{1 - 2 \, \lambda_k} \Big ) \Bigg] \,\,,
\label{eq:4.21}
\eea
for the heat kernel expansion (\ref{eq:4.10}). Hereby 
\be
s_l(a):=\sum_{-[a]}^{[a]} \left( 1 - \left( \frac n a \right)^2 \right)^l \, , \quad l=1,2,3,
\ee
with $a=kL/2\pi$ and these functions are normalised such that $s_l(a)=1$ for $a<1$. These 
three sums can be performed explicitly, the corresponding expressions are given in Eq.\ 
(\ref{eq:D.40}) in Appendix~D.5.

For the case of the heat kernel expansion (\ref{eq:4.9}) one obtains  instead of the expressions 
above for $B_0$ and $B_1$: 
\bea
B_0 (\lambda_k;kL) &=& \frac 1 { \pi} \,\,\, \frac {1}{k \,\, L} \,\, 
\sum_{-[k L/2 \pi]}^{[k L/2 \pi]} \,\, \Bigg[ \Big ( 1 - \, \omega_n^2\,\, \Big )^2 \,\,\, 
\Bigg ( - 5 \,\,\, \Big ( \frac {1}{1 - 2 \, \lambda_k} \Big )^2 - 1 \Bigg ) \nonumber \\
&+&
\Big ( 1 -\, \omega_n^2 +\, \omega_n^2 \,\, \ln \, \omega_n^2\Big ) \,\, 
\Bigg ( \frac {15}{4} \,\, \Big ( \frac {1}{1 - 2 \, \lambda_k} \Big ) - \frac {5}{2} \Bigg ) \Bigg ] \,\,,  
\label{eq:4.20a} \\
B_1 (\lambda_k;kL) &=& \frac 1 \pi  \,\,\, \frac {1}{k \,\, L} \,\, 
\sum_{-[k L/2 \pi]}^{[k L/2 \pi]} \,\, \Bigg[ \Big ( 1 - \, \omega_n^2\,\,  \Big )^3 \,\,\, 
\Bigg ( - \frac {5}{6} \,\,\, \Big ( \frac {1}{1 - 2 \, \lambda_k} \Big )^2 \Bigg ) 
\nonumber  \\
&+&
\frac {15}{16} \, \Big ( 1 - \,\omega_n^4 + 2 \,\, \omega_n^2 \,\, \ln \, \omega_n^2  \Big) \,\, 
\Big ( \frac {1}{1 - 2 \, \lambda_k} \Big ) \Bigg] \,\,,
\label{eq:4.21a}
\eea
where the definition (\ref{eq:4.13}) for $\omega_n^2$ has been used. The functions $A_0$ and 
$A_1$ do not change.


\section{$\beta$-Functions and Effective Coupling}

For the convenience of the reader the expressions for the 
$\beta$-functions and the anomalous 
dimension are repeated ({\it cf.}, Eqs.\ (\ref{eq:2.68}),  (\ref{eq:2.71}) and (\ref{eq:2.72})):
\begin{align}
\partial_t \, \lambda_k &= \beta_{\lambda} \, ( g_k , \lambda_k ) = \eta_{Nk} \, \lambda_k - 2 \, 
\lambda_k + g_k \,\,\, 
\Big( A_0 \, (\lambda_k\,;\, kL) - \eta_{Nk} \,\, A_1 \, (\lambda_k\,;\, kL) \Big) \,\,\,
\label{eq:4.24}  
\qquad \\
\partial_t \, g_k &= \beta_g \, ( g_k , \lambda_k ) = ( D - 2 + \eta_{Nk} ) \, g_k \,\,, \quad {\mathrm {with}}
\quad
\eta_{Nk} = \frac {g_k \,\,B_0 (\lambda_k\,;\, kL)}{1 + g_k \,\, B_1 (\lambda_k\,;\, kL)} \,\,.
\label{eq:4.25}
\end{align}
In the approximate background field flow
one has $A_1=B_1=0$ with $A_0$ and $B_0$ given in the case of 
dimensional reduction in Eqs.\ (\ref{eq:4.15}) and (\ref{eq:4.16}). For the case starting from 
the exact functional identity the
four functions are given in Eqs.\ (\ref{eq:4.18}) - (\ref{eq:4.21}).  

A decisive observation is now that in both considered cases, these four functions 
are proportional to $1/kL$ for small $kL$, for a more detailed discussion of the 
limits of these functions see the next subsection. And as one sees from Eqs.\ 
(\ref{eq:4.24}) and (\ref{eq:4.25}) these functions appear always exactly with 
precisely one factor $g_k$. This makes plain that a relation between an effective
running (dimensionless) coupling in four space-time dimensions to the one in five
space-time dimensions has to take the form 
\begin{equation}
\boxed{g_k^{\,\,4D} = \frac {g_k^{\,\,5D}}{kL} \quad {\mathrm{for}} \quad kL\ll 1 .}
\label{eq:gk4d5d}
\end{equation}

This identification is corroborated by the following dimensional analysis. Going back to the
dimensionful Newton coupling $G_N$ with canonical dimension $2-D$ ({\it i.e.}, $G_N^{\,\,5D}$ has 
dimension -3 and $G_N^{\,\,4D}$ dimension -2) it is obvious that dimensional reduction cannot
consist only in taking the limit $L\to 0.$ Instead one needs to replace the coupling
$G_N^{\,\,5D}$ by a coupling of the correct canonical dimension for four dimensions, and because
$L$ is the only scale available one needs to identify $G_N^{\,\,5D}/L$ with $G_N^{\,\,4D}$.
Inserting this identification into Eq.\ (\ref{eq:gk}) one arrives also at the relation 
Eq.\ (\ref{eq:gk4d5d}).

Of course, one does not want to change the coupling ``by hand'' or in some other discontinuous 
way. In order to obtain a more appropriate description for the physics of dimensional reduction
one is looking for an effective coupling
$g_{\mathrm{k,eff}}$ fulfilling the following requirements:
\begin{itemize}
\item
It should be well-defined and finite in both limits $L\to \infty$ and $L\to 0$. 
\item
It should connect smoothly  both limits.
\item
In the (semi-)classical regime it should behave like $k^2$ ({\it i.e.}, like a four-dimensional
running coupling) for $k\ll 1/L$ and like $k^3$ ({\it i.e.}, like a five-dimensional
running coupling) for $k\gg 1/L$.
\item 
The crossover from four- to five-dimensional behaviour should occur for scales $k \approx 1/L$.
\end{itemize}

A definition which fulfils these requirements is given by
\be
g_{\mathrm{k,eff}} = \; g_k \,\, \frac {B_0 \, (\lambda_k\,;\, kL)}{B_0^{\infty}} \quad \mathrm 
{with} \quad B_0^{\infty} = \lim_{L \to \infty} \,\, B_0 \, (\lambda_k\,;\, kL) \,\,\,.
\label{eq:4.28}
\ee
By construction $g_{\mathrm{k,eff}} \to  g_k$ for large $L$,  {\it i.e.}, it becomes identical to
the running  (dimensionless) coupling  in five space-time dimensions for $L\gg 1/k$. In 
the opposite limit $L\to 0$ the effective coupling $g_{\mathrm{k,eff}} $ stays (on a general 
RG trajectory)
non-vanishing and finite due to the $1/kL$ divergence of the function $B_0$. 

In Chapter~3 it was recognised that the influence of the function $B_1$ on the RG
trajectories is of absolutely of minor importance except when a trajectory passes 
closely to the boundary line $1/\eta_{Nk}=0$. As those trajectories are not in the 
class of  ADD model trajectories I am looking for (see also below) 
let us neglect for the sake of a qualitatively correct and quantitatively reasonably accurate
 discussion the function $B_1$ in 
the anomalous dimension derived from the exact functional identity
(in the corresponding approximate background field flow
expression it is anyhow not present). Then one can define an effective 
anomalous dimension $\eta_{Nk,\mathrm{eff}}$ which deviates from $\eta_{Nk}$ only
by the scale derivative of the logarithm of the function $B_0$:
\be
\eta_{Nk,\mathrm{eff}} = \; \underbrace{B_0^{\infty} \,\, g_{\mathrm{k,eff}}}_{\eta_{Nk}} + \frac 
{\partial_t \,\, B_0 \, (\lambda_k\,;\, kL)}{B_0 \, (\lambda_k\,;\, kL)} \,\,\,.
\label{eq:4.29}
\ee
Exploiting the fact that for small $kL$ one has $B_0 \propto 1/kL$ and thus
\be
\partial_t \,\, B_0 \, (\lambda_k\,;\, kL) = - B_0 \, (\lambda_k\,;\, kL) + {\cal O} (kL)
\ee
the relation 
\be
\eta_{Nk,\mathrm{eff}} = \eta_{Nk} -1 \quad {\mathrm{for}} \quad kL\ll 1
\ee
is evident, {\it i.e.}, one obtains the lowering of the anomalous dimension by one
as it is required by the general formula $\eta_{Nk} = 2-D$. Therefore, based on the
definition of $g_{\mathrm{k,eff}}$ as provided in Eq.\ (\ref{eq:4.28}) the five- and the 
four-dimensional limit, $L\to \infty$ and $L\to 0$, resp., are both well-defined.

Changing now the perspective from a varying $L$ at fixed $k$ to a varying RG scale $k$
at fixed size $L$ of the extra dimension one can get two different types of RG flows for
the effective coupling $g_{\mathrm{k,eff}}$:
\begin{itemize}
\item In the UV, {\it i.e.}, at large RG scale $k$, one starts close to the five-dimensional 
UV fixed point with $g_{\mathrm{k,eff}} \approx g_k \approx g_\ast $. 
Still staying in the quantum regime
$k$~becomes smaller than $1/L$ and one crosses to the four-dimensional  UV fixed point, and 
only later on in the flow one runs into a four-dimensional (semi-)\\ classical regime. Such 
a trajectory lacks, to the best of my knowledge, a physical interpretation.  
In any case they are not of the ADD type.
\item Also in the second case one starts in the UV very close to the UV fixed point but the 
five-dimensional
(semi-)classical running of $g_{\mathrm{k,eff}} \approx g_k$ sets in for a scale very much
larger than $1/L$, {\it i.e.}, for $k\gg 1/L$. 
For several orders of magnitude one follows the  five-dimensional (semi-)classical 
trajectory until $k\approx 1/L$. Then, more or less suddenly,  a cross-over from the
 five-dimensional  to the four-dimensional (semi-)classical  RG flow takes place.
Such a RG trajectory is then a realisation of the ADD model. 
\end{itemize}


\section{Discussion of the Limits}

In this section a discussion of the limits $kL\to \infty$ and $kL\to 0$ will be 
given. Note that the $\beta$-functions and the anomalous dimension depend 
only on the dimensionless product $kL$. Therefore from the mathematical 
point of view the discussion of the limits of the respective functions
in terms of the product $kL$ is complete. Of course,
the physical interpretation changes depending whether the limits are taken for
fixed RG scale $k$ or fixed size of the compact dimensions $L$, {\it cf.}, the 
discussion in the last section.

For the approximate background field flow and  $kL\gg 1$ one expands 
the hyperbolic functions  in Eqs.\ (\ref{eq:4.15}) and (\ref{eq:4.16}) to obtain
\bea
A_0 (\lambda_k;kL)  &=& \frac 1 {8 \,\, \pi} \Bigg [ \frac {15}  {\sqrt{ 1 - 2\, \lambda_k}} \,\, 
\Big(1+2e^{-kL\sqrt{1 - 2\, \lambda_k} }\Big) \,\,
- \,\, 10 \,\,\Big( 1+2e^{-kL} \Big) \nonumber \\
&&  \qquad \qquad  + {\cal O}( e^{-2kL}) \Bigg ]\,\,, \label{eq:4.32}  \\ 
B_0 (\lambda_k;kL)  &=&  - \,\, \frac 1 {32 \,\, \pi}  \Bigg [ 
\frac{25}{(1 - 2\, \lambda_k)^{3/2} }  \,\, 
\Bigg(1+2\Big( 1+kL\sqrt{1 - 2\, \lambda_k} \Big)e^{-kL\sqrt{1 - 2\, \lambda_k} } \Bigg) \,\, 
\nonumber \\
&&  \qquad \qquad + 18 \,\, \Bigg( 1+ 2 \Big( 1+kL \Big) e^{-kL} \Bigg) \,\,  
+ {\cal O}( e^{-2kL}) \Bigg ] \,\,.
\label{eq:4.33}
\eea
A more direct way is to use the Euler-MacLaurin formula for sums ($B_{2l}$ are the Bernoulli
numbers) 
\be
\sum_{n=a}^b f(n)  = \int _a^b f(x) + \frac 1 2 (f(a)+f(b)) + \sum_{l=1}^\infty 
\frac{B_{2l}}{(2l)!} \Big(f^{(2l-1)} (b) -  f^{(2l-1)} (a)\Big) \, \, .
\label{eq:4.34}
\ee
and to perform the limits at the level of the sums.
This, of course, provides the same results. In case that the heat kernel expansion (\ref{eq:4.9}) 
is used one obtains the same limit for $A_0$ but the limit for $B_0$ is now based on Eq.\ 
(\ref{eq:4.17}). The leading order term as calculated with the help of the expansion 
(\ref{eq:4.34}) is then given by\footnote{All the limits given in this section have been checked
by taking numerically, and when possible analytically, the limits in the sums over
Kaluza-Klein modes.}
\be
B_0 (\lambda_k;kL) = - \,\, \frac 1 {32 \,\, \pi}  \Bigg [ 
\frac{30}{(1 - 2\, \lambda_k)^{3/2} }  \,\,  + \frac {44} 3 + {\cal O}( e^{-kL}) \Bigg ] \,\,.
\label{eq:4.35}
\ee
As one sees the limits (\ref{eq:4.33}) and (\ref{eq:4.35}) differ slightly which has no impact on 
the qualitative and practically no influence on the quantitative behaviour. However, the five-
dimensional UV fixed point is only then precisely assumed if the form Eq.\  (\ref{eq:4.17}) with the 
limiting value (\ref{eq:4.35}) is employed.

For the approximate background field flow
and $kL\ll 1$ one can either expand the hyperbolic functions or simply 
exploit the fact that only the zero mode contributes to the Kaluza-Klein mode sums.
One obtains
\bea
A_0 (\lambda_k;kL)  &=&\,\,\,\,\,\ \frac 1 {kL} \,\,\, \frac 1 {8 \,\, \pi} 
\Bigg [ \frac {30}  {1 - 2\, \lambda_k} \,\, 
- \,\, 20 \,\, + \frac 5 6 (kL)^2 + {\cal O}( (kL)^4) \Bigg ] \,\,,\label{eq:4.36}  \\ 
B_0 (\lambda_k;kL)  &=  & - \,\,  \frac 1 {kL} \,\,\,  \frac 1 {8 \,\, \pi}  \Bigg [ 
\frac{25}{(1 - 2\, \lambda_k)^{2} }  \,\, 
+ 18 \,\, + {\cal O}( (kL)^4) \Bigg ] \,\,.
\label{eq:4.37}
\eea
Needless to say that these limits do not reproduce the four-dimensional UV fixed point due to the
fact that several factors entering the calculation are in a non-trivial manner dependent on
the number of space-time dimensions. But again the deviations are of no qualitative and
minor quantitative importance. To be complete let me mention that  the corresponding 
leading-order limit  of Eq.\ (\ref{eq:4.17}) is also given by the expression (\ref{eq:4.37}).

For the case of the expressions derived from the exact functional identity
the limit \underline{$k L \to \infty$}} will be discussed first. In this limit the  sums 
in Eqs.\ (\ref{eq:4.18} - \ref{eq:4.21a}) become Riemann integrals, the respective values are:
\begin{align}
\int_{-1}^{1}& \,\, \frac {dp}{2 \,\, \pi} \,\, (1 - p^2) = \frac {2}{3\, \pi} \,\,, \nonumber& 
\int_{-1}^{1}& \,\, \frac {dp}{2 \,\, \pi} \,\, (1 - p^2)^2 = \frac {8}{15\, \pi} \,\,, \nonumber& \\ 
\int_{-1}^{1}& \,\, \frac {dp}{2 \,\, \pi} \,\, (1 - p^2)^3 = \frac {16}{35\, \pi} \,\,, \\ 
\int_{-1}^{1}& \,\, \frac {dp}{2 \,\, \pi} \,\, (1 - p^2 + p^2 \, \log \, p^2) = \frac {4}{9\, \pi} \,\,, \qquad&
\int_{-1}^{1}& \,\, \frac {dp}{2 \,\, \pi} \,\, (1 - p^4 + 2 \, p^2 \, \log \, p^2) = \frac {16}{45\, \pi} \,\,. 
\nonumber
\end{align}
For the case of the heat kernel expansion (\ref{eq:4.9}) 
this results as anticipated  in the expressions (\ref{eq:3.15}) for the four functions 
$A_0$, $A_1$, $B_0$ and $B_1$. Correspondingly, the five-dimensional UV fixed point is
precisely reproduced. For the case of the heat kernel expansion (\ref{eq:4.10}) 
the limit changes slightly as one has to substitute in the second line of Eq.\  (\ref{eq:4.20}),
resp.,  Eq.\  (\ref{eq:4.20a}) $4/9\pi \to 2/3\pi$ and  in the second line of Eq.\  (\ref{eq:4.21}),
resp.,  Eq.\  (\ref{eq:4.21a}) $16/45\pi \to 8/15\pi$, {\it i.e.}, these two terms obtain a weight 
increased by 50\%. Correspondingly the fixed point values are slightly shifted,
one has then $g_\ast=2.7650$ 
and $\lambda_\ast= 0.26292$.

The limit \underline{$kL\to 0$} is identical for both cases of the heat kernel expansion.
In this limit the expressions with the Litim regulator show an 
interesting feature. For $kL<2\pi$ the Kaluza-Klein
mode sums degenerate to the zero mode term only.   
The corresponding  $n = 0$ terms yield:
\be
\left.
\begin{aligned}
A_0 \, (\lambda_k \,;\, kL <2\pi) &= \,  \frac {1}{k \,\, L} \,\,\, \frac {1}{2 \, \pi} \,\,   
\Big (\frac {15}{2} \,\, c_k - 5 \Big ) \,, \\[2mm]
A_1 \, (\lambda_k \,;\, kL <2\pi) &= \frac {1}{k \,\, L} \,\,\, \frac {1}{2 \, \pi} \,\, \frac {5}{4} \,\, c_k  \,,
\\[3mm] \label{eq:4.39} 
B_0 \, (\lambda_k \,;\, kL <2\pi) &= \frac {1}{k \,\, L} \,\,\, \frac {1}{\pi} \,\, \Big ( - 5 \,\, c_k^2 + \frac 
{15}{4} \,\, c_k \,-\, \frac {7}{2} \, \Big )\,,
\\[2mm] 
B_1 \, (\lambda_k \,;\, kL<2\pi) &= \frac {1}{k \,\, L} \,\,\, \frac {1}{\pi} \,\, \Big ( - \frac {5}{6} \,\, c_k^2 
+ \frac {15}{16} \,\, c_k \, \Big )\,,
\end{aligned}
\right. 
\ee
with $c_k$ as defined in Eq.\ (\ref{eq:3.11}).

And as expected, due to the truncation and to the fact that I use a five-dimensional regulator, 
these are not exactly the expressions needed to reproduce the four-dimensional UV fixed point
in $g_{\mathrm{k,eff}} $, {\it cf.},  Eqs.\ (\ref{eq:3.13}).


\goodbreak

\section{Renormalisation Group Flows and Phase Diagrams  in $D = 4 + 1$}

In this section numerical results for the solution of the RG equations with four extended and 
one compactified dimension will be presented. Some details of the numerical treatment are
described in Appendix~F.

\subsection{Results for the Background Field Flow}

In the case without dimensional reduction the $\beta$-functions contain only dimensionless 
quantities, see {\it e.g.}, Eqs.\ (\ref{eq:2.68}) and  (\ref{eq:2.71}). They constitute a system
of differential equations for the dimensionless running coupling constants $g_k$ and 
$\lambda_k$. Due to this it is evident that the numerical values of the RG scale $k$ are 
completely irrelevant. In the case of dimensional reduction the nature of the system of 
differential equations changes drastically. 
\begin{figure}[h]
 \centering
 \includegraphics[width=0.49\textwidth]{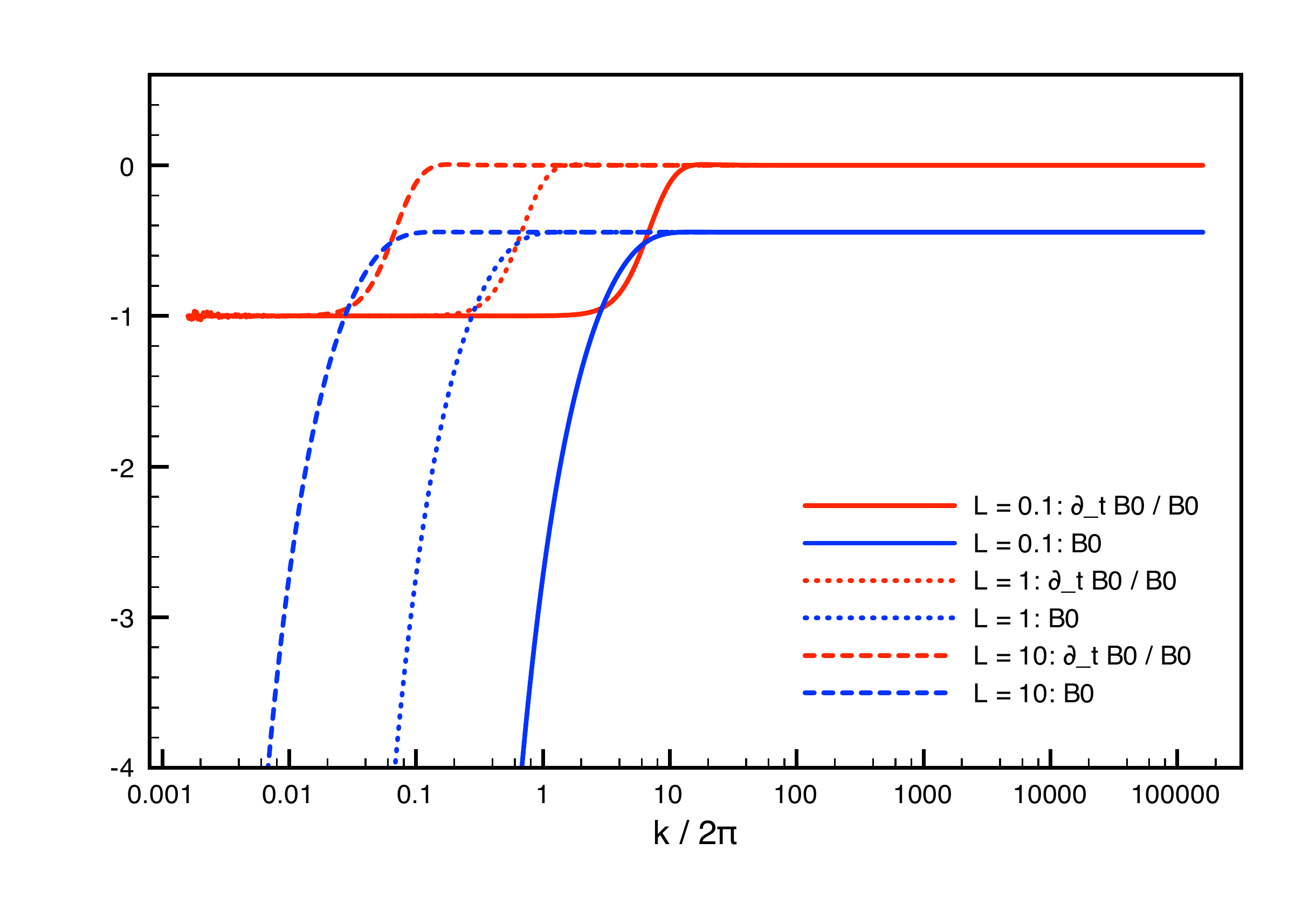}
 \includegraphics[width=0.49\textwidth]{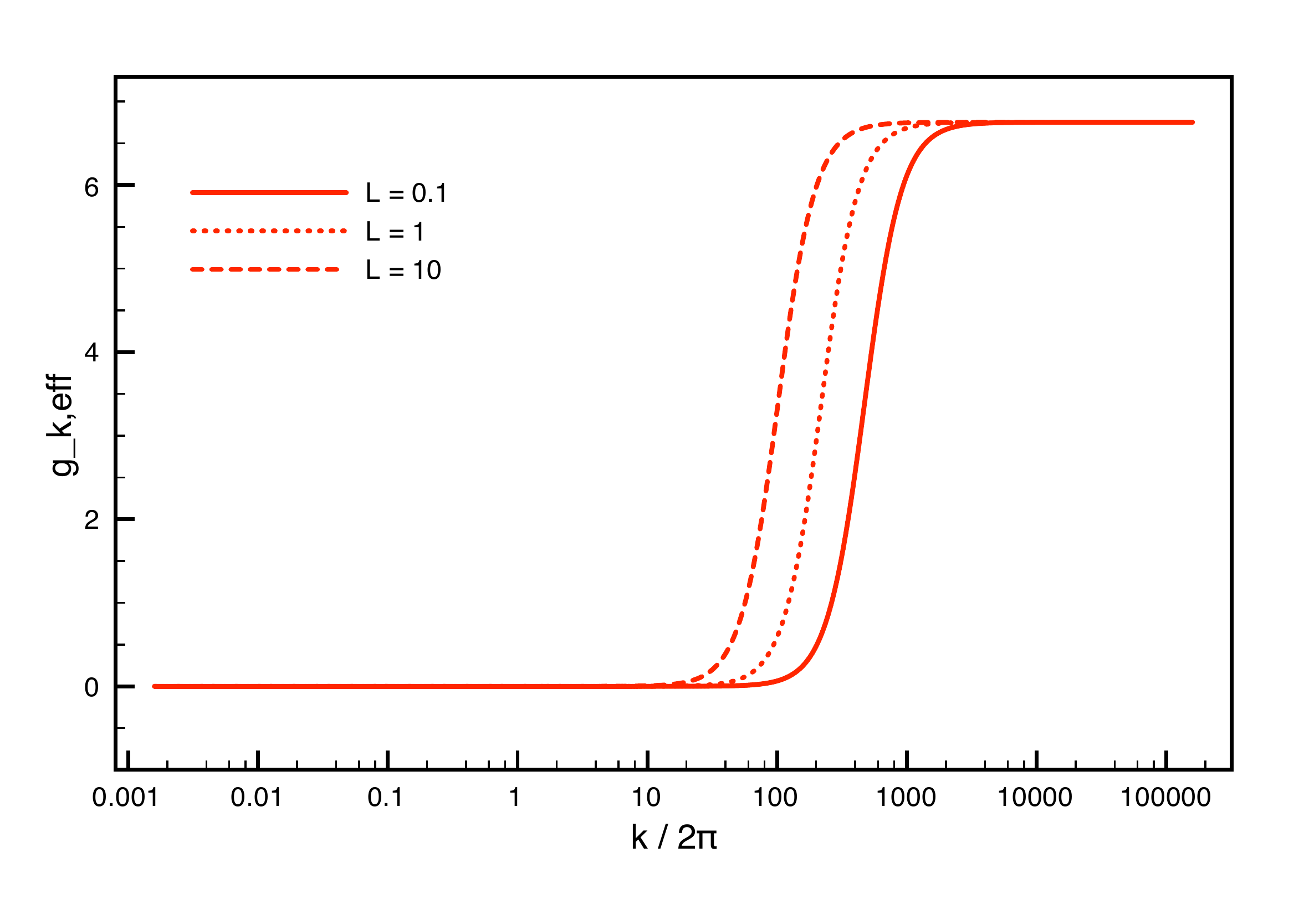} 
 \includegraphics[width=0.6\textwidth]{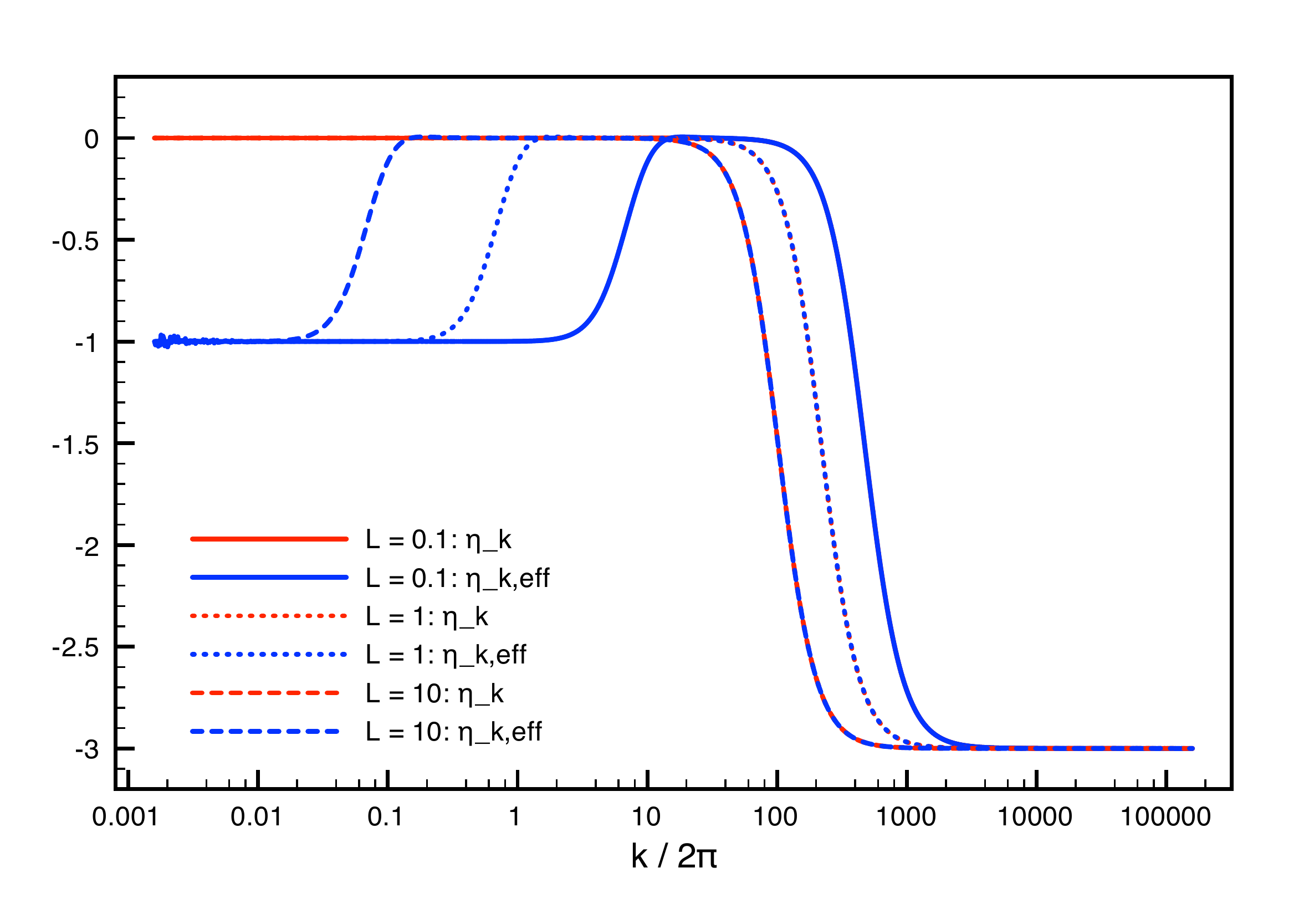} 
\vspace{-2mm}
\caption[Comparison between three different values of $L$.]{\label{fig:4.1}
Comparison between three different values of $L$ (0.1 - full lines, 1~-~dotted lines and 
10 - dashed lines) and all other parameters 
kept identical. The upper left panel shows the function $B_0$ (\ref{eq:4.16}) 
(blue lines) and its scale derivative
(red lines),
the upper right panel $g_{\mathrm{k,eff}}$ (\ref{eq:4.28}) and the lower panel the anomalous 
dimensions $\eta_{Nk}$ (\ref{eq:2.72}) 
(red lines) and $\eta_{Nk,\mathrm{eff}}$ (\ref{eq:4.29})  (blue lines). For
simplicity the cosmological constant $\lambda_k$ has been set to zero in this test 
calculation.}
 \end{figure}
Now $L$ provides a length scale, and therefore 
the cases $k>1/L$ and $k<1/L$ become different as it is also exemplified by the discussion 
of the two different limits in the last section. However, there should be a remnant of the 
original scale invariance: The numerical value of $L$ is irrelevant to the investigated system
in the sense that the results for two different values of $L$ can be mapped onto each other 
by a corresponding rescaling of the RG scale $k$. I explicitly tested this scenario for the 
approximate background field flow, 
see Fig.\ \ref{fig:4.1}, where one clearly sees due to the linear-log presentation the scaling of
the results for three different values of $1/L$.  
From now on and without loss of generality I will use $L=1$.

In Fig.\ \ref{fig:4.2}  two examples for the RG scale dependence of the coupling constants 
are shown. Both sets reach the UV fixed point at different RG scales $k$ as measured in units of 
$1/L$. To see the approach to the fixed point $g_k$,  $g_{\mathrm{k,eff}}$, 
$\lambda_k$,  $\eta_{Nk}$ or  $\eta_{Nk,\mathrm{eff}}$ are equally suitable. For discriminating 
the different regions, however, $\eta_{Nk,\mathrm{eff}}$ is most efficient.  For both sets 
$\eta_{Nk,\mathrm{eff}}=2-D=-3$ in fixed point regime and $\eta_{Nk,\mathrm{eff}}=-1$ for $k<1/L$ 
indicating a (semi-)classical regime according to $D-1=4$ dimensions. The both sets are 
different with respect to the existence of  (semi-) classical regime with five-dimensional 
running: For the set displayed by the full lines there are more than three orders of magnitude 
between $1/L$ and the scale where the UV fixed point behaviour starts. Accordingly, there exists 
for approximately two orders of magnitude a five-dimensional classical regime. 
Therefore the respective RG trajectory constitutes a realisation of the ADD model. 
For the set  shown with the dashed line the UV fixed point is reached for $k\approx 8\pi /L$, and
as one sees from the behaviour of $\eta_{Nk,\mathrm{eff}}$ the separation of scales is too 
small for a five-dimensional classical regime to exist.

\begin{figure}[H]
\centering
\includegraphics[width=0.98\textwidth]{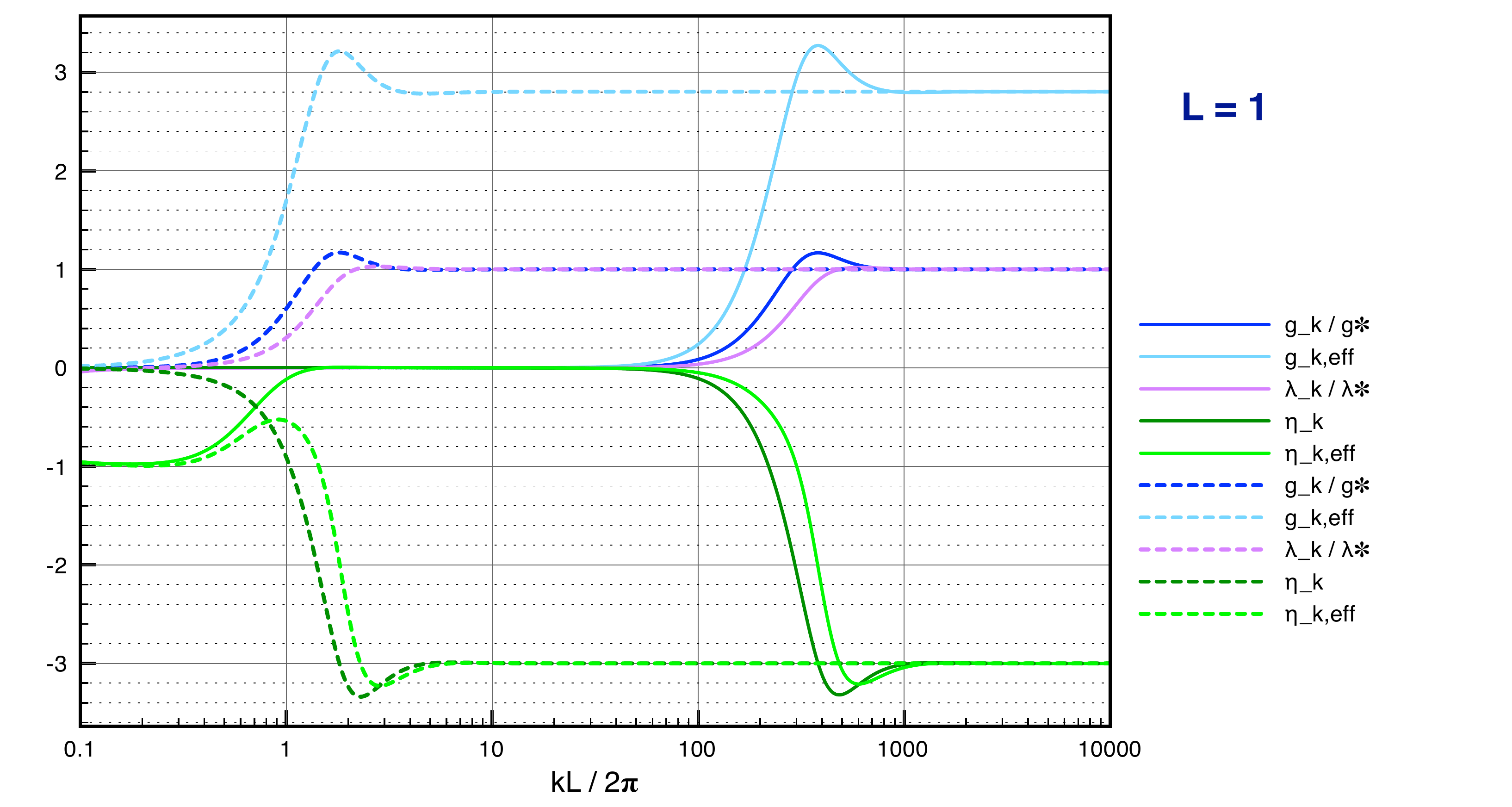}
\caption[Two examples for the RG scale dependence of  coupling constants.]
{\label{fig:4.2}  Two examples for the RG scale dependence of  coupling constants resulting 
from the Eqs.\ (\ref{eq:2.64}) and (\ref{eq:2.66})
with (\ref{eq:4.15}) and (\ref{eq:4.16}) as input: Shown is
$g_k$ (normalised by $g_\ast$),  $g_{\mathrm{k,eff}}$  (\ref{eq:4.28}), 
$\lambda_k$ (normalised by $\lambda_\ast$),  $\eta_{Nk}$ (\ref{eq:2.62})
and  $\eta_{Nk,\mathrm{eff}}$
 (\ref{eq:4.29}).
For the set displayed by the dashed lines the UV fixed point is reached for $k\approx 8\pi /L$, for the
set with the full lines there are more than three orders of magnitude between $1/L$ and 
the scale where the UV fixed point behaviour starts.  }
\end{figure}

\begin{figure}[h]
\vspace{-3mm}
\begin{center}
\includegraphics[width=\textwidth]{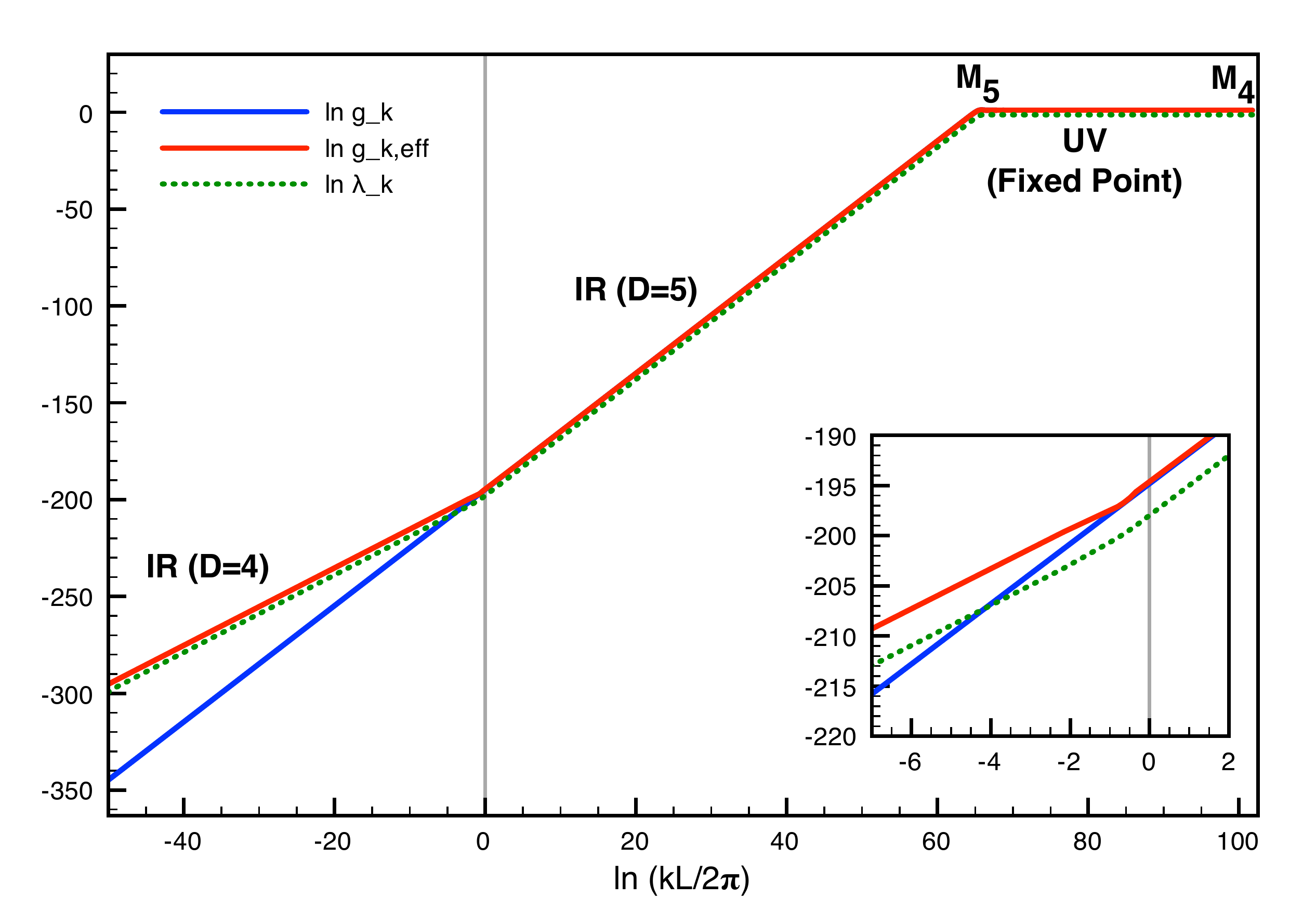}
\end{center}
\caption[A further example for the RG scale dependence of  coupling constants.]
{\label{fig:4.3}  A further example for the RG scale dependence of  coupling constants:
The scale where the UV fixed point is reached is many orders of magnitude larger than $1/L$.
$g_k$ and $\lambda_k$ are solutions of the differential equations  (\ref{eq:2.64}) and (\ref{eq:2.66})
with the functions (\ref{eq:4.15}) and (\ref{eq:4.16}) as input;
$g_{\mathrm{k,eff}}$  is defined in (\ref{eq:4.28}).}
\end{figure}

In Fig.\ \ref{fig:4.3} an example for a RG trajectory being a realisation of the 
ADD model  is given. In the beginning of this chapter we estimated that for $n=1$ 
one has the relation $M_5 \approx 10^{32} /L$ to obtain $M_4\approx 10^{16} M_5$.
In the displayed example the UV fixed point region starts at 
$k\approx e^{66} \cdot 2\pi /L \approx  5\cdot 10^{28} \cdot 2\pi /L$,
{\it i.e.}, one has $M_5 \approx 10^{30} /L$ and $M_4\approx 10^{15} M_5$. 
In between $2\pi/L$ and the fixed 
point, both $g_k$ and $g_{\mathrm{k,eff}}$  grow like $k^3$ from a value 
$e^{-195} \approx 10^{-85}$ to the fixed point value. For $k<1/L$ one
sees a clear difference in the power laws: $g_{\mathrm{k,eff}}$ grows like $k^2$ according
to the four-dimensional behaviour. The dimensionless cosmological constant $\lambda_k$ 
shows an interesting behaviour: It possesses also a crossover from a $k^2$ to a $k^3$ power 
law.\footnote{For a starting value of $\lambda_k$  very much larger than 
$g_{\mathrm{k,eff}}$,  see Fig.~\ref{fig:4.7} below.}
This can be understood from the explicit expression for $\beta_\lambda$ and the 
function $A_0$ as well as the fact that $\eta_k$ can be neglected completely in 
$\beta_\lambda$ for all $k\ll M_5$, {\it i.e.}, before the fixed point region is reached.
 For $\lambda_k<g_{\mathrm{k,eff}}$ one has for $kL\ll1$ 
\be
\partial_t\lambda_k\approx g_kA_0 = const. \,\,  \times \frac {g_k}{kL} = const.^\prime \, 
\times g_{\mathrm{k,eff}}
\ee
where $const.^\prime$ is a number of order one. Therefore,  $\lambda_k \propto k^2$ with 
a fixed ratio to $g_{\mathrm{k,eff}}$. For $kL$ large but still $k\ll M_5$ one has 
$g_k \approx g_{\mathrm{k,eff}} \propto k^3$ and $A_0$ being a constant of order one.
Thus also $\lambda_k$ assumes the power law $k^3$ until the fixed point region is reached.
In the inset of Fig.\ \ref{fig:4.3} the transition region is shown. Hereby the vertical grey line 
denotes $kL=2\pi$. One sees that the merger of $g_k$ and  $g_{\mathrm{k,eff}}$ as well as 
crossover in power laws for $g_{\mathrm{k,eff}}$  and $\lambda_k $ happen at $kL\approx 2.5$ 
and therefore a factor two to three lower than na\"ively expected.


\subsection{Results following from the Exact Functional Identity}

In Fig.\ \ref{fig:4.6}  the RG scale dependence of  coupling constants is displayed for 
the case of the Litim regulator using the same set of starting conditions as for the 
full lines in Fig.\ \ref{fig:4.2}.  Hereby one sees strong similarities between the results 
of the approximate background field flow and the expressions derived from the exact functional 
identity but also two differences: 
At $k\gtrsim 2\pi$  the anomalous dimension $\eta_{Nk,\mathrm{eff}}$ shows oscillations 
characteristic to the ``onset'' of Kaluza-Klein modes.  This effect is purely due to the 
regulator and of minor importance.  The other difference, namely a much more drastic
overshooting of $g_k$ before it reaches its UV fixed point value,  might be of more significance
{\it cf.}, the discussion on phenomenological applications  in the next chapter.

\begin{figure}[tbh]
\flushright
\includegraphics[width=0.95\textwidth]{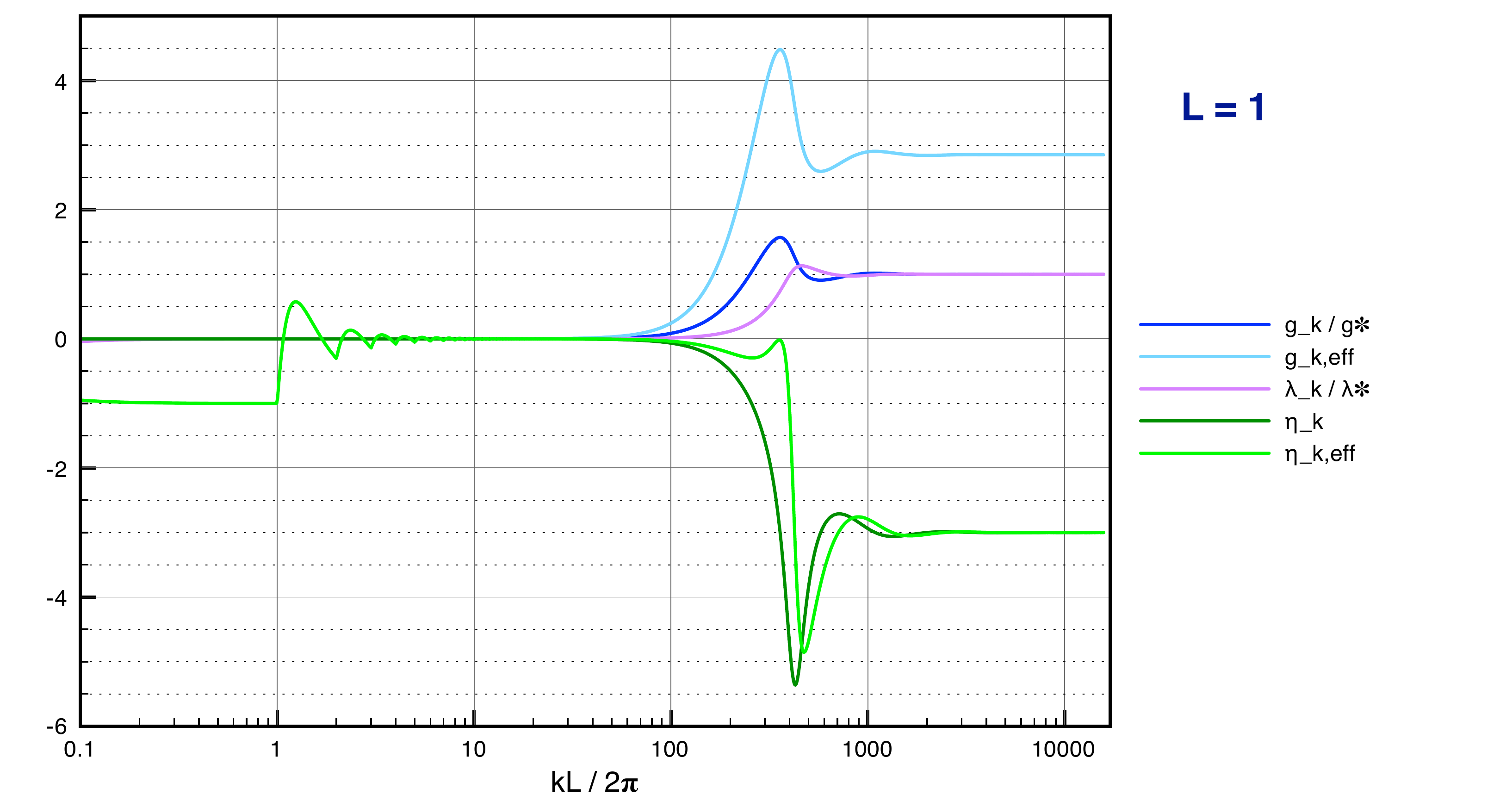}
\caption[Same example as in Fig.\ \ref{fig:4.2} for the RG scale dependence of  
coupling constants but now for the $\beta$-functions derived from the exact functional identity.]
{\label{fig:4.6}  Same example as the full lines in Fig.\ \ref{fig:4.2} for the RG scale dependence 
of  coupling constants but now for the $\beta$-functions (\ref{eq:2.68}) and (\ref{eq:2.71}) 
derived from the exact functional identity with the functions (\ref{eq:4.18}) - (\ref{eq:4.21})
as input. }
\end{figure}

In Fig.\ \ref{fig:4.7} the coupling constants derived from the exact functional identity 
are  displayed.  If the same starting conditions had been chosen as in Fig.\ \ref{fig:4.3}  
for the approximate background field flow  the two versions of results would be 
practically indistinguishable with the given resolution of these figures. 
Therefore, an even smaller value of $g_k$ was chosen as starting point in the IR.   
In the examples of Fig.~\ref{fig:4.7}  one has even larger ratios of
scales as  in Fig.\ \ref{fig:4.3}: $1/L$ and $M_5$ differ by 32 to 33 orders of magnitude.
This demonstrates that also for the 
$\beta$-functions derived from the exact functional identity one
can obtain RG trajectories with a separation of scales  as required by the $n=1$ ADD model.

\begin{figure}[tbh]
\includegraphics[width=\textwidth]{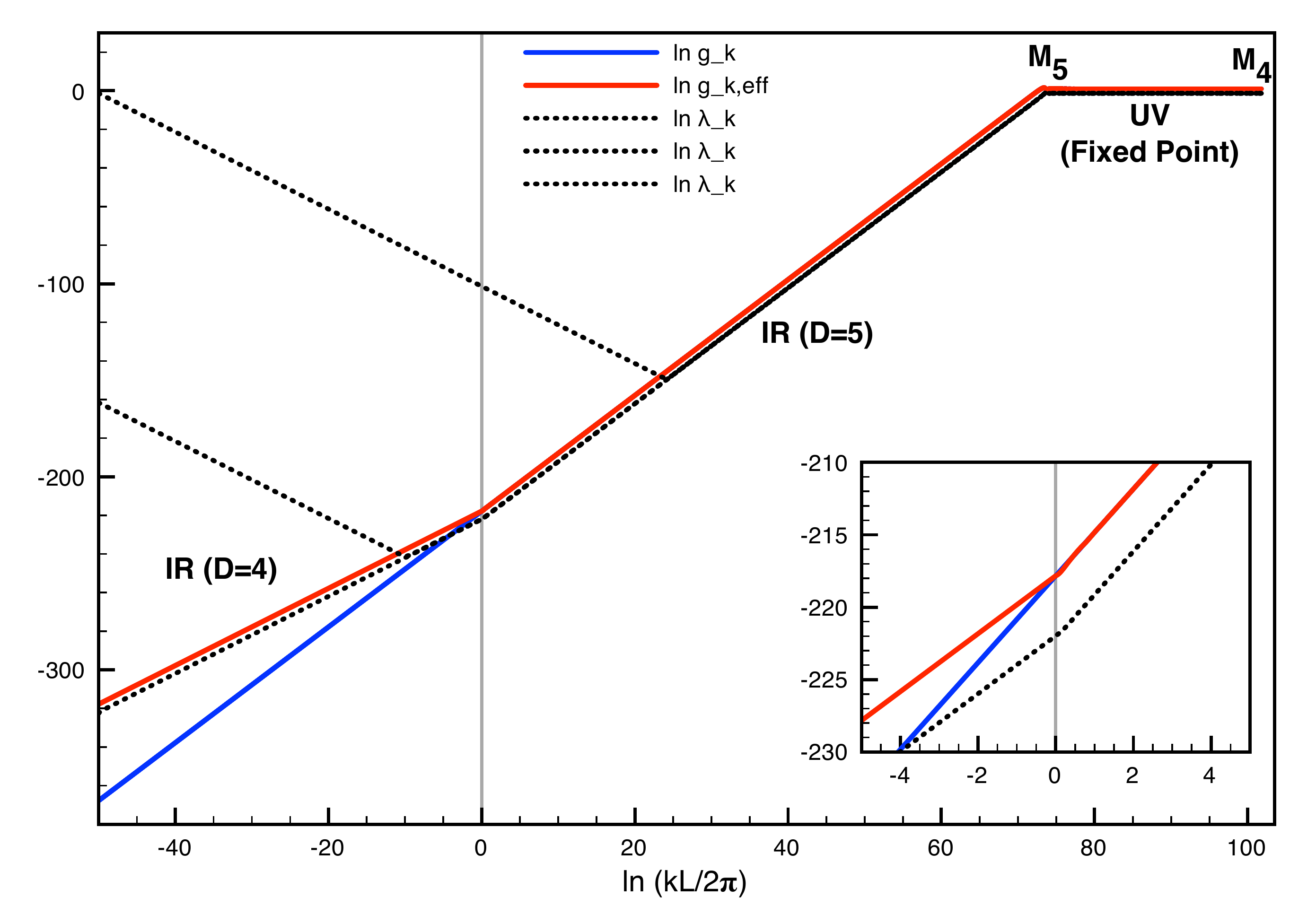}
\caption[Similar to Fig.\ \ref{fig:4.3} but now for the coupling constants derived
from the exact functional identity, and for $\lambda_k$ several starting conditions are 
displayed.]
{\label{fig:4.7} Similar to Fig.\ \ref{fig:4.3} but now for the coupling constants derived
from the exact functional identity, and for $\lambda_k$ several starting conditions are 
displayed. $g_k$ and $\lambda_k$ are solutions of the differential 
equations  (\ref{eq:2.68}) and (\ref{eq:2.71}) with the functions (\ref{eq:4.18}) - (\ref{eq:4.21})
as input. 
 $g_{\mathrm{k,eff}}$  is defined in (\ref{eq:4.28}).}
\end{figure}

For Fig.\ \ref{fig:4.7}  three different starting values of  the dimensionless cosmological constant 
$\lambda_k$ have been chosen. Hereby it is important to note that the influence on the running 
of $g_k$ is absolutely negligible: The three curves for $g_k$ as well as the three curves for
$g_{\mathrm{k,eff}}$ are exactly on top of each other.
If the absolute value of $\lambda_k$ is chosen as small as 
$g_{\mathrm{k,eff}}$ or even smaller one sees the  crossover from a $k^2$ to a 
$k^3$ power  law as in  Fig.\ \ref{fig:4.3}.  
If the starting value of $\lambda_k$ is chosen very much larger than 
$g_{\mathrm{k,eff}}$ then it   shows for increasing $k$ first a power-law decrease with $1/k^2$
as then the leading term in $\beta_\lambda$  is such that 
$\partial_t\lambda_k\approx -2 \lambda_k$. 
After it  falls below the value of  $g_{\mathrm{k,eff}}$ it follows the power-law  increase of  
$g_{\mathrm{k,eff}}$ until both reach together their respective UV fixed point value. 
This is demonstrated here with starting values of $\lambda_k=0.4$ and $\lambda_k=10^{-70}$,
the difference between the both cases being that the latter has a part where it increases like 
$k^2$ in the four-dimensional classical regime. In the former case the increase of $\lambda_k$ starts
only in the five-dimensional classical regime. In the calculations also starting conditions 
 $\lambda_k=-0.4$ and $\lambda_k=-10^{-70}$ have been chosen. Plotting then the logarithm 
 of the absolute values of $\lambda_k$ produces indistinguishable curves from the respective cases 
 of positive $\lambda_k$ except the locations of the respective minima: In the cases when starting 
 with  negative values for $\lambda_k$ one has at exactly these locations the zero crossing.

However, the most important observation is the following:
Comparing the two respective insets in the two figures  \ref{fig:4.3} and \ref{fig:4.7}
(which display a zoom around the 
value $k=2\pi/L$ indicated in both figures by the vertical grey line) shows a significant 
deviation. For the Litim regulator the merger of $g_k$ and  $g_{\mathrm{k,eff}}$ as well as 
crossover in power laws for $g_{\mathrm{k,eff}}$  and $\lambda_k $ happens quite precisely
at the na\"ively expected scale $k=2\pi/L$.

The difference of an approximate factor two in between these two calculations is certainly 
noteworthy. One has to conclude  that within the employed truncations,
for a given extension $L$ of the compact dimension, the RG results for the precise 
scale of the dimensional crossover  depends on the regulator.

\begin{figure}[tbh]
\includegraphics[width=\textwidth]{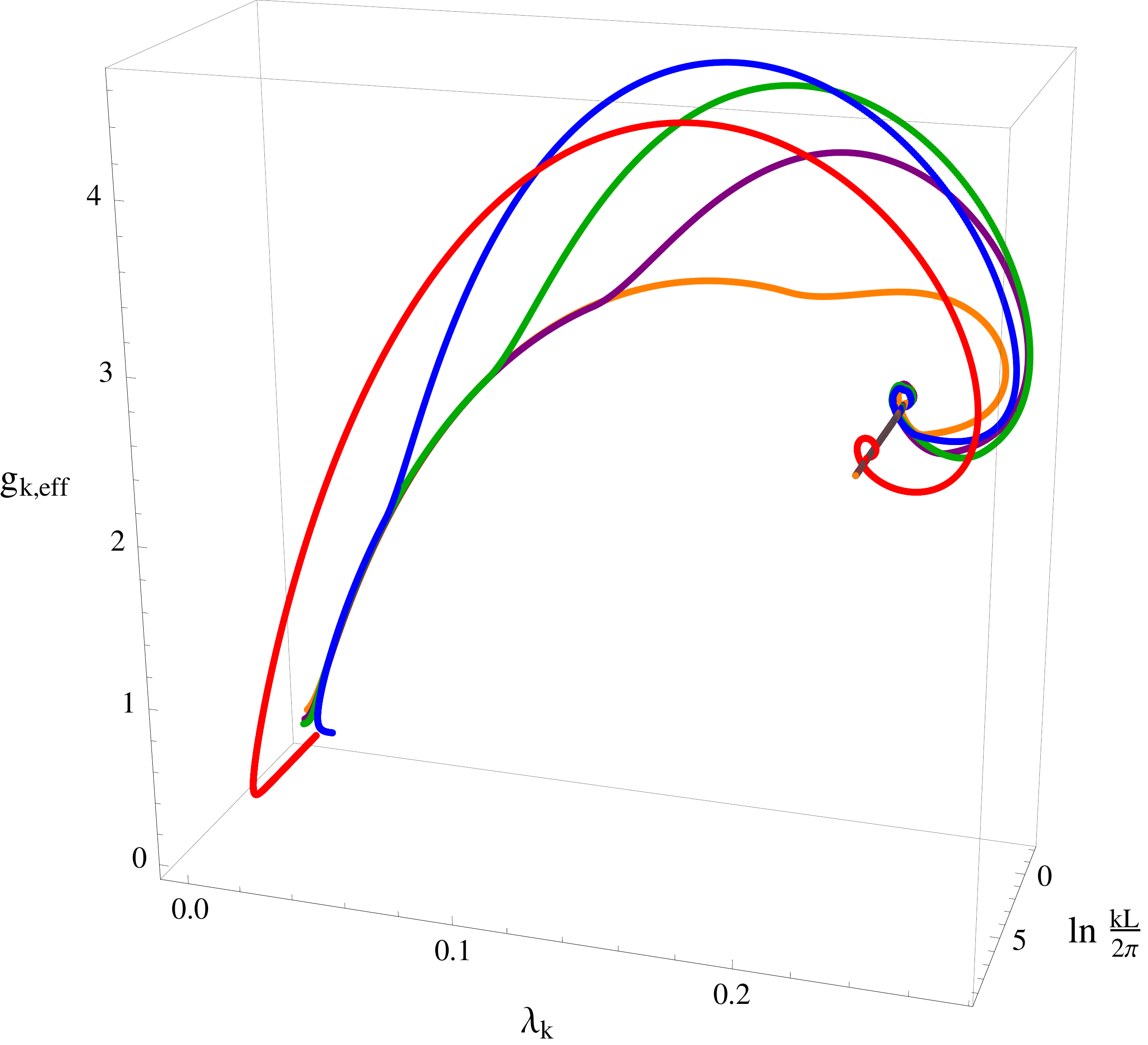} 
\vspace{-5mm}
\caption[Phase diagram for the running gravitational coupling $g_{\mathrm{k,eff}}$ and the cosmological 
constant $\lambda_k$ in $4 + 1$ dimensions with physical trajectories in function 
of $\ln (kL/2\pi)$.]
{\label{fig:4.10}Shown is the phase diagram for the running gravitational coupling $g_{\mathrm{k,eff}}$ 
and  the cosmological constant $\lambda_k$ in $4 + 1$ dimensions  in function 
of $\ln (kL/2\pi)$. 
Displayed are RG trajectories corresponding to different values for $M_5L$:
7000 (red line), 31 (blue line), 28 (green line), 24 (purple line) and 23 (orange line).
The red line shows the separatrix and, in the part visible in this linear representation, 
the five-dimensional running.
These trajectories are physical and except the separatrix 
show a clearly visible four- to five-dimensional
crossover occurring at $kL=2\pi$.
They are solutions of the differential 
equations  (\ref{eq:2.68}) and (\ref{eq:2.71}) with the functions
(\ref{eq:4.18}), (\ref{eq:4.19}),  (\ref{eq:4.20a}) and (\ref{eq:4.21a})
as input. 
 $g_{\mathrm{k,eff}}$  is defined in (\ref{eq:4.28}). }
\end{figure}

In Fig.\ \ref{fig:4.10} the phase diagram for the running  gravitational coupling $g_{\mathrm{k,eff}}$ and the 
cosmological constant $\lambda_k$ in $4 + 1$ dimensions within the approach based on the 
exact functional identity is shown. The third axis is $\ln (kL/2\pi)$. 
Here only trajectories have been displayed which are possible realisations of the ADD 
model.  The separatrix is given by the red line. In the part which is visible in
this linear plot it follows the five-dimensional running, for this trajectory one has 
$M_5L \approx 7000$. Therefore that part of the other
RG trajectories running parallel to the separatrix flows according to the five-dimensional 
$\beta$-functions.  All displayed trajectories follow the five-dimensional  (semi-)classical 
running before they enter the fixed point regime. Note, however, that the separation of scales is 
quite small to allow for a linear representation:
$M_5L \approx 31$ for the blue, $\approx 28$ for the green, $\approx 24$ for the purple and 
$\approx 23$ for the orange lines.
Choosing larger separations of scales  the crossover behaviour would be invisible in a 
linear plot, {\it cf.}, the log-log plot in Fig.\ \ref{fig:4.7}.  At $k= 2\pi /L$ the trajectories depart 
from the four-dimensional running.   One recognises 
that there is a regime where the change from four- to five-dimensional (semi-) classical running 
takes place. As one can see ({\it cf.}, also Fig.\ \ref{fig:4.6}) this intermediate regime is up to values
of $k \approx 3 \cdot 2\pi /L$. 

From this section one can infer the following most relevant information: For RG trajectories
being a realisation of the ADD model one has a quite sharp crossover from four- to 
five-dimensional behaviour which, when employing the Litim regulator, takes place
in the interval $2\pi < kL < 6\pi$.  However, the calculated position of the crossover
is regulator dependent as the comparison to the approximate background field flow 
demonstrates. Reaching from the IR the  fixed point regime the gravitational coupling 
is rapidly changing and oscillates around the fixed point value
before it freezes to it. The width of this region is approximately one order 
of magnitude. The values of Newton's coupling and of the cosmological constant are 
to a high precision the ones of the five-dimensional RG equations because the corresponding 
Matsubara sums are quite accurately presented already by their limiting values.



\chapter{{\color{BrickRed}Conclusions}}
~\vspace{-10mm}

\rightline{\footnotesize{\it{``Imagination is more important than knowledge."}}}
\rightline{\footnotesize{\it{(A.\ Einstein)}}}
\minitoc

\section{Phenomenological Implications}

The basic idea of the ADD model is to bring the scale of gravity down to the 
EW scale, or at least down to only slightly above the EW 
scale.  As already mentioned in the introduction besides many other implications,
particle production processes may then be significantly enhanced by virtual graviton 
exchange.

Of course, in this context the value and scale dependence of the gravitational
coupling play a significant role. Hereby one can identify for RG trajectories representing
a realisation of the ADD model two transition regions. The first one is the one where 
the running of the gravitational coupling and the cosmological constant changes from a 
(semi-)classical four-dimensional to a  (semi-)classical higher-dimensional  behaviour. The second one
is the transition from the (semi-)classical to the quantum, {\it i.e.}, fixed point, regime. If the
latter is brought down to the electroweak scale this opens up the possibility that quantum
gravity might be tested at the LHC.  But also the first transition region can, at least in principle, 
be searched for in experiments. The most significant change in the behaviour of gravity would 
be the modification from an inverse-square law of the gravitational force to a power-law
$F\propto 1/r^{2+n}$ in $4+n$ space-time dimensions. Here, however, one has to note that 
an observed deviation from Newton's gravitational law is not immediately identical to
a precise measurement of the size of the compact dimension(s). In this thesis I even found quite 
a sizeable regulator dependence of the scale of the dimensional crossover even for a given
size $L$ of the compact dimension.  

On the other hand,  at the higher-dimensional Planck scale 
the values of the running gravitational constant could have been
directly calculated from the higher-dimensional $\beta$-functions. This becomes evident
already from the fact that for the RG trajectories of interest the Matsubara sums are 
in the fixed point region already extremely close to their limiting values. 
In the best investigated examples of this thesis the starting conditions of the RG flow 
were chosen such that  a separation of $1/L$ and $M_5$ by 30 to 33 orders of magnitude 
resulted. Then the mode sums are to any achievable numerical precision identical 
to their limiting values. In view of the 
fact that  also for larger values of the number of extra dimensions, 
$1/L$ and $M_D\approx m_{EW}$ stay significantly separated,
this conclusion seems robust. Therefore, 
for an analysis of the phenomenological implications  of the asymptotic safety 
scenario at highest achievable accelerator energies it is sufficient to 
simply investigate the running of the gravitational constant in $D=4+n$ dimensional 
quantum gravity. Given the very general appearance of complex scaling exponents 
in the  asymptotic safety scenario the resulting rapid changes of the gravitational constant will 
provide an unique ``fingerprint''.


\section{Concluding Remarks and Outlook}

To summarise, in this thesis I discussed the application of the  asymptotic safety scenario 
to EH quantum gravity in the case of four extended and one compactified
dimensions. All investigations were done in the EH theory and in deDonder
gauge. As tools I used an approximate background field flow and an exact
functional equation with the Litim regulator.

In the main chapter I presented results for corresponding  RG trajectories. Some of them 
possess the separation of scales as required by the ADD model  
\cite{ArkaniHamed:1998rs,ArkaniHamed:1998nn}: These RG trajectories can
be considered as realisations of this model.  Especially, they provide an explicit
example for an UV completion of the ADD model. 

The investigations presented in this thesis can be considered only as first 
steps leading to studies of the ADD model assumptions within the
asymptotic safety  scenario of EH quantum gravity. The case of only one compact dimension
while the true Planck scale is at or only slightly above the EW scale
is excluded by experiment: The compact extra dimension would have a size 
comparable to the solar system. However, an ADD model with two and more 
compactified dimensions is still viable. Introducing several compact dimensions
lead immediately to the question of the topological properties of the compactified
space. In a first investigation one certainly will assume the easiest case, namely
the topology of a torus with $n=D-4$ independent periodic boundary 
conditions. This will lead then to nested Matsubara sums. The techniques 
developed in this thesis should allow to treat the cases of a few ({\it e.g.}, two to four)
nested sums in a efficient manner. 

Nevertheless, the bottom line is that  if there are ``large'' 
extra dimensions in nature this, first, might open up the exciting possibility of testing
quantum gravity at the LHC  \cite{Litim:2007iu,Litim:2007ee,Gerwick:2011jw}. 
Should a strong scale-dependence of the 
gravitational constant be measured  then one can interpret this as evidence for
the  asymptotic safety scenario at work in quantum gravity. 
Second, a deviation from the inverse-square law of the gravitational force should 
becomes visible at the $\mu$m to nm scale.  Hereby, the length scale of the crossover
in between the two power laws will be related to the size $L$ of the compact dimensions.
The crossover length scale is by definition observable, the size $L$ is directly related to the 
excitation energies of Kaluza-Klein modes and thus also a physical length scale. Therefore,
these both length scales (and automatically the ratio in between them) are expected to 
be independent of the renormalisation scheme and scale. However, in this theses  the
RG scale where the crossover happens was different for the employed two different regulators.
This result clearly requires a better understanding and deserves therefore a more profound
investigation.


\newpage
\thispagestyle{empty}

\appendix

\chapter{{\color{BrickRed}Notations and Definitions}}

In this thesis I used 
 units such that
  $\hbar=1$ and $c=1$.  One has then $G_N = 1/m_{\mathrm Pl} ^2$ with $m_{\mathrm Pl} 
= 1.22093(7) \cdot 10^{19}$ GeV \cite{Beringer:1900zz}.

The definition of the expectation value of the full metric is given in Eq.\ (\ref{eq:2.9}). With 
this metric one obtains for the Ricci scalar, the Ricci tensor, the Riemann tensor and the 
Christoffel symbols the following relations:
\be
\overbrace{R}^\text{Ricci scalar} = g^{\mu \nu} \underbrace{R_{\mu \nu}}_\text{Ricci tensor}
\ee
with 
\be
\overbrace{R^{\mu}_{\,\,\,\, \nu \lambda \rho}}^\text{Riemann tensor} = \partial_{\lambda} \, 
\Gamma^{\mu}_{\,\,\, \nu \rho} - 
\partial_{\rho} \, \Gamma^{\mu}_{\,\,\, \nu \lambda} + 
\Gamma^{\mu}_{\,\,\, \sigma \lambda} \, \Gamma^{\sigma}_{\,\,\, \nu \rho} - 
\Gamma^{\mu}_{\,\,\, \sigma \rho} \, \Gamma^{\sigma}_{\,\,\, \nu \lambda} \,\, ,
\ee
and contracting first and third indices yields:
\be
\underbrace{R_{\nu \rho}}_\text{Ricci tensor} = \partial_{\lambda} \, \Gamma^{\lambda}_{\,\,\, \nu 
\rho} - 
\partial_{\rho} \, \Gamma^{\lambda}_{\,\,\, \nu \lambda} + 
\Gamma^{\lambda}_{\,\,\, \sigma \lambda} \, \Gamma^{\sigma}_{\,\,\, \nu \rho} - 
\Gamma^{\lambda}_{\,\,\, \sigma \rho} \, \Gamma^{\sigma}_{\,\,\, \nu \lambda} 
\ee
where
\be
\underbrace{\Gamma^{\mu}_{\,\,\, \nu \lambda}}_\text{Christoffel symbols} = \frac {1}{2} \, g^{\mu 
\rho} \big( \partial_{\nu} \, g_{\rho \lambda} + \partial_{\lambda} \, g_{\rho \nu} - \partial_{\rho} \, 
g_{\nu \lambda} \big ) \,\, .
\ee

The variations of the Riemann and the Ricci tensor are given by 
\be
\delta \, R^{\mu}_{\,\,\,\, \nu \lambda \rho} = \partial_{\lambda} \, \delta \, \Gamma^{\mu}_{\,\,\, \nu 
\rho} - 
\partial_{\rho} \, \delta \, \Gamma^{\mu}_{\,\,\, \nu \lambda} + 
\delta \, \Gamma^{\mu}_{\,\,\, \sigma \lambda} \, \Gamma^{\sigma}_{\,\,\, \nu \rho} + 
\Gamma^{\mu}_{\,\,\, \sigma \lambda} \, \delta \, \Gamma^{\sigma}_{\,\,\, \nu \rho} - 
\delta \, \Gamma^{\mu}_{\,\,\, \sigma \rho} \, \Gamma^{\sigma}_{\,\,\, \nu \lambda} - 
\Gamma^{\mu}_{\,\,\, \sigma \rho} \, \delta \, \Gamma^{\sigma}_{\,\,\, \nu \lambda}
\ee
and
\be
{\delta \, R_{\mu \nu}} = \frac {1}{2} \, g^{\lambda \eta} 
\big( D_{\nu} \, D_{\mu} \, \delta \, g_{\lambda \eta} - 
D_{\lambda} \, D_{\nu} \, \delta \, g_{\eta \mu} - 
D_{\lambda} \, D_{\mu} \, \delta \, g_{\eta \nu} + 
D_{\lambda} \, D_{\eta} \, \delta \, g_{\mu \nu} \big ) \,\, ,
\ee
where $D_\nu$ is the covariant derivative such that $D_\nu V^\mu = \partial _\nu V^\mu +
 \Gamma^{\mu}_{\,\,\, \nu \lambda} V^\lambda$.

\goodbreak

The covariant gauge fixing functional can be written as \cite{Reuter:1996cp} 
\be
\Gamma_{k,gf} = 
\kappa^2 \, \mathrm Z_{Nk} \, \frac {1}{\alpha} \, \int d^D x \, \sgb
\,\, \bar {g}^{\mu \nu} \, ({\cal F}_{\mu}^{\,\alpha \beta} \, g_{\alpha \beta})\, \, ({\cal F}_{\nu}^{\,\rho 
\sigma} \, g_{\rho \sigma}) \,\,,
\label{eq:2.35}
\ee
where  $\alpha$ is a  gauge parameter
and  the  tensor ${\cal F}_{\mu}^{\,\alpha \beta}$ depends on the 
background covariant  derivative,
\be
{\cal F}_{\mu}^{\,\alpha \beta} = \delta_{\mu}^{\beta} \, \,\bar{g}^{\alpha \gamma} \, \, \bar {D}
_{\gamma} \, - \, \frac{1}{2} \, \bar{g}^{\alpha \beta} \, \bar{D}_{\mu} \,\,.
\ee
An important property of the  tensor ${\cal F}_{\mu}^{\,\alpha \beta}$ is that it vanishes when 
contracted with the background metric
\be
{\cal F}_{\mu}^{\,\alpha \beta} \, \bar g_{\alpha \beta} =0 \,\,.
\ee

Landau-deWitt gauge corresponds to the limit $\alpha \to 0$, deDonder gauge to $\alpha=1$. 
In this thesis I used deDonder gauge. 


\chapter{{\color{BrickRed}Derivation of the Functional Renormalisation Group Equations}}
\minitoc

Throughout this thesis I employed two forms of the FRG, an exact functional identity and an
approximated background field flow.
Both are functional differential equations for the effective average action.  
Here, I provide a brief presentation of the derivation of the Wetterich equation and 
a discussion of the Litim  regulator as well as
an alternative derivation of the approximated background field flow in proper-time regularisation. 



\section{An exact functional identity for the effective average action}

As the Wetterich equation is an exact equation an appropriate 
starting point is the generating functional $Z[J]$ where $J(x)$ is a source coupled to the 
field(s).  Taking the logarithm provides the generating functional of the connected Green 
functions $W[J]$. A Legendre transform leads to the effective action $\Gamma [\bar \Phi]$
 which is the  generating functional of the one-particle irreducible Green functions. In the 
Wilsonian RG approach  one also performs these three steps, however, with a  $k$-dependent 
regulator function included, {\it i.e.}, one starts with\footnote{In this thesis only  
Euclidean Quantum Field Theories are considered, {\it i.e.},  it is always assumed that a Wick rotation has been
performed.  In addition, I will use the deWitt notation $J\cdot \Phi = \int d^Dx J(x) \Phi (x)$.}
\be
Z_k[J] = \int {\cal D} \Phi \exp \left( -S[\Phi] -\Delta S_k[\Phi] +J\cdot \Phi \right) \,\,.
\ee
In this functional integral the term $\Delta S_k[\Phi]$ is included to provide a smooth 
momentum cutoff such that the UV modes are unchanged and the IR modes are suppressed.
The standard choice is a quadratic cutoff which, in momentum space, reads
\be
\Delta S_k[\Phi] =  \frac 1 2 \int \dq \, \Phi(-q) \, R_k(q^2) \, \Phi(q) \,\,.
\ee
For the UV modes one thus requires 
\be
\lim_{q^2/k^2\to \infty}  R_k(q^2) = 0 
\ee
and for the IR modes 
\be
R_k(q^2)  \,\,\, \xrightarrow{q^2/k^2\to 0} \,\,\, \infty \,\,,
\ee
{\it cf.} also the discussion in Ref.\ \cite{Litim:2008tt}.

The flow for the generating functional $W_k$ is then straightforwardly determined to be
\be
\partial_t W_k[J] = \partial _t \ln Z_k[J] = - \langle \partial_t \Delta S_k \rangle 
= - \frac 1 2 \int \dq \, \partial_t R_k(q^2) \, \langle \Phi(q) \Phi(-q) \rangle \,\,.
\ee 
Next one introduces the Legendre transformed functional 
\be
\Gamma_k[\bar \Phi] = \sup_J \left( J\cdot \bar \Phi - \ln Z_k [J] - \Delta S_k[\bar \Phi] \right)
\ee
where $\bar \Phi$ is the expectation value of $\Phi$ for a fixed external current $J$,
$\bar \Phi = \langle \Phi \rangle _J$. Returning for clarity for the time being to a real-space
representation 
one employs then the definition of the connected two-point function
\be
G(x,y) := \frac{\delta^2\ln Z_k [J]}{\delta J(x) \, \delta J(y) }
\ee
and the fact that the quantum equation of motion 
\be
J(x) = \frac{\delta \Gamma_k [\bar \Phi]}{\delta \bar \Phi (x)} +
(R_k\cdot \bar \Phi ) (x)
\ee
leads to
\be
G(x,y) = \Bigg(  
\frac{\delta^2 \Gamma_k [\bar \Phi]}{\delta \bar \Phi (x)\,\delta \bar \Phi (y) }
+ R_k (x-y) \Bigg)^{-1}
\ee
to show that in momentum space  the  equation 
\be
\partial _t \Gamma_k [\bar \Phi] = \frac 1 2 \int \dq \Bigg(  
\frac{\delta^2 \Gamma_k [\bar \Phi]}{\delta \bar \Phi (q)\,\delta \bar \Phi (-q) }
+ R_k (q^2) \Bigg)^{-1} \partial_t R_k (q^2) \,\,
\ee
is fulfilled.
In case discrete indices are present they have to be summed over. Denoting by 
$\mathrm{Tr}$ these sums as well as the integrals leads then to 
the compact notation for the Wetterich equation \cite{Wetterich:1992yh}
\be
\boxed{
\partial _t \Gamma_k = \frac 1 2 {\mathrm{Tr}} 
{\Bigg(\left( \Gamma_k^{(2)} 
+ R_k  \right)^{-1} \partial_t R_k \Bigg)  } }
\label{eq:B.23}
\ee
with 
\be
\Gamma_k^{(2)} 
:= \frac{\delta^2 \Gamma_k [\bar \Phi]}{\delta \bar \Phi (q)\,\delta \bar \Phi (-q) }
\,\,.
\ee

Throughout this thesis the Litim regulator
\be
\boxed{
R_ {k} (q^2) = ({k}^2 - q^2)\, \Theta \, ({k}^2 - q^2)} 
\label{eq:B.15}
\ee 
has been used. Its scale derivative can be straightforwardly calculated
\be
\partial_tR_ {k} (q^2) = 2 \,{k} ^2 \,\Theta ({k} ^2 - q^2)  \,\,.
\ee


\section{Background field flow in proper-time regularisation}

Here an alternative derivation of the background field flow in proper-time regularisation is 
summarised following the presentation in Ref.\ \cite{Bonanno:2004sy}.
To arrive at  this flow one may start with the observation that in every Quantum Field Theory the one-loop 
contribution to the effective action can expressed as a logarithm of a determinant. Denoting 
all the fields of the theory as $\Phi$ and  with $S^{(2)}_\Phi$ the matrix of the second 
functional  derivatives of the action $S_\Phi$ w.r.t.\ $\Phi$  one obtains
\be
\Gamma ^{\mathrm {(1-loop)}} [ \Phi] = \frac 1 2 \ln \det S^{(2)}_\Phi - \ln \det S^{(2)}_{gh}
\label{eq:B.1}
\ee
where $S_\Phi$ is assumed to contain potential gauge fixing terms. $S^{(2)}_{gh}$ denotes 
the second derivative of the ghost term associated to this gauge fixing term. 

In the next step one introduces the proper time presentation 
\be
 \ln \det S^{(2)} = {\mathrm {Tr}} \ln S^{(2)} \to 
 - \int _{1/\Lambda^2}^\infty \frac{ds}{s} \, {\mathrm {Tr}} \exp 
 \left( -sS^{(2)}  \right) \,\,.
\ee
UV singularities are then regularised by introducing a corresponding cutoff $\Lambda$. In 
the simplest case one substitutes the lower limit of the $s$-integral by $1/\Lambda^2$.  
Also in this form of 
the RG one wants to follow the idea of the Wilsonian RG and introduces a scale $k$ which 
acts as an IR cutoff. Again, in the simplest case one substitutes then the upper limit of the
$s$-integral by $1/k^2$. 
In this thesis I will use, however, a quite general class of proper time regulator 
functions in order to allow for a comparison to \cite{Liao:1994fp,Bonanno:2004sy}: 
\be
{\mathrm {Tr}} \Big(\ln S^{(2)}\Big)_{\mathrm {reg.}} 
= - \int _0^\infty \frac{ds}{s} \,\, f^m_k(s)  \, {\mathrm {Tr}} \exp \left( -sS^{(2)}  \right) \,\,.
\ee
The regulator functions $f^m_k(s)$ depend on an additional parameter $m$ and are for all 
$m$ chosen such that they interpolate smoothly from zero for $s\gg 1/k^2$ to one for 
small $s$. For the calculations in this thesis the family of regulator functions
\be
f_k^m (s) = e^{(- \mathrm Z \, s \, k^2)} \, \sum_{\mu=0}^m \, \frac{1}{\mu !} \, (\mathrm Z \, s \, 
k^2)^{\mu}
\label{eq:B.4}
\ee
have been used.
Here $Z$ denotes a wave function renormalisation function. The advantage of introducing 
it into the regulator will become evident in Appendix~D. 
In the following also the scale derivative of these functions will be needed. Employing the 
usual definitions
\be
t= \ln k, \quad k \frac \partial {\partial k} = \frac \partial {\partial t} =: \partial _t \,\,,
\ee
one gets 
\be
\partial_t f_k^m (s) = -2 \, \frac{1}{m!} \, (\mathrm Z \, s \, k^2)^{m+1} \,\, e^{- \mathrm Z \, s \, k^2} 
\,\,.
\label{eq:B.6}
\ee
In
\be
\partial_t  {\mathrm {Tr}} \Big( \ln S^{(2)} \Big)_{\mathrm {reg.}} 
= - \int _0^\infty \frac{ds}{s} \,\, \partial_t  f^m_k(s) \, {\mathrm {Tr}} \exp \left( -sS^{(2)}  \right) 
\ee
the integral can be performed resulting in
\be
\partial_t  {\mathrm {Tr}} \Big( \ln S^{(2)} \Big)_{\mathrm {reg.}} 
=   2 \, {\mathrm {Tr}}  \Bigg( \frac{Z\, k^2}{S^{(2)}+ Z\, k^2} \Bigg)^{m+1} \,\,.
\ee

Regularising the determinants in Eq.\ (\ref{eq:B.1})  as described above makes the one-loop 
effective action $\Gamma_k [\Phi] :=S_\Phi + \Gamma_k ^{\mathrm {(1-loop)}}[\Phi]$ dependent on 
the scale $k$:
\be
\partial_t  {\Gamma_k}  [\Phi] = \partial_t  \Gamma_k ^{\mathrm {(1-loop)}}[\Phi]
= - \frac{1}{2} \, {\mathrm {Tr}} \int_0^{\infty} \, \frac{ds}{s} \,\, 
\partial_t f_k^m (s) \,\, (e^ {-s \,  {S}_\Phi^{(2)}} -2 \,\, e^{-s \, S_{gh}^{(2)}})  \,\,.
\ee
This is still a one-loop expression. The underlying idea to implement the RG improvement
is now to turn this into a 
self-consistent and thus non-perturbative equation by replacing on the r.h.s.\ 
${S}_\Phi^{(2)}$ by $\Gamma_k^{(2)}$, the second derivative of the fully dressed and scale-
dependent $\Gamma_k$ which then has to be 
calculated self-consistently:
\be
\boxed{
\partial_t {\Gamma_k}  [\Phi]= - \frac{1}{2} \, {\mathrm {Tr}} \int_0^{\infty} \, \frac{ds}{s} \,\, 
\partial_t f_k^m (s) \,\, \Big(e^ {-s \,  {\Gamma}_k^{(2)}} -2 \,\, e^{-s \, S_{gh}^{(2)}}\Big) }\,\,.
\label{eq:B.10}
\ee
This form of the RG equation (with ghost quantum corrections neglected, see also 
Subsection~2.2.2) will be used in the following section for EH gravity.


\chapter{{\color{BrickRed}Second Variation of the Effective Action}}
\minitoc

\section{Basic Variations}

For the first and second variations of the effective average action, there are a few terms 
which should be computed first. Looking to Eqs. (\ref{eq:2.32}) and (\ref{eq:2.35}) one 
can see that when doing a variation  over $\hat {S_k} [g;\bar g]$ terms like $\delta \sg$ and  $
\delta (\sg\, \,R)$ will appear. For this reason I will start with them.

\bigskip
\noindent
For the variation of the first term one gets:
\bea
\delta \sg &=& \delta \,\,e^{\frac{1}{2}\, \mathrm {tr} \ln g} \nonumber\\
&=& \frac {1}{2} \, \sg \,\,g^{\mu \, \nu} \, h_{\mu \, \nu} \,\,.
\eea
And for the variation of the second term:
\bea
\delta (\sg\, \,R) &=& (\delta \sg) \, R + \sg \,\,(\delta \, R) \nonumber\\
&=& (\delta \,\,e^{\frac{1}{2}\, \mathrm {tr} \ln g}) \, R + \sg \,\, (\delta \,(g^{\mu \nu} R_{\mu \nu})) 
\nonumber\\
&=& \frac{1}{2} \, \sg \,\, h_{\mu}^{\,\,\, \mu} \, R \nonumber\\
&+& \sg \,\, (- h^{\mu \nu} \, R_{\mu \nu} + D^{\alpha} \,\,D^{\mu}\,\, h_{\alpha \mu} - D^2 \,\,h_{\mu}
^{\,\,\, \mu} ) \,\,.
\eea
For the second variation of $\delta \sg$ after some computations one finds:
\be
\delta^2 \, \sg = \frac {1}{4} \, \sg \,\,h_{\mu}^{\,\, \mu}\,\, h_{\mu}^{\,\, \mu} - \frac {1}{2} \,\sg\,\,  
h^{\mu \nu}\,\, h_{\mu \nu} \,\,.
\ee
And for the second variation of $\delta (\sg\, \,R)$ :
\bea
\delta^2 \, (\sg\, \,R) &=& \frac {1}{4} \sg\,\, h_{\mu}^{\,\, \mu}\,\, h_{\mu}^{\,\, \mu} \, R - \frac {1}{2} 
\sg \,\,h^{\mu \nu}\,\, h_{\mu \nu} \, R \nonumber\\
&+& \sg \,\, h_{\mu}^{\,\, \mu}\, ( - h^{\mu \nu}\, R_{\mu \nu} + D^{\alpha} \,\,D^{\mu}\, h_{\alpha \mu} 
- D^2 \,h_{\mu}^{\,\, \mu} ) \nonumber\\
&+&  \sg\,\, [ \,2\, h^{\mu \rho} \,\,h_{\rho}^{\,\, \nu}\, R_{\mu \nu} - \frac {1}{2} \,\,h^{\mu \nu} \,
( D^{\alpha} \,D_{\mu} \,\,h_{\alpha \nu} + D^{\alpha}\, D_{\nu}\,\, h_{\alpha \mu} \nonumber\\ 
&-& D^2 \,\,h_{\mu \nu} - D_{\nu}\,\, D_{\mu}\,\, h_{\alpha}^{\,\, \alpha} ) - \frac {1}{2}\,\, h_{\lambda}
^{\,\, \lambda}\,\, D^{\alpha} \,\,D^{\mu} \,\,h_{\alpha \mu} \nonumber\\
&+& \frac {1}{2} \,\,h_{\rho}^{\,\, \rho} \,\,D^{\alpha} \,\,D_{\alpha} \,\,h_{\lambda}^{\,\, \lambda} \,] \,\,.
\eea

Varying the gauge fixing action (\ref{eq:2.35})
provides the gauge fixing condition 
\be
F_{\nu} := \sqrt {2} \,\, \kappa \,\, {\cal F}_{\nu}^{\alpha \beta} \, [\bar {g}] \, h_{\alpha \beta}  
\stackrel{!}{=} 0 \,\,.
\ee
For further use I note that
\be
\frac{\partial F_{\nu}}{\partial h_{\alpha \beta}} = \sqrt {2} \,\, \kappa \,\, {\cal F}_{\nu}^{\alpha 
\beta} \, [\bar {g}] \,\,.
\ee
Finally, the ghost action is given by
\be
 \Gamma_{gh} [h,c,\bar {c};\bar {g}] = - \, \kappa^{-1} \int d^{D} x \, \sgb
 \,\, \bar  {c}_{\mu} \,\, \bar {g}^{\mu \nu} \,\, \frac{\partial F_{\nu}}{\partial h_{\alpha \beta}} \,\, {\cal 
 L}_c \, (\bar{g}_{\alpha \beta} + h_{\alpha \beta}) \,\,,
\ee
where the Lie derivative ${\cal L}_c$ can be written as:
\be
{\cal L}_c \, \gamma_{\alpha \beta} = \gamma_{\beta \nu} \, D_{\alpha} \, c^{\nu} \, + \,
\gamma_{\alpha \nu} \, D_{\beta} \, c^{\nu} \,\,,
\ee
with $\gamma_{\alpha \beta}$ being the full metric. As usual one can reexpress the ghost 
action in terms of the Faddeev-Popov operator 
\bea
 \Gamma_{gh} [h,c,\bar {c};\bar {g}] &=& - \sqrt{2} \, \int d^{D} x \, \sgb
 \,\, \bar  {c}_{\mu} \,\, {\cal M} [\gamma;\bar {g}]^{\mu}_{\,\,\,\nu} \, c^{\nu}
 \,\, , \\
{\cal M} [\gamma;\bar {g}]^{\mu}_{\,\,\,\nu} &=& (\bar {g}^{\mu \beta} \, \bar {g}^{\alpha \gamma} \, \bar 
{D}_{\gamma} \, \gamma_{\beta \nu} \, D_{\alpha} \, + \bar {g}^{\mu \beta} \, \bar {g}^{\alpha 
\gamma} \, \bar {D}_{\gamma} \, \gamma_{\alpha \nu} \, D_{\beta} \, - \bar {g}^{\mu \lambda} \, \bar 
{g}^{\sigma \rho} \, \bar {D}_{\lambda} \, \gamma_{\rho \nu} \, D_{\sigma} ) \,\,.
\nonumber
\eea

\section{Tensor Decomposition of $\Gamma_k^{(2)}$}

In both, the background field flow (\ref{eq:2.1}) and the exact functional equation (\ref{eq:2.2}), 
the second
variation of the effective average action appears on the respective right hand sides. The
 first and second  variations of the two terms in the EH action are derived in Appendix~C.1. 
Based on these results the second variation of the effective average action 
$ {\Gamma_k} [g;\bar g]$ within the deDonder gauge ($\alpha =1$) can be straightforwardly 
calculated.  As explained in the last paragraph of Subsection~2.2.1 the 
background metric has to be identified with the physical metric, {\it i.e.},  in Eqs.\  (\ref{eq:2.1}) 
and (\ref{eq:2.2}) the second variation of $ {\Gamma_k} [\bar g;\bar g]$
enters. It is given by
\bea
{ {\frac {1}{2} \,\, \delta^2 \,\, \Gamma_k}} \,[\bar g;\bar g] &=& \kappa^2 \, \mathrm Z_{Nk} \, \int 
d^D x \, \sgb \, \Bigg [ \Bigg( \frac{1}{4} \,\,  h_{\mu}^{\,\, \mu} \,\,  h_{\mu}^{\,\, \mu} - \frac{1}{2} \,
\,h^{\mu \nu} h_{\mu \nu} \Bigg) \Big( - \bar {R} + 2 \bar{\lambda}_k \Big) \nonumber \\
&+& h_{\mu}^{\,\, \mu} \,\, h^{\mu \nu} \,\, \bar {R}_{\mu \nu} - h^{\mu \rho} \,\, h_{\rho}^{\,\, \nu} \,\, 
\bar {R}_{\mu \nu} + h^{\mu \nu} \,\, \bar {R}^{\alpha}_{\,\, \,\,\mu \nu}~^{\!\lambda} \,\, h_{\alpha 
\lambda} \nonumber \\
&-& \frac{1}{2} \,\, h^{\mu \nu} \,\, \bar {D}^2 \,\, h_{\mu \nu} + \frac {1}{4} \,\, h_{\mu}^{\,\, \mu} \,\, 
\bar {D}^2 \,\, h_{\mu}^{\,\, \mu} \Bigg ] \,\,.
\eea
Next I decompose the fluctuating metric into a traceless part and a trace part. The idea is 
to diagonalise partly the effective average action. For this one defines
\be
h_{\mu \nu} =: \hat {h}_{\mu \nu} + \frac{1}{D} \, \bar{g}_{\mu \nu} \, h \,\,,
\qquad h := h_{\mu}^{\,\, \, \mu} 
\label{eq:2.46} 
\ee
where the $\hat {h}_{\mu \nu}$ is representing the traceless part.
Consequently, $\bar {g}^{\mu \nu} \, \hat {h}_{\mu \nu} = 0$, and one may rewrite the second 
variation as
\bea 
{ {\frac {1}{2} \,\, \delta^2 \,\, \Gamma_k}} \,[\bar g;\bar g] &=& \kappa^2 \, \mathrm Z_{Nk} \, \int 
d^D x \, \sgb \,\, \Bigg [ \frac {1}{2} \, \hat {h}^{\mu \nu} \, \Big( - \bar {D}^2 + \bar R - 2 \, \bar 
{\lambda}_k \Big) \, \hat {h}_{\mu \nu} \nonumber \\
&-& \Bigg ( \frac {D - 2}{4 \, D} \Bigg ) \, h \, \Bigg ( - \bar {D}^2 - 2 \, \bar {\lambda}_k + \bar R \, 
\Bigg ( \frac {D - 4}{D} \Bigg ) \Bigg ) \, h \nonumber \\
&+& \Bigg ( \frac {D - 4}{D} \Bigg ) \, \hat {h}^{\mu \nu} \, \bar {R}_{\mu \nu} \, h + \hat {h}^{\mu \nu} 
\, \hat {h}_{\alpha \lambda} \, \bar {R}^{\alpha}_{\,\, \,\,\mu \nu}~^{\!\lambda} -  \hat {h}^{\mu \rho} \, 
\hat {h}_{\rho \nu} \, \bar {R}_{\mu}^{\,\,\,\, \nu} \Bigg ] \,\,.
\label{eq:2.47}
\eea
Until now the computations were background independent. To simplify further I will exploit 
the relations  (\ref{eq:2.33}) and (\ref{eq:2.34}) (valid for a maximally symmetric space)
to simplify Eq.\ (\ref{eq:2.47}),
\bea
{ {\delta^2 \,\, \Gamma_k}} \,[\bar g;\bar g] &=& \kappa^2 \, \mathrm Z_{Nk} \, \int d^D x \, \sgb \, 
\Bigg [ \hat {h}^{\mu \nu} \Big ( - \bar {D}^2 - 2 \, \bar {\lambda}_k + C_T \, \bar {R} \Big ) \, \hat {h}
_{\mu \nu} \nonumber \\
&-& \Bigg ( \frac {D - 2}{2 \, D} \Bigg ) \, h \, \Big ( - \bar {D}^2 - 2 \, \bar {\lambda}_k + C_S \, \bar {R} 
\Big ) \, h \Bigg ] \,\,,
\label{eq:2.48}
\eea
where I introduced a short notation for the following terms:
\be
C_T = \Bigg ( \frac {D^2 - 3 \, D + 4}{D \, (D - 1)} \Bigg )
\ee
and
\be
C_S = \Bigg ( \frac {D - 4}{D} \Bigg ) \,\,.
\ee

\goodbreak

For the second variation of the ghost action one obtains 
\be
S_{gh}^{(2)} [g;\bar {g}]^{\mu}_{\,\,\,\nu} = - \sqrt{2} \,\, {\cal M} [g;\bar {g}]^{\mu}_{\,\,\,\nu}\,\,.
\ee
 The Faddeev-Popov operator ${\cal M} [\bar {g};\bar {g}]^{\mu}_{\,\,\,\nu}$ 
in this equation also simplifies in the maximally symmetric space:
\be
{\cal M} [\bar {g};\bar {g}]^{\mu}_{\,\,\,\nu} = - \,C_V \, \delta^{\mu}_{\,\,\,\, \nu} \, \bar {R} \,+  
\delta^{\mu}_{\,\,\,\, \nu} \, \bar {D}^2 
\label{eq:2.52}
\ee
with
\be
C_V = - \, \Bigg ( \frac {1}{D} \Bigg ) \,\,.
\ee


\chapter[{\color{BrickRed}Trace Evaluations and Heat Kernel Expansion}]{{\color{BrickRed}
Trace Evaluations \\
and Heat Kernel Expansion}}
\minitoc

As a first step for deriving the $\beta$-functions one notes that 
in both, the background field flow (\ref{eq:2.1}) and the exact functional equation (\ref{eq:2.2}),
the l.h.s.\  can be written within the EH theory as
\be
\partial_t \Gamma_k  [\bar g;\bar g] 
= 2 \, \kappa^2 \, \int d^D x \, \sgb \, \Big( (\partial_t \, \mathrm Z_{Nk}) \, (- \bar {R}(\bar {g}) + 2 \, 
\bar \lambda_k) + 2 \, \mathrm Z_{Nk} \, \partial_t \,\bar \lambda_k \Big) \,\,.
\ee

To evaluate the traces  in Eqs.\  (\ref{eq:2.1}) and (\ref{eq:2.2})
 one needs to consider the cases of the approximate background field flow and of the 
 exact functional identity separately.

\section{Background Field Flow}

The computation of the r.h.s. of Eq.\  (\ref{eq:2.1}) requires to identify the three different 
regulator functions in Eqs.\  (\ref{eq:B.4}) and (\ref{eq:B.6}):
\begin{center}
\begin{tabular}{rl}
traceless tensor: &$Z \to \mathrm Z_{Nk} \, \kappa^2$ \,\,,\\[3mm]
scalar (trace part): &$Z \to - \frac {D - 2}{2 \, D} \,\, \mathrm Z_{Nk} \, \kappa^2$ \,\,,\\[3mm]
vector (ghost part): &$Z \to \sqrt {2}$ \,\,. \\
\end{tabular}
\end{center}

\bigskip

Plugging in the respective three versions of Eq.\  (\ref{eq:B.6}) into Eq.\  (\ref{eq:2.1}) and using 
Eq.\  (\ref{eq:2.48}) as well as Eq.\  (\ref{eq:2.52}), one obtains
\bea
&& \partial_t \Gamma_k  [\bar g;\bar g] = \nonumber \\
&=& {\mathrm {Tr}_T} \int_0^{\infty} \, \frac{ds}{s} \, \, 
\frac{1}{m!} \, \, ( \mathrm Z_{Nk} \, \kappa^2 \, s \, k^2 )^{m+1} \,\,  
\exp \Big ({- \mathrm Z_{Nk} \, \kappa^2 \, s \, 
( - \bar {D}^2 + k^2 - 2 \, \bar {\lambda}_k + C_T \, \bar {R} )} \Big ) \nonumber \\
&+& {\mathrm {Tr}_S} \int_0^{\infty} \, \frac{ds}{s} \, \, \frac{1}{m!} \, \, 
\Bigg ( - \frac {D - 2}{2 \, D} \,\, \mathrm Z_{Nk} \, \kappa^2 \, s \, k^2 \Bigg )^{m+1} \,\, \times
\nonumber \\
&& \hspace{30mm} \times
\exp \Bigg ({\mathrm Z_{Nk} \, \kappa^2 \, \frac {D - 2}{2 \, D} \, s \, 
( - \bar {D}^2 + k^2 - 2 \, \bar {\lambda}_k + C_S \, \bar {R} )} \Bigg ) \nonumber \\
&-& 2 \, {\mathrm {Tr}_V} \int_0^{\infty} \, \frac{ds}{s} \, \, \frac{1}{m!} \, \, 
(\sqrt {2} \, s \, k^2 )^{m+1} \,\,  
\exp \Big ({- s \, \sqrt {2} \, ( - \bar {D}^2 +k^2 + C_V \, \bar {R} )} \Big ) \,\,,
\eea
and expanding up to linear order in $\bar R$
\bea
&& \partial_t \Gamma_k  [\bar g;\bar g] \approx \nonumber \\
&\approx& {\mathrm {Tr}_T} \int_0^{\infty} \, \frac{ds}{s} \, \, \frac{1}{m!} \, \, 
( \mathrm Z_{Nk} \, \kappa^2 \, s \, k^2 )^{m+1} \,\, 
\exp \Big ({- \mathrm Z_{Nk} \, \kappa^2 \, s \, ( - \bar {D}^2 + k^2 - 2 \, \bar {\lambda}_k )} \Big ) 
\times \nonumber \\
&& \hspace{70mm} \times
( 1 - \mathrm Z_{Nk} \, \kappa^2 \, s \, C_T \, \bar {R} ) \nonumber \\
&+& {\mathrm {Tr}_S} \int_0^{\infty} \, \frac{ds}{s} \, \, \frac{1}{m!} \, \,
\Bigg ( - \frac {D - 2}{2 \, D} \,\, \mathrm Z_{Nk} \, \kappa^2 \, s \, k^2 \Bigg )^{m+1} \,\, \times 
\label{eq:C.2}
\\
&& \hspace{10mm} \times 
\exp \Bigg ({\frac {D - 2}{2 \, D} \,\, \mathrm Z_{Nk} \, \kappa^2 \, s \, 
( - \bar {D}^2 + k^2 - 2 \, \bar {\lambda}_k )} \Bigg ) \,\,
\Bigg ( 1 + \frac {D - 2}{2 \, D} \,\, \mathrm Z_{Nk} \, \kappa^2 \, s \, C_S \, \bar {R} \Bigg )  
\nonumber \\
&-& 2 \, {\mathrm {Tr}_V} \int_0^{\infty} \, \frac{ds}{s} \, \, \frac{1}{m!} \, \, 
(\sqrt {2} \, s \, k^2 )^{m+1} \,\, 
\exp \Big ({- s \, \sqrt {2} \, ( - \bar {D}^2 +  k^2 )} \Big ) \,\, 
( 1 - s \, \sqrt {2} \, C_V \, \bar {R} ) \,\,. \nonumber
\eea

Next I compute the trace of the three exponentials separately using the formulae for the heat 
kernel expansion \cite{Reuter:1996cp,Bonanno:2004sy}
\bea
\mathrm {Tr} [W( - \bar {D}^2)] &=& (4\pi)^{-D/2} \mathrm{tr} (I) \Bigg(  Q_{D/2}[W]  
\int d^D \, x \,\, \sgb  \nonumber \\
&&+ \frac 1 6 \, Q_{D/2-1}[W]  \int d^D \, x \,\, \sgb
\,\, \bar {R}  + {\cal {O}} (\bar {R}^2) \Bigg)  
\label{eq:C.3}\\
{\mathrm{and}}\qquad \qquad && \nonumber \\
Q_n[W] &=& \frac 1 {\Gamma(n)} \int _0 ^\infty dz \, z^{n-1} \, W(z) \,\,,
\label{eq:C.4}
\eea
where $W$ is an appropriately chosen function.\footnote{Here it is helpful to note that the 
functional $Q_n[W]$ is directly
related to an integral over covariant background momenta $q$:
\be\textstyle{
 \int \frac {d^D \, q}{(2 \, \pi)^D} \,\, W (q^2) \,\, =\,\,\, \frac {(4 \, \pi)^{-D / 2} }{\Gamma \, (D/2)} \,\, 
\int_0^{\infty} \, dz \,\, z^{D/2 - 1} \,\, W(z) \,\, = (4 \, \pi)^{-D / 2} \, Q_{D/2}[W]  \, . } \nonumber
\ee
The condition to be fulfilled by the otherwise arbitrary function $W$ is the requirement that the 
integral above is well-defined.
}
For the tensor trace this yields
\bea
&& {\mathrm {Tr}_T} \, \exp \Big ({- \mathrm Z_{Nk} \, \kappa^2 \, s \, 
( - \bar {D}^2 + k^2 - 2 \, \bar {\lambda}_k )} \Big ) = \nonumber \\
&=& (4 \, \pi)^{-D/2} \,\, \mathrm{tr}_T (I) \times
 \nonumber \\
&&\times \Bigg (\, \frac {1}{\Gamma \, (D/2)} \, \int_0^{\infty} \, dz \, z^{D/2 \,- 1} \, \exp \Big ({- 
\mathrm Z_{Nk} \, \kappa^2 \, s \, ( z + k^2 - 2 \, \bar {\lambda}_k )} \Big ) \, \int d^D \, x \,\, \sgb 
\nonumber \\
&&+ \frac {1}{6} \,\, \frac {1}{\Gamma \, (D/2 \,- 1)} \, \int_0^{\infty} \, dz \, z^{D/2 \,- 2} \, \exp \Big ({- 
\mathrm Z_{Nk} \, \kappa^2 \, s \, ( z + k^2 - 2 \, \bar {\lambda}_k )} \Big ) \, \int d^D \, x \,\, \sgb \,\, 
\bar {R} 
\nonumber \\
&&+ {\cal {O}} (\bar {R}^2) \, \Bigg ) \,\,,
\eea
with $\mathrm{tr}_T (I) = \frac {1}{2} \, (D-1) \, (D+2)$. Analogously for the scalar trace
\bea
&& {\mathrm {Tr}_S} \, 
\exp \Bigg ({\frac {D - 2}{2 \, D} \,\, \mathrm Z_{Nk} \, \kappa^2 \, s \, 
( - \bar {D}^2 + k^2 - 2 \, \bar {\lambda}_k )} \Bigg ) = \nonumber \\
&=& (4 \, \pi)^{-D/2} \,\, \mathrm{tr}_S (I) \times
 \nonumber \\
&&
\times \Bigg [\, \frac {1}{\Gamma \, (D/2)} \, 
\int_0^{\infty} \, dz \, z^{D/2 \,- 1} \, 
\exp \Bigg ({\frac {D - 2}{2 \, D} \,\, \mathrm Z_{Nk} \, 
\kappa^2 \, s \, ( z +  k^2 - 2 \, \bar {\lambda}_k )} \Bigg ) 
 \int d^D \, x \,\, \sgb \nonumber \\
&&+ \frac {1}{6} \,\, \frac {1}{\Gamma \, (D/2 \,- 1)} \, \int_0^{\infty} \, dz \, z^{D/2 \,- 2} 
\, \exp \Bigg ({\frac {D - 2}{2 \, D} \,\, \mathrm Z_{Nk} \, \kappa^2 \, s \, 
( z +  k^2 - 2 \, \bar {\lambda}_k )} \Bigg ) \times \nonumber \\ 
&& \hspace{50mm} \times \int d^D \, x \,\, \sgb \,\, \bar {R} + {\cal {O}} (\bar {R}^2) \, \Bigg ] \,\,,
\eea
with $\mathrm{tr}_S (I) = 1$  and 
\bea
&&{\mathrm {Tr}_V} \, 
\exp \Big ({- s \, \sqrt {2} \, ( - \bar {D}^2 +  k^2 )} \Big ) = 
\nonumber\\
&=& (4 \, \pi)^{-D/2} \,\, \mathrm{tr}_V (I) \times
\nonumber\\
&&\times \Bigg [\, \frac {1}{\Gamma \, (D/2)} \, 
\int_0^{\infty} \, dz \, z^{D/2 \,- 1} \, \exp \Big ({- s \, \sqrt {2} \, ( z +  k^2 )} \Big ) \,\,
 \int d^D \, x \,\, \sgb \nonumber \\
&&+ \frac {1}{6} \,\, \frac {1}{\Gamma \, (D/2 \,- 1)} \, 
\int_0^{\infty} \, dz \, z^{D/2 \,- 2} \, \exp \Big ({- s \, \sqrt {2} \, ( z +  k^2 )} \Big ) \,
 \int d^D \, x \,\, \sgb \,\, \bar {R} \nonumber\\
&&+ {\cal {O}} (\bar {R}^2) \, \Bigg ] \,\,,
\eea
with $\mathrm{tr}_V (I) = D$ .

\goodbreak

Putting all terms together, performing first the integral over $s$ and afterwards the integral 
over $z$, the resulting equation can be written as
\bea
&&2 \, \kappa^2 \, \int d^D x \, \sgb \, 
\Big(\partial_t \, \mathrm Z_{Nk} \, (- \bar {R}(\bar {g}) + 2 \, \bar \lambda_k) + 2 \, 
\mathrm Z_{Nk} \, \partial_t \,\bar \lambda_k\Big) \approx \nonumber\\
&\approx& (4 \, \pi)^{-D/2} \,\,\, \frac {\Gamma (m+1 - D/2)}{\Gamma (m+1)} \,\,\, (k^2)^{m+1} \times 
\nonumber \\
&& 
\times \Bigg ( \frac {1}{(k^2 - 2 \, \bar {\lambda}_k )^{(m+1 - D/2)}} \,\,\,
\frac {1}{2} \,\, D(D+1) \, - 2 \, D \,\,\, \frac {1}{(k^2)^{(m+1 - D/2)}} \, \Bigg ) \,\, \int 
d^D \, x \,\, \sgb \nonumber \\
&+& (4 \, \pi)^{-D/2} \,\,\, \frac {\Gamma (m+2 - D/2)}{\Gamma (m+1)} \,\,\, 
(k^2)^{m+1} \times \nonumber \\
&& 
\times \Bigg ( \frac {1}{(k^2 - 2 \, \bar {\lambda}_k )^{(m+2 - D/2)}} \,\,\, \frac {1}{12} \,\,\,
(-5\, D + 7) \, D \, - \frac {1}{3} \,\, (D + 6) \,\,\, \frac {1}{(k^2)^{(m+2 - D/2)}} \, \Bigg ) \times
\nonumber \\
&& 
\times \int d^D \, x \,\, \sgb \,\, \bar {R} \,\,. 
\label{eq:C.8}
\eea

This equation can be split into two independent equations, one for the part independent of the 
Ricci scalar and one for the part linear in the Ricci scalar:
\bea
&& \hspace{-12mm}
2 \, \kappa^2 \, \Big ( ( \partial_t \, \mathrm Z_{Nk} ) \, 2 \, \bar \lambda_k + 
2 \, \mathrm Z_{Nk} \, ( \partial_t \,\bar \lambda_k) \, \Big ) =
\nonumber \\
&&= (4 \, \pi)^{-D/2} \,\,\, \frac {\Gamma (m+1 - D/2)}{\Gamma (m+1)}  \,\,\, k^D   \Bigg ( \frac {\frac 1 2 D(D+1)}{(1 - 2 \, \bar {\lambda}_k / k^2)^{(m+1 - D/2)}} \,\,\, 
 \, - 2 \, D  \, \Bigg ) ~~~~
\eea
and
\bea
&&\hspace{-10mm} - 2 \, \kappa^2 \, ( \partial_t \, \mathrm Z_{Nk} ) =
 \nonumber \\ &&\hspace{-10mm} =
(4 \, \pi)^{-D/2} \,\,\, \frac {\Gamma (m+2 - D/2)}{\Gamma (m+1)} \,\,\, 
k^{D-2}  \Bigg ( \frac {\frac {1} {12} \,\,\,
 (-5\, D + 7) \, D }{(1 - 2 \, \bar {\lambda}_k /k^2)^{(m+2 - D/2)}} \,\,\, 
  - \frac {1}{3} \,\, (D + 6) \, \Bigg ).  \nonumber \\
\eea
Rewriting the last two equations in terms of the dimensionless Newton's constant
$
g_k \equiv k^{D-2} \,\,\, G_k \equiv  k^{D-2} \,\,\, \mathrm Z_{Nk}^{-1} \,\,\,  {G_N} 
$
and the dimensionless cosmological constant
$
\lambda_k \equiv  k^{-2} \,\,\, \bar {\lambda}_k 
$
leads to the following system of equations
\bea
\hspace{-10mm} \partial_t \, \Big ( \mathrm Z_{Nk} \, {\lambda}_k{k^2} \Big ) &=& 
8 \, \pi \, g_k k^2 \,\, \mathrm Z_{Nk} \,\, (4 \, \pi)^{-D/2} \,\,\, 
\frac {\Gamma (m+1 - D/2)}{\Gamma (m+1)}  \times \nonumber 
\\
&&
\times \Bigg ( \frac {\frac {1}{2} \,\, D(D+1) }{(1 - 2 \, \lambda_k )^{(m+1 - D/2)}} \,\,\, 
 - 2 \, D  \, \Bigg )
\eea
and
\bea
\partial_t \, \mathrm Z_{Nk} &=& - 16 \, \pi \, g_k\,\, 
\mathrm Z_{Nk} \,\, (4 \, \pi)^{-D/2} \,\,\, 
\frac {\Gamma (m+2 - D/2)}{\Gamma (m+1)}  \times \nonumber \\
&&
\times \Bigg ( \frac {\frac {1}{12} \,\,\, (-5\, D + 7) \, D}{(1 - 2 \lambda_k )^{(m+2 - D/2)}} 
 \, - \frac {1}{3} \,\, (D + 6) \,\,\,   \, \Bigg ) \,\,,
\eea
which relates directly to the $\beta$-functions as given in Sec.~2.3


\section{Exact Functional Identity}

Before performing the analogous evaluation of the traces for the Wetterich equation 
a note on the regulator function needs to be given.   Starting from 
\be
\tilde {R}_k \, (q^2) = q^2 \, r \, (y) \qquad \mathrm {with} \qquad y = 
\frac {q^2}{k^2}\,\,,
\ee
and
\be
\partial_t \, \tilde {R}_k \, (q^2) = - \, 2 \, y \,  r^{\prime} \, (y) \, q^2 
\ee
one obtains the Litim regulator ({\it cf.},  Eq.\ (\ref{eq:2.4})) if 
 the dimensionless regulator function $r \, (y)$ is given by
\be
r_{opt} \, (y) = \Bigg ( \frac {1}{y} - 1 \Bigg ) \,\, \Theta \,\, ( 1 - y ) \,\,.
\ee
The corresponding scale derivative reads
\be
\partial_t \, \tilde {R}_k \, (q^2) = 2 \, k^2 \, \Theta \,\, ( 1 - y ) \,\,.
\ee
The full regulator including the factor $Z_k$, {\it i.e.}, the  running wave function 
renormalisation constant, is given by:
\be
R_k = \mathrm {Z_k} \, \tilde {R}_k \, (- \bar {D}^2) = \mathrm {Z_k} \, (- \bar {D}^2) \, r_{opt} \, (y) 
\qquad \mathrm {with} \qquad y = \frac {- \bar {D}^2}{k^2}\,\,,
\label{eq:C.13}
\ee
which results in a scale derivative
\be
\partial_t \, R_k = - \eta_{Nk,gh} \, R_k +  \mathrm {Z_k} \, 2 \, k^2 \, 
\Theta \,\, ( 1 - y ) \,\,.
\label{eq:C.14}
\ee
The Wetterich equation for the Einstein-Hilbert theory explicitly reads:
\bea&&2 \, \kappa^2 \, \int d^D x \, \sgb \, 
\Big(\partial_t \, \mathrm Z_{Nk} \, (- \bar {R}(\bar {g}) + 2 \, \bar \lambda_k) + 
2 \, \mathrm Z_{Nk} \, \partial_t \,\bar \lambda_k\Big) = \nonumber \\
&& = \frac {1}{2} \,\, \mathrm {Tr_T} \, \Bigg [ \Bigg ( \kappa^2 \, \mathrm Z_{Nk} \, 
( - \bar {D}^2 - 2 \, \bar {\lambda}_k + C_T \, \bar {R}) + \mathrm Z_{Nk} \, \kappa^2 \, 
( - \bar {D}^2 ) \, r_{opt} \, (y) \Bigg )^{-1} \times \nonumber \\
&&~ ~ ~ ~~~~~~~~~\times  \Bigg ( - \eta_{Nk} \, \mathrm {Z}_{Nk} \, \kappa^2 \, 
( - \bar {D}^2 ) \, r_{opt} \, (y) + \mathrm {Z}_{Nk} \, \kappa^2 \, 2 \, k^2 \, 
\Theta \,\, ( 1 - y ) \Bigg ) \, \Bigg ] \nonumber  \\
&&+ ~~\, \frac {1}{2} \,\, \mathrm {Tr_S} \, \Bigg [ \Bigg ( \kappa^2 \, \mathrm Z_{Nk} \, 
c_D\, ( - \bar {D}^2 - 2 \, \bar {\lambda}_k + C_S \, \bar {R})
+ c_D \, 
\mathrm Z_{Nk} \, \kappa^2 \, ( - \bar {D}^2 ) \, r_{opt} \, (y) \Bigg )^{-1} \times 
\nonumber \\
&& \qquad \qquad \times \, \Bigg ( - \eta_{Nk} \, 
c_D\, \mathrm {Z}_{Nk} \, \kappa^2 \,
( - \bar {D}^2 ) \, r_{opt} \, (y) +
c_D\, \mathrm {Z}_{Nk} \, \kappa^2 \, 2 \, k^2 \,
\Theta \,\, ( 1 - y ) \Bigg ) \, \Bigg ] \nonumber \\
&& - \, 2 \,\, \frac {1}{2} \,\, \mathrm {Tr_V} \, \Big [ \Big ( \sqrt {2} \,
( - \bar {D}^2 + C_V \, \bar {R} ) + \sqrt {2} \, ( - \bar {D}^2 ) \,  r_{opt} \, (y) \Big )^{-1} 
\times \nonumber \\
&&~ ~ ~ ~~~~~~~~~\times  
\,\, \Big ( - \eta_{gh} \, \sqrt {2} \, ( - \bar {D}^2 ) \,  r_{opt} \, (y) + \sqrt {2} \,\, 2 \, k^2 
\, \Theta \,\, ( 1 - y ) \Big ) \, \Big ] \,\,,
\label{eq:D.20}
\eea
where $c_D=-(D-2)/2D$ and for the employed truncation one puts additionally $\eta_{gh}=0$.

Once again one expands up to linear order in $\bar {R}$ and employs the heat kernel expansion.
Using the abbreviations
\bea
\label{eq:barck}
\bar c_k & :=&  \frac {1}{k^2 - 2 \, \bar {\lambda}_k} \, ,\nonumber \\
t_k(z)&:=& \Big ( - \eta_{Nk} \,\, (k^2 - z) + 2 \, k^2 \Big ) \,\, \Theta \,\, \Big ( 1 - \frac {z}{k^2} 
\Big ) \, , 
\eea
one obtains
\bea 
&& 2 \, \kappa^2 \, \int d^D x \, \sgb \, \Big(\partial_t \, \mathrm Z_{Nk} \, 
(- \bar {R}(\bar {g}) + 2 \, \bar \lambda_k) + 2 \, \mathrm Z_{Nk} \, \partial_t 
\,\bar \lambda_k\Big)=\nonumber \\
&&=\frac {1}{2} \,\,\, (4 \, \pi)^{-D / 2} \,\,\, \frac {1}{2} \, (D-1) \, (D+2) \,\,\, 
\Bigg \{ \frac {1}{\Gamma \, (D/2)} \,\,\, \int_0^{\infty} \, dz \,\, z^{D/2 - 1} \times
\nonumber \\
&& \hspace {3cm}  
\times \Big [ t_k(z)  \,\,\, 
\Big (\bar c_k  \,\, - 
\,\, \bar c_k^2 C_T \, \bar {R} \Big ) \, \Big ] \,\,  \,\, \int d^D \, x \,\, \sgb \nonumber \\
&& \hspace {8mm}  + \, \, \frac {1}{6} \,\,\, \frac {1}{\Gamma \, (D/2 - 1)} \,\,\, 
\int_0^{\infty} \, dz \,\, z^{D/2 - 2}  \,\,\,
 t_k(z) \bar c_k  \, 
\int d^D \, x \,\, \sgb \, \bar {R} \, \Bigg \} \nonumber \\
&&+\frac {1}{2} \,\,\, (4 \, \pi)^{-D / 2} \,\,\, \Bigg \{ \frac {1}{\Gamma \, (D/2)} \,\,\, 
\int_0^{\infty} \, dz \,\, z^{D/2 - 1} 
 \Big [t_k(z)  \,\,\, \Big( \bar c_k\,\, -
 \,\, \bar c_k^2 C_S \, \bar {R} \Big ) \, \Big ] \times
\nonumber \\
&& \hspace {10cm} \times 
  \int d^D \, x \,\, \sgb\nonumber \\
&&+ \, \, \frac {1}{6} \,\,\, \frac {1}{\Gamma \, (D/2 - 1)} \,\,\, 
\int_0^{\infty} \, dz \,\, z^{D/2 - 2}  \,\,\, 
 t_k(z)  \,\,\,  \bar c_k \,\,\, \int d^D \, x \,\, \sgb \, \bar {R} \, \Bigg \} \nonumber \\
&&- \, \,  (4 \, \pi)^{-D / 2} \,\,\, D \,\,\, \Bigg \{ \frac {1}{\Gamma \, (D/2)} \,\,\, 
\int_0^{\infty} \, dz \,\, z^{D/2 - 1} \,\,\, \Bigg [ 2 \, k^2 \,\, \Theta \,\, \Big ( 1 - \frac {z}{k^2} \Big ) 
\,\,\, \Big ( \frac {1}{k^2} \,\, - \,\, \frac {C_V \, \bar {R}}{k^4} \Big ) \, \Bigg ] \times
\nonumber \\
&& \hspace {10cm} \times 
\int d^D \, x \,\, \sgb\nonumber \\
&&+ \, \, \frac {1}{6} \,\,\, \frac {1}{\Gamma \, (D/2 - 1)} \,\,\, 
\int_0^{\infty} \, dz \,\, z^{D/2 - 2} \,\,\, \Bigg [ 2 \, k^2 \,\, \Theta \,\, \Big ( 1 - \frac {z}{k^2} \Big ) 
\,\,\, \Big ( \frac {1}{k^2} \Big ) \, \Bigg ] \,\,
\int d^D \, x \,\, \sgb \, \bar {R} \, \Bigg \}  \nonumber \\
\label{eq:C.16}
\eea
where the integrals over $z$ can  be performed easily. 

After all the computations one arrives at the following 
flow equation:
\bea
&& 2 \, \kappa^2 \, \int d^D x \, \sgb \, \Big(\partial_t \, \mathrm Z_{Nk} \, 
(- \bar {R}(\bar {g}) + 2 \, \bar \lambda_k) + 2 \, \mathrm Z_{Nk} \, \partial_t 
\,\bar \lambda_k\Big)=\nonumber \\
&=& (4 \, \pi)^{-D/2} \,\, \frac {k^D}{\Gamma \, (D/2)} \,\,\, \Bigg ( \,\frac {(D + 1)}{(D + 2)} 
\,\,\,\, \frac {(- \eta_{Nk} + D + 2) \, k^2}{(k^2 - 2 \, \bar {\lambda_k})} \,\, - 4 \, \Bigg ) 
\, \int d^D \, x \,\, \sgb \nonumber \\
&+& (4 \, \pi)^{-D/2} \,\, \frac {k^D}{\Gamma \, (D/2)} \,\,\, 
\Bigg ( - \, \frac {(D - 1)}{(D + 2)} \,\,\,\, (- \eta_{Nk} + D + 2) \,\,\, 
\frac {k^2}{(k^2 - 2 \, \bar {\lambda}_k)^2}  
\nonumber \\
&&~ ~ ~ ~~~~~~~~~ +
\frac {(D + 1)}{12} \,\, (- \eta_{Nk} + D) \,\, \frac {1}{(k^2 - 2 \, \bar {\lambda_k})} - 
2 \, D \, k^2 \,\, \bigg ( \frac {2}{D^2} + \frac {1}{6} \, \bigg ) \, \Bigg ) \times
\nonumber \\
&&~ ~ ~ ~~~~~~~~~ \qquad \qquad \times 
\,\, \int d^D \, x \,\, \sgb \, \bar {R} \,\,\,.
\label{eq:C.17}
\eea
Again it is straightforward to read off the $\beta$-functions as given in Sec.~2.3.

\section{Tracing with a Compact Dimension}

From $D$ dimensions one is chosen compact, and thus the periodic boundary conditions as in 
Eq.\ (\ref{eq:4.1}) arise.  Therefore $D-1$ components of the momentum vector are continuous,
one is discrete:
\be
q_{D,n} = \frac{2\pi n}{L} \,\, .
\ee
The traces involve then a $(D-1)$-dimensional integral and a sum, and as explained in Sect.~4.1
one substitutes
\be
\int \frac{d^Dq}{(2\pi)^D} \rightarrow \frac{1}{L} \, \sum_n \, \int \frac{d^{D-1}q}{(2\pi)^{D-1}} \,\,.
\label{eq:subst}
\ee

In Chapter~2 and in the derivations in this Appendix, the identity 
\be
\int \frac {d^D \, q}{(2 \, \pi)^D} \,\, W (q^2) = (4 \, \pi)^{-D / 2} \,\,\, \frac {1}{\Gamma \, (D/2)} \,\, 
\int_0^{\infty} \, dz \,\, z^{D/2 - 1} \,\, W (z) 
\label{eq:int}
\ee
has been implicitly used, {\it i.e.}, all momentum integrals have been already expressed as 
integrals over $z=q^2$. Hereby, $W(z)$ is an arbitrary function except the requirement that the
integral should be well-defined and finite. 
Likewise, the integrals (\ref{eq:C.4}) can be equivalently written as momentum 
integrals, {\it e.g.},
\be 
(4 \, \pi)^{-D / 2} \, Q_{D/2}[W] = \int \frac {d^D \, q}{(2 \, \pi)^D} \,\, W (q^2) \,\,.
\ee
The (back-)substitution of $Q_{D/2-1}[W]$ is slightly more subtle,
\be
(4 \, \pi)^{-D / 2} \, Q_{D/2-1}[W] = (D/2 - 1) \int \frac {d^D \, q}{(2 \, \pi)^D} \,\, \frac {1}{q^2} \,\, W 
(q^2) \,\,.
\ee
In total this amounts to rewrite the heat kernel expansion as
\bea
\mathrm {Tr} [W( - \bar {D}^2)] &=& \mathrm{tr} (I) \Bigg(   \int \frac {d^D \, q}{(2 \, \pi)^D} \,\, W 
(q^2) 
\int d^D \, x \,\, \sgb  \nonumber \\
&&+ \frac 1 6 \bigg(\frac D 2 - 1\bigg) \int \frac {d^D \, q}{(2 \, \pi)^D} \,\, \frac {1}{q^2} \,\, W (q^2) 
 \int d^D \, x \,\, \sgb
\,\, \bar {R}  \nonumber \\
&&+ {\cal {O}} (\bar {R}^2) \Bigg)  \,\,,
\eea
in which then the substitution (\ref{eq:subst}) can be straightforwardly applied
\bea
\mathrm {Tr} [W( - \bar {D}^2)] &=& \mathrm{tr} (I) \Bigg( \frac{1}{L} \, \sum_n \, 
\int \frac{d^{D-1}q}{(2\pi)^{D-1}} W (q^2) 
\int d^D \, x \,\, \sgb  \nonumber \\
&&+ \frac 1 6 \bigg(\frac D 2 - 1\bigg)  \frac{1}{L} \, \sum_n \, 
\int \frac{d^{D-1}q}{(2\pi)^{D-1}}  \,\, \frac {1}{q^2} \,\, W (q^2) 
 \int d^D \, x \,\, \sgb
\,\, \bar {R}  \nonumber \\
&&+ {\cal {O}} (\bar {R}^2) \Bigg)  \,\,.
\eea
Doing the angular integrals, {\it i.e.}, applying Eq.\ (\ref{eq:int}) for $D\to D-1$, one gets
\bea
\mathrm {Tr} [W( - \bar {D}^2)] &=& \mathrm{tr} (I) (4 \, \pi)^{-(D-1) / 2} \,\,\,
 \frac {1}{\Gamma \, (D/2-1/2)} \times \nonumber \\
&& \times \Bigg( \frac{1}{L} \, \sum_n \, 
\int_0^{\infty} \, dz \,\, z^{D/2 - 3/2} \,\,  W(z+q_{D,n} ^2)
\int d^D \, x \,\, \sgb  \nonumber \\
&&+ \frac 1 6 \bigg(\frac D 2 - 1\bigg)  \frac{1}{L} \, \sum_n \, 
 \int_0^{\infty} \, dz \,\, z^{D/2 - 3/2} \,\,  \frac 1 {z+q_{D,n} ^2} W(z+q_{D,n} ^2)
\times   \nonumber \\
&& \times 
\int d^D \, x \,\, \sgb \,\, \bar {R}  + {\cal {O}} (\bar {R}^2) \Bigg)  \,\,.
\label{eq:4.9}
\eea
As described in Chapter~4 the term $\displaystyle{\frac 1 {z+q_{D,n} ^2} }$ in 
the part of this expression linear in $\bar R$ leads to logarithms in the 
sums over $n$. Decreasing for this part the dimensionality of the trace simplifies
the expressions enough such that the sums can be
performed in a closed form. The corresponding approximate formula for the heat kernel 
expansion reads
\bea
\mathrm {Tr} [W( - \bar {D}^2)] &=& \mathrm{tr} (I) (4 \, \pi)^{-(D-1) / 2} \,\,\,
 \frac {1}{\Gamma \, (D/2-1/2)} \times \nonumber \\
&& \times \Bigg( \frac{1}{L} \, \sum_n \, 
\int_0^{\infty} \, dz \,\, z^{D/2 - 3/2} \,\,  W(z+q_{D,n} ^2)
\int d^D \, x \,\, \sgb  \nonumber \\
&&+ \frac 1 6 \bigg(\frac D 2 - 1\bigg)  \frac{1}{L} \, \sum_n \, 
 \int_0^{\infty} \, dz \,\, z^{D/2 - 5/2} \,\,  W(z+q_{D,n} ^2) \times \nonumber \\
&& \times \int d^D \, x \,\, \sgb \,\, \bar {R}  + {\cal {O}} (\bar {R}^2) \Bigg)  \,\,.
\label{eq:4.10}
\eea


%

\section{Dimensionally Reduced Background Field Flow}

\noindent
Starting from Eq.\ (\ref{eq:C.2}) and using Eq.\ (\ref{eq:4.10}) for the evaluation of the
traces yields
\be
\hspace{-4cm} 
2 \, \kappa^2 \, \int d^D x \, \sgb \, \Big(\partial_t \, \mathrm Z_{Nk} \, (- \bar {R}(\bar {g}) + 2 \, \bar 
\lambda_k ) + 2 \, \mathrm Z_{Nk} \, \partial_t \,\bar \lambda_k \Big) = \\ 
\ee
\be
\hspace*{-5mm}
\begin{aligned}
&= \mathrm{tr}_T (I) \,\,\,\frac{1}{L} \,\,\, \sum_n \,\,\, u_D \, 
\int_0^{\infty} \, dz \, z^{(D-3)/2} \, (d_k(z;\bar \lambda_k))^{m+1} \, \int d^D \, x \,\, \sgb\\
&- \mathrm{tr}_T (I) \,\,\,C_T\,\,\, \frac{m+1}{k^2}\,\,\,\frac{1}{L} \,\,\, \sum_n \,\,\, u_D\, \int_0^{\infty} \, dz \, z^{(D-3)/2} \, (d_k(z;\bar \lambda_k))^{m+2} \, \int d^D \, x \,\, \sgb \,\, \bar R \\
&+\mathrm{tr}_T (I) \,\,\,\frac{(D/2 - 1)}{6}\,\,\,\frac{1}{L} \,\,\, \sum_n \,\,\, u_D\, \int_0^{\infty} \, dz \, z^{(D-5)/2} \,(d_k(z;\bar \lambda_k))^{m+1} \, \int d^D \, x \,\, \sgb\,\, \bar R \\
\end{aligned}
\nonumber
\ee
\be
\hspace*{-5mm}
\begin{aligned}
&+ \mathrm{tr}_S (I) \,\,\,\frac{1}{L} \,\,\, \sum_n \,\,\, u_D\, 
\int_0^{\infty} \, dz \, z^{(D-3)/2} \,(d_k(z;\bar \lambda_k))^{m+1} \, \int d^D \, x \,\, \sgb \\
&- \mathrm{tr}_S (I) \,\,\,C_S\,\,\, \frac{m+1}{k^2}\,\,\,\frac{1}{L} \,\,\, \sum_n \,\,\, u_D\, \int_0^{\infty} \, dz \, z^{(D-3)/2} \, (d_k(z;\bar \lambda_k))^{m+2} \, \int d^D \, x \,\, \sgb \,\, \bar R \\
&+ \mathrm{tr}_S (I) \,\,\,\frac{(D/2 - 1)}{6}\,\,\,\frac{1}{L} \,\,\, \sum_n \,\,\,u_D\, \int_0^{\infty} \, dz \, z^{(D-5)/2} \, (d_k(z;\bar \lambda_k))^{m+1} \, \int d^D \, x \,\, \sgb \,\, \bar R \\
\end{aligned}
\nonumber
\ee
\be
\hspace*{-5mm}
\begin{aligned}
&- 2 \,\,\, \mathrm{tr}_V (I) \,\,\,\frac{1}{L} \,\,\, \sum_n \,\,\, u_D
\, \int_0^{\infty} \, dz \, z^{(D-3)/2} \, (d_k(z;0))^{m+1} \, \int d^D \, x \,\, \sgb \\
&+ 2 \,\,\, \mathrm{tr}_V (I) \,\,\,C_V\,\,\, \frac{m+1}{k^2}\,\,\,\frac{1}{L} \,\,\, \sum_n \,\,\,u_D\, \int_0^{\infty} \, dz \, z^{(D-3)/2} \, (d_k(z;0))^{m+2} \, \int d^D \, x \,\, \sgb \,\, \bar R \\
&- 2 \,\,\,\mathrm{tr}_V (I) \,\,\,\frac{(D/2 - 1)}{6}\,\,\,\frac{1}{L} \,\,\, \sum_n \,\,\, u_D \, 
\int_0^{\infty} \, dz \, z^{(D-5)/2} \, (d_k(z;0))^{m+1} \, \int d^D \, x \,\, \sgb \,\, \bar R \,\,\,.\\
\end{aligned}
\nonumber
\ee
where the abbreviations 
\bea
\label{eq:uk}
u_D&:=& \frac {(4 \,  \pi)^{-(D-1)/2}}{\Gamma \, ((D-1)/2)} \, ,\nonumber  \\
d_k(z;\bar \lambda_k) &:=&  \Bigg( \frac{k^2}{z + \big(\frac{2\,\pi\,n}{L}\big)^2 + k^2 - 2\,
\bar{\lambda_k}} \Bigg)\, ,
\eea
have been used.

The integrals over $z$ can be done straightforwardly:
\be
\hspace{-4cm} 
2 \, \kappa^2 \, \int d^D x \, \sgb\, \Big(\partial_t \, \mathrm Z_{Nk} \, (- \bar {R}(\bar {g}) + 2 \, \bar 
\lambda_k ) + 2 \, \mathrm Z_{Nk} \, \partial_t \,\bar \lambda_k \Big) = \nonumber \\ 
\ee
\be
\begin{aligned}
&= (\mathrm{tr}_T (I) +\mathrm{tr}_S (I)) \,\,\,(4 \, \pi)^{-(D-1)/2} \,\,\,
\frac{1}{L} \,\,\, \sum_n \,\,\,  \frac{\Gamma \, (-(D-3)/2 + m)}{\Gamma \, (m+1)} \,\,\, 
(k^2)^{m+1} \,\times\\
&   \times 
\Bigg( \Big(\frac{2\,\pi\,n}{L}\Big)^2 + k^2 - 2\,\bar{\lambda_k} \Bigg)^{(D-3)/2 \,-\, m}
 \, \int d^D \, x \,\, \sgb \\
&- (\mathrm{tr}_T (I) \, C_T +\mathrm{tr}_S (I) \, C_S) \,\,\,(4 \, \pi)^{-(D-1)/2} \,\,\,
\frac{1}{L} \,\,\, \sum_n \,\,\, \frac{\Gamma \, (-(D-5)/2 + m)}{\Gamma \, (m+1)} \,\,\, 
(k^2)^{m+1} \,\times \\
&   \times 
\Bigg( \Big(\frac{2\,\pi\,n}{L}\Big)^2 + k^2 - 2\,\bar{\lambda_k} \Bigg)^{(D-5)/2 \,-\, m} \, \int d^D \, x 
\,\, \sgb\,\, \bar R \\
&+ (\mathrm{tr}_T (I) + \mathrm{tr}_S (I)) \,\,\,(4 \, \pi)^{-(D-1)/2} \,\,\,
\frac{1}{L} \,\,\, \sum_n \,\,\, \frac{D-2}{6 \, (D-3)} \,\,\,
 \frac{\Gamma \, (-(D-5)/2 + m)}{\Gamma \, (m+1)} \,\,\, (k^2)^{m+1} \,\times \\
&   \times 
\Bigg( \Big(\frac{2\,\pi\,n}{L}\Big)^2 + k^2 - 2\,\bar{\lambda_k} \Bigg)^{(D-5)/2 \,-\, m} \, 
\int d^D \, x \,\, \sgb\,\, \bar R \\
&- 2 \,\,\, \mathrm{tr}_V (I) \,\,\,(4 \, \pi)^{-(D-1)/2} \,\,\,
\frac{1}{L} \,\,\, \sum_n \,\,\, \frac{\Gamma \, (-(D-3)/2 + m)}{\Gamma \, (m+1)} \,\,\, 
(k^2)^{m+1} \,\times  \\
&   \times 
\Bigg( \Big(\frac{2\,\pi\,n}{L}\Big)^2 + k^2 \Bigg)^{(D-3)/2 \,-\, m} \, 
\int d^D \, x \,\, \sgb \\
&+ 2 \,\,\, \mathrm{tr}_V (I) \, C_V \,\,\,(4 \, \pi)^{-(D-1)/2} \,\,\,
\frac{1}{L} \,\,\, \sum_n \,\,\, \frac{\Gamma \, (-(D-5)/2 + m)}{\Gamma \, (m+1)} \,\,\, 
(k^2)^{m+1} \,\times
 \\
&   \times 
\Bigg( \Big(\frac{2\,\pi\,n}{L}\Big)^2 + k^2 \Bigg)^{(D-5)/2 \,-\, m} \, 
\int d^D \, x \,\, \sgb\,\, \bar R \\
&- 2 \,\,\, \mathrm{tr}_V (I) \,\,\,(4 \, \pi)^{-(D-1)/2} \,\,\,
\frac{1}{L} \,\,\, \sum_n \,\,\, \frac{D-2}{6 \, (D-3)} \,\,\, \frac{\Gamma \, (-(D-5)/2 + m)}
{\Gamma \, (m+1)} \,\,\, (k^2)^{m+1} \,\times  
\\
&   \times 
\Bigg( \Big(\frac{2\,\pi\,n}{L}\Big)^2 + k^2 \Bigg)^{(D-5)/2 \,-\, m} \, \int d^D \, x \,\, \sgb \,\, \bar R \,\,.\
\end{aligned}
\ee
\goodbreak
This equation can be simplified
\be
\label{eq:C.22}
\hspace{-3cm} 
2 \, \kappa^2 \, \int d^D x \, \sgb\, 
\Big(\partial_t \, \mathrm Z_{Nk} \, (- \bar {R}(\bar {g}) + 2 \, \bar \lambda_k )
 + 2 \, \mathrm Z_{Nk} \, \partial_t \,\bar \lambda_k \Big) = \nonumber  \\ 
\ee
\be
\begin{aligned}
&= \int d^D \, x \,\, \sgb\,\,\,\Bigg\{ (4 \, \pi)^{-(D-1)/2} \,\,\, \frac{(k^2)^{(D-1)/2}
\Gamma \, (-\frac{(D-3)}{2} + m)}{\Gamma \, (m+1)} \,\,\,
\times  \\
&\qquad\qquad \times \ \frac{1}{L} \,\,\, \sum_n \,\,\,  \quad \Bigg[ \frac{1}{2} \,\, D \, (D+1) \,\, \big( \omega_n^2 + 1 - 2\,  \lambda_k \big)^{(D-3)/2 \,-\, m}
\\ 
& \qquad\qquad \qquad\qquad \qquad
 - 2 \, D \,\, \big( \omega_n^2 + 1 \big)^{(D-3)/2 \,-\, m} \Bigg] \Bigg\} \\
\end{aligned}
\ee
\be
\begin{aligned}
&+ \int d^D \, x \,\, \sgb \,\, \bar R \,\,\,\Bigg\{ (4 \, \pi)^{-(D-1)/2} \,\,\, \frac{(k^2)^{(D-3)/2}
\Gamma \, (-\frac{(D-5)}{2} + m)}{\Gamma \, (m+1)} \,\,\,
\times \nonumber \\
&\qquad\qquad \times \,\,\,\frac{1}{L} \,\,\, \sum_n \,\,\, \Bigg[ \frac{-\,D\,\,(5\,D^2 - 23\,D + 20)}{12 \, (D-3)} \,\, 
\big( \omega_n^2 + 1 - 2\, \lambda_k \big)^{(D-5)/2 \,-\, m}  \\ 
& \qquad\qquad \qquad\qquad + \Bigg( \frac{-D^2 - 4D + 18}{3 \, (D-3)} \Bigg) \,\, \big( \omega_n^2 
+ 1 \big)^{(D-5)/2 \,-\, m} \Bigg] \Bigg\} 
\end{aligned}
\ee
where the definition (\ref{eq:4.13}) for $\omega_n^2$ has been used.

\section[Exact Functional Identity: Dimensional Reduction]
{Exact Functional Identity: Dimensional \\Reduction}

Starting point is the expression (\ref{eq:D.20}) but now the traces are performed
with the help of either Eq.\ (\ref{eq:4.9}) or (\ref{eq:4.10}).  After some algebra 
and using the definitions  (\ref{eq:uk}) or (\ref{eq:barck})
as well as 
\be
\tilde t_k(z,n) := -\eta_{Nk} (k^2-(2\pi n/L)^2-z)+2k^2 \quad {\mathrm {and}} \quad
\tilde q^2 := k^2-(2\pi n/L)^2
\ee
one  arrives then at
\be
\hspace{-4cm} 
2 \, \kappa^2 \, \int d^D x \, \sgb \, \Big(\partial_t \, \mathrm Z_{Nk} \, (- \bar {R}(\bar {g}) + 2 \, \bar 
\lambda_k ) + 2 \, \mathrm Z_{Nk} \, \partial_t \,\bar \lambda_k \Big) = 
\nonumber  \\ 
\ee
\be
\hspace*{-5mm}
\begin{aligned}
=&\int d^D \, x \,\, \sgb\,\,  \,\,\, \frac{1}{L} \,\,\, \sum_{n^2\le (kL/2\pi)^2} \,\,\, 
u_D \,  \int_0^{\tilde q^2} \, dz \, z^{(D-3)/2} \, 
\,  \left(  \frac 1 4 D(D+1)  
\tilde t_k(z,n)\bar c_k  -2D   \right)  \\
+&\int d^D \, x \,\, \sgb\,\, \bar R \,\,\,\,\, 
 \frac{1}{L} \,\,\, \sum_{n^2\le (kL/2\pi)^2} \,\,\, 
 u_D \,  \int_0^{\tilde q^2} \, dz \, z^{(D-3)/2} \,  \\ &\hspace{40mm}
 \Bigg[ \left((-\frac 1 4 (D-1)(D+2)C_T-\frac 12 C_S )  \tilde t_k(z,n)\bar c_k ^2 +2DC_V 
 \right) 
  \\ 
 & ~ \hspace*{30mm} +  \frac 1 {z+\textcolor{red}{(2\pi n/L)^2}}
\,\, \left(  (\frac 1 {48} D(D+1)(D-2) )  \tilde t_k(z,n)\bar c_k -\frac 1 6 D(D-2) 
 \right) \Bigg]  \\
\end{aligned}
\ee
where the difference in the two types of traces (\ref{eq:4.9}) or (\ref{eq:4.10}) is in keeping 
or neglecting the term in red. Setting $D=5$ and performing the integrals over $z$ one arrives 
straightforwardly at the expressions for given in Sect.~4.2.  There the following definition has been used
\be
s_l(a)=\sum_{-[a]}^{[a]} \left( 1 - \left( \frac n a \right)^2 \right)^l \, , \quad l=1,2,3,
\label{eq:D.39}
\ee
where $a=kL/2\pi$ and $s_l(a)=1$ for $a<1$. The sums can be performed:
\bea
s_1(a)&=&(1+2[a]) \left ( 1 - \frac {[a]} {3a^2} (1+[a])\right) \, ,
\nonumber \\ 
s_2(a) &=&  (1+2[a])  \left ( 1 - \frac {[a]} {15a^4}  (1+[a]) \Big(1+10a^2-3[a](1+[a])\Big)\right) \, ,
\nonumber \\
s_3(a) &=& (1+2[a])  \Bigg( 1 - \frac {[a]}  {105a^6}  (1+[a]) \times  \label{eq:D.40} \\
&&\hspace{10mm} \times \Big( 5+21(a^2+5a^4)+3[a](1+[a])
\big(-5-21a^2+5[a](1+[a])\big) \Big) \Bigg) \, , \nonumber \eea
the limits for $a\to \infty$ are
\bea
s_1(a)&\to&~\frac 4 3 \, a \, , \nonumber \\
s_2(a)&\to& \frac {16}{15} \, a \, , \nonumber \\
s_3(a)&\to& \frac {32}{35} \, a\, . \label{eq:D.41}
\eea


\newpage
\thispagestyle{empty}

\chapter{{\color{BrickRed} Threshold Functions}}

Following Refs.\ \cite{Berges:2000ew,Litim:2001up} in this Appendix the so-called threshold
functions are introduced and some of their properties are discussed. First, the case without
dimensional reduction is considered.

The basic threshold function in  $D$ dimensions is defined as 
\be
l_0^D(w) := \frac 1 4 v_D^{-1} k^{-D} \int \frac{d^Dq}{(2\pi)^D} \,\, \frac{\partial_t (R_k(q^2)/Z_k)}
{q^2+R_k(q^2)/Z_k+k^2w}
\ee
where $v_D^{-1}=2(4\pi )^{D/2} \Gamma(D/2)$ and the regulator $R_k$ is defined in Eq.\ (\ref{eq:C.13}).
Comparing this expression with the ones given in Appendix~D.2 reveals that in the cases treated 
in this thesis one has $w=-2\lambda_k$.
The substitution $y=q^2/k^2$ and performing the angular integrals yields
\be
l_0^D(w) = \frac 1 2 \int_0^\infty dy\, y^{D/2-1} \frac {y \,\partial_t \,r(y)}{y(1+r(y))+w}.
\ee
For the Litim regulator, {\it i.e.}, for $r_{opt} \, (y) =  (  {1}/{y} - 1 ) \,\, \Theta \,\, ( 1 - y ) $, one obtains
\be
l_0^D(w) = \frac 1 {1+w}  \int_0^1 dy\, y^{D/2-1} = \frac 2 D  \frac 1 {1+w} .
\ee
Higher threshold functions can be defined by taking derivatives 
\be
\partial_w l_m^D(w) = -(\delta_{m,0}+m) \, l_{m+1}^D(w)
\ee
or equivalently
\be
l_{m}^D(w) = \frac 1 2 (\delta_{m,0}+m)  \int_0^\infty dy\, y^{D/2-1} 
\frac {y \,\partial_t \,r(y)}{(y(1+r(y))+w)^{m+1}}.
\ee

In the case of dimensional reduction applying  the substitution (\ref{eq:4.3}),
\be
\int \frac{d^Dq}{(2\pi)^D} \rightarrow \frac{1}{L} \, \sum_n \, \int \frac{d^{D-1}q}{(2\pi)^{D-1}} \,\, ,
\ee
motivates to introduce the following generalised threshold function\footnote{In Ref.\ \cite{Litim:2001up}
those functions have been introduced for non-vanishing temperatures $T$. Substituting $T\to 1/L$ and 
$ \tau \to 2\pi /kL$ in the corresponding expressions of Ref.\ \cite{Litim:2001up} provides the ones
given here.}
\be
l_0^D(w,L)= \frac {v_{D-1}}{v_D} \frac 1 {kL} \sum_n \int_0^\infty dy \, y^{D/2-3/2} 
\frac{-(y+\omega_n^2)^2\, r^\prime (y+\omega_n^2)}{(y+\omega_n^2)(1+r(y+\omega_n^2))+w}
\ee
where $\partial_t r(y) = -2yr^\prime (y)$ has been substituted and the definition (\ref{eq:4.13}), {\it i.e.},
$\omega_n = 2\pi n /kL$ has been used.
The higher threshold functions are again related to $l_0^D(w,L)$ by deriving it $n$ times. By construction their limiting values are given by:
\bea
l_m^D(w,L\to \infty ) &=& l_m^D(w) \, ,\nonumber \\
l_m^D(w,L\to 0 ) &=&  \frac {v_{D-1}}{v_D}  \frac 1 {kL}  l_m^{D-1} (w)\, .
\eea
The latter limit is evident from the fact that for $L\to 0$ only the zero mode contributes to the sum.

For the Litim regulator the dependence on $L$ factorizes,
\be
l_0^D(w,L) = \frac{2}{D-1} \frac 1 {1+w}   \frac {v_{D-1}}{v_D} \frac 1 {kL} \sum_n (1-\omega_n^2)^{D/2-1/2}
\Theta (1-\omega_n^2)
\ee
and for higher threshold functions
\be
l_m^D(w,L) = \frac{D}{D-1} l_m^D(w) \frac {v_{D-1}}{v_D} \frac 1 {kL} \sum_n (1-\omega_n^2)^{D/2-1/2}
\Theta (1-\omega_n^2) .
\ee
For $D=5$ the sum can be performed explicitly,
\be
\sum_n (1-\omega_n^2)^2 \Theta (1-\omega_n^2) = 
\sum_{n=-[a]}^{[a]} (1-n^2/a^2)^2  =s_2(a)
\ee
with $a=kL/2\pi$ and $s_2(a)$ being defined in Eq.\ (\ref{eq:D.39}), see also the explicit form 
(\ref{eq:D.40}).


\chapter{{\color{BrickRed}Numerical treatment}}

In Chapters 3 and 4 the UV fixed points have
been determined from the roots of the $\beta$-functions
with the SOLVE function of MATHEMATICA. The RG flows have been
calculated   by solving the differential equations by 
either using the function NDSOLVE of MATHEMATICA or by a FORTRAN code containing a
self-written fourth-order Runge-Kutta algorithm. In the cases both methods have been
applied to the same parameter set the results agreed very well. Also other numerical
checks have been passed with high precision, {\it e.g.}, the numerical
solution of the differential equations reproduced the UV fixed point values typically with a 
precision of ${\cal O}(10^{-8})$. Where available I compared my results to those 
available in  the literature, and was therefore able to verify them.

In most cases the flow has been integrated from the IR to the UV. It has been checked 
that the opposite direction of the flow generates the same trajectory as long as one does
not start too close to the UV fixed point. Typically the flow has been started at $k_{min}=0.1$ and 
terminated at $k_{max}=10^5$ with ${\cal O} (10^4)$ steps in between,
except for Figs.\ \ref{fig:4.3} and \ref{fig:4.7} where $k_{min}=10^{-21}$ and $k_{max}=10^{45}$
has been used with $10^7$ Runge-Kutta steps. 

In cases the sums over Kaluza-Klein modes could not be expressed in a closed form
they have been performed numerically.
As the sums originating from the  exact functional identity
are finite this has been straightforwardly possible. For $kL>10^8$ the sums then have been 
simply been substituted by the limiting expressions. For the sums within approximate 
background field flow  convergence 
has been tested and a numerical accuracy at the level of $10^{-8}$ was obtained by 
adjusting the numerical limits of the sums accordingly.

\newpage
\thispagestyle{empty}


\cleardoublepage
\phantomsection
\addcontentsline{toc}{chapter}{{\color{BrickRed}References}}


\newpage

\thispagestyle{empty}

\cleardoublepage
\phantomsection
\addcontentsline{toc}{chapter}{{\color{BrickRed}Acknowledgements}}

\chapter*{{\color{BrickRed}Acknowledgements}}

Many persons have contributed to the work of this thesis, I am grateful to all of them.

First of all, with great pleasure I thank my advisor, Daniel F. Litim, who introduced me 
to such an overwhelmingly and outstandingly interesting topic! He always made the most 
complicated 
problems look easier to me. It was delightful to have the opportunity to work with somebody 
like Daniel! I really enjoyed all our discussions, being in Sussex, Graz, or anywhere else.

Secondly, I thank my co-advisor, Bernd-Jochen Schaefer, for the helpful discussions, hints and the 
critical reading of my thesis.

During the work of this thesis I had gladly many occasions to discuss with very interesting physicists,
and for some of them I want to take the chance to mention explicitly here: Holger Gies, Jan Martin Pawlowski, 
Frank Saueressig and my good old friend, John Swain. Here, specially, I would like to thank Jan for  many conversations and advices.

My thanks also go to my colleagues from the SICQFT group in Graz, but one person in particular, H\`{e}lios Sanchis-Alepuz (right now at the University of Giessen), with whom I had 
joyful discussions about classical gravity.

As part of this work was also done at the University of Sussex, where I spent several very pleasant visits, I would 
like to thank the Department of Physics and Astronomy and my colleagues there for their kind hospitality. 
In special, I want to mention Kevin Geoffrey Falls and Edouard Marchais for their very helpful  and interesting discussions in 
the beginning of my thesis. It was always a lot of fun to talk with them!

Financial support from the Paul-Urban Foundation at the University of Graz is acknowledged.

Last, but not least, I would like to thank my family for all their support and trust! And of course, my beloved husband, who was always there for me, specially when I was deeply drowning in my higher dimensional world... Thank you for all the energy, support and love when I most needed! 

\newpage
\thispagestyle{empty}


\end{document}